\documentclass[preprint,11pt,3p,times]{elsarticle}

\usepackage[group-separator={,},exponent-product = \cdot]{siunitx}
\usepackage{amsmath}
\usepackage[mathscr]{eucal}
\usepackage{graphicx}
\usepackage{svg}
\usepackage{caption}
\usepackage{subfigure}
\usepackage{hyphenat}
\usepackage{bm}
\usepackage{upgreek}
\usepackage{adjustbox}
\usepackage[ruled]{algorithm2e}

\usepackage{multirow}

\usepackage{url}
\usepackage{color}
\definecolor{colUniBwOr}{rgb}{0.929,0.431,0.0} 
\definecolor{colUniBwGr}{RGB}{113,112,114} 
\definecolor{matlabBlue}{rgb}{0 0.4470 0.7410} 
\definecolor{matlabOrange}{rgb}{0.8500 0.3250 0.0980} 
\definecolor{matlabYellow}{rgb}{0.9290 0.6940 0.1250} 
\definecolor{matlabGreen}{rgb}{0.4660 0.6740 0.1880} 

\usepackage{tikz}
\usepackage{tikz-3dplot}
\usepackage{pgfplots}
\usetikzlibrary{shapes,arrows,calc,fit,backgrounds,shapes.multipart}

\usepackage{pgfplots}
\pgfplotsset{compat = newest}

\usepackage{amsthm}
\theoremstyle{definition}

\newtheoremstyle{remarkstyle} 
  {}{}{}{}{\bfseries}{.}{.5em}{{\thmname{#1 }}{\thmnumber{#2}}{\thmnote{ (#3)}}}
\theoremstyle{remarkstyle}
\newtheorem{definition}{Definition}[section]
\newtheorem{remark}{Remark}[section]

\graphicspath{{figures/}}

\newcommand{\secref}[1]{Section~\ref{#1}} 
\newcommand{\secsref}[2]{Sections~\ref{#1} and~\ref{#2}} 
\newcommand{\secsrangeref}[2]{Sections~\ref{#1} -- \ref{#2}} 
\newcommand{\defref}[1]{Definition~\ref{#1}} 
\newcommand{\figref}[1]{Figure~\ref{#1}} 
\newcommand{\figsref}[2]{Figures~\ref{#1} and~\ref{#2}} 
\newcommand{\Figref}[1]{Figure~\ref{#1}} 
\newcommand{\tabref}[1]{Table~\ref{#1}} 
\newcommand{\algref}[1]{Algorithm~\ref{#1}} 

\newcommand{\teo}[1]{\ensuremath{\boldsymbol{#1}}} 
\newcommand{\mao}[1]{\ensuremath{\mathbf{#1}}} 
\newcommand{\tet}[1]{\ensuremath{\boldsymbol{#1}}} 
\newcommand{\mat}[1]{\ensuremath{\mathbf{#1}}} 

\newcommand{\lagMultDiscVec}{\bm{\uplambda}} 

\newcommand{\id}{\ensuremath{I}} 
\newcommand{\idmat}{\mat{\id}} 

\newcommand{\mr}[1]{\ensuremath{\mathrm{#1}}} 

\newcommand{\indexedIter}[2]{#1^{#2}} 
\newcommand{\indexedRow}[2]{#1_{#2}} 
\newcommand{\indexedRowCol}[3]{#1_{#2#3}} 

\newcommand{\indexArbitraryOne}{a}
\newcommand{\indexArbitraryTwo}{b}

\newcommand{\indexBeam}{\mathcal{B}}
\newcommand{\indexSimoReissner}{\mathrm{SR}}
\newcommand{\indexSolid}{\mathcal{S}}
\newcommand{\indexSolidBeam}{\mathcal{BS}}
\newcommand{\indexTorsionFree}{\mathrm{TF}}

\newcommand{\indexedDomain}[2]{#1^{#2}}
\newcommand{\indexedDomainSweep}[3]{#1^{#2,#3}}
\newcommand{\indexedDomainMultiSweep}[4]{#1^{#2,#3,#4}}
\newcommand{\indexedBeam}[1]{\indexedDomain{#1}{\indexBeam}}
\newcommand{\indexedBeamSweep}[2]{\indexedDomainSweep{#1}{\indexBeam}{#2}}
\newcommand{\indexedBeamMultiSweep}[3]{\indexedDomainMultiSweep{#1}{\indexBeam}{#2}{#3}}
\newcommand{\indexedSolid}[1]{\indexedDomain{#1}{\indexSolid}}
\newcommand{\indexedSolidSweep}[2]{\indexedDomainSweep{#1}{\indexSolid}{#2}}

\newcommand{\indexedDispCoupling}[1]{{#1}^{\mathcal{V}}}
\newcommand{\indexedRotCoupling}[1]{{#1}^{\mathcal{R}}}
\newcommand{\indexedSolidBeam}[1]{\indexedDomain{#1}{\indexSolidBeam}}

\newcommand{\inv}[1]{#1^{-1}} 
\newcommand{\trans}[1]{#1^{\mr{T}}} 

\newcommand{\diagElement}{d}
\newcommand{\diagEntry}[1]{\diagElement_{#1}}
\newcommand{\diag}[1]{\mathrm{diag}\left(#1\right)}

\newcommand{\textfrac}[2]{#1/#2} 

\newcommand{\abs}[1]{\left |#1\right |}
\newcommand{\norm}[1]{\left\|#1\right\|} 
\newcommand{\normOne}[1]{\norm{#1}_{1}} 
\newcommand{\normTwo}[1]{\norm{#1}_{2}} 
\newcommand{\normFrobenius}[1]{\norm{#1}_{\mathrm{F}}} 

\newcommand{\anyQuantity}{\left(\bullet\right)} 
\newcommand{\anyMatrix}{H} 

\newcommand{\volume}{V}
\newcommand{\beamRadius}{R}
\newcommand{\geometryRatio}{\beamRadius/\thickness}
\newcommand{\stiffnessRatio}{\indexedBeam{\youngs} / \indexedSolid{\youngs}}

\newcommand{\disp}{d} 
\newcommand{\youngs}{E} 
\newcommand{\poisson}{\nu} 

\newcommand{\gap}{g} 
\newcommand{\penaltyParam}{\epsilon} 
\newcommand{\penaltyParamDisp}{\indexedDispCoupling{\penaltyParam}} 
\newcommand{\penaltyParamRot}{\indexedRotCoupling{\penaltyParam}} 

\newcommand{\indLinIter}{n} 
\newcommand{\indSweep}{k} 
\newcommand{\indSpaiSweep}{m}
\newcommand{\indRow}{i} 
\newcommand{\indCol}{j} 
\newcommand{\indBlockRow}{\iota} 
\newcommand{\indBlockCol}{\zeta} 
\newcommand{\indBaseVecOne}{\xi}
\newcommand{\indBaseVecTwo}{\eta}

\newcommand{\numBlockRows}{N_{\mathrm{R}}}

\newcommand{\nproc}{n^\mr{proc}} 
\newcommand{\nRows}{M}

\newcommand{\tSetup}{T_{\mathrm{setup}}}
\newcommand{\tSolve}{T_{\mathrm{solve}}}
\newcommand{\tTotal}{T_{\mathrm{total}}}

\newcommand{\sol}{x} 
\newcommand{\rhs}{b} 

\newcommand{\residualNonlinear}{f}
\newcommand{\residualLinear}{r}

\newcommand{\approxMat}[1]{\widehat{#1}}

\newcommand{\matBeam}{\mat{A}}
\newcommand{\matBeamElement}{a}
\newcommand{\matBeamEntry}[2]{\indexedRowCol{\matBeamElement}{#1}{#2}} 
\newcommand{\matSolid}{\mat{C}}
\newcommand{\matBeamSolid}{\mat{B}_{1}}
\newcommand{\matSolidBeam}{\mat{B}_{2}}
\newcommand{\matIdentity}{\mat{I}}
\newcommand{\matZero}{\mat{0}}

\newcommand{\linOp}{\boldsymbol{\mathcal{A}}}

\newcommand{\matFactD}{\mathcal{D}}
\newcommand{\matFactL}{\mathcal{L}}
\newcommand{\matFactU}{\mathcal{U}}

\newcommand{\myPrec}{P}
\newcommand{\precMatrix}{\mat{\myPrec}}
\newcommand{\schur}{S}
\newcommand{\schurMatrix}{\mat{\schur}}

\newcommand{\graph}{\mathcal{J}}
\newcommand{\graphOf}[1]{\graph\left(#1\right)}

\newcommand{\graphElement}{j}
\newcommand{\graphElementOf}[3]{\graphElement\left(#1\right)_{#2#3}}

\newcommand{\nnz}{\mathrm{nnz}}
\newcommand{\nnzOf}[1]{\nnz\left(#1\right)}

\newcommand{\thresholdingOf}[3]{\mathcal{F}_{#3}\left(#1,#2\right)}
\newcommand{\threshMat}[1]{\underline{#1}}

\newcommand{\spaiRefinementOp}[2]{\mathcal{R}\left(#1,#2\right)} 

\newcommand{\spaiMinimizeOp}[2]{\mathcal{M}\left(#1,#2\right)} 

\newcommand{\spaiDropTol}{\sigma}
\newcommand{\spaiIndRow}{\indRow}
\newcommand{\spaiRefinementLevel}{\ell}

\newcommand{\spaiOf}[1]{{#1}^{\ast}}

\newcommand{\iluDropTol}{\tau}
\newcommand{\iluLevelFill}{p}
\newcommand{\iluOverlap}{\delta}

\newcommand{\thickness}{t} 
\newcommand{\distributedLoad}{q} 

\newcommand{\beamsolid}{beam\hyp{}solid}
\newcommand{\blockLU}{Block\hyp{}LU}

\newcommand{\fiberfluid}{fiber\hyp{}fluid}
\newcommand{\fibersolid}{fiber\hyp{}solid}

\newcommand{\fixedpoint}{fixed\hyp{}point}

\newcommand{\LTwo}{L\textsubscript{2}}

\newcommand{\mixeddimensional}{mixed\hyp{}dimensional}
\newcommand{\mortartype}{mortar\hyp{}type}
\newcommand{\multidimensional}{multi\hyp{}dimensional}
\newcommand{\multigrid}{multigrid}

\newcommand{\multilevel}{multi\hyp{}level}

\newcommand{\multiphysics}{multi\hyp{}physics}

\newcommand{\nonlinear}{nonlinear}

\newcommand{\nonsymmetric}{non\hyp{}symmetric}
\newcommand{\nonzero}{non\hyp{}zero}
\newcommand{\nonzeros}{{\nonzero}s}
\newcommand{\oneDthreeD}{1D/3D}
\newcommand{\oneDtwoD}{1D/2D}
\newcommand{\outofthebox}{out-of-the-box}

\newcommand{\speedup}{speed\hyp{}up}

\newcommand{\subblock}{sub\hyp{}block}
\newcommand{\subblocks}{{\subblock}s}

\newcommand{\threeDthreeD}{3D/3D}

\newcommand{\torsionfree}{torsion\hyp{}free}

\newcommand{\GaussSeidel}{Gau{\ss}--Seidel}
\newcommand{\GreenLagrange}{Green--Lagrange}

\newcommand{\PiolaKirchhoff}{Piola--Kirchhoff}
\newcommand{\KirchhoffLove}{Kirchhoff--Love}
\newcommand{\RugeStueben}{Ruge--St\"uben}
\newcommand{\SimoReissner}{Simo--Reissner}
\newcommand{\StVenantKirchhoff}{St.-Venant--Kirchhoff}

\newcommand{\SoftwarePackage}[1]{\textsc{#1}} 
\newcommand{\baci}{\SoftwarePackage{4C}}
\newcommand{\meshpy}{\SoftwarePackage{MeshPy}}
\newcommand{\muelu}{\SoftwarePackage{MueLu}}
\newcommand{\superlu}{\SoftwarePackage{SuperLU}}
\newcommand{\trilinos}{\SoftwarePackage{Trilinos}}

\newcommand{\ie}{i.e.,}
\newcommand{\etal}{\emph{et al.}}
\newcommand{\eg}{e.g.,}

\newcommand{\cf}{cf.}
\newcommand{\wrt}{w.r.t.}

\usepackage[colorlinks=true,
            linkcolor=black,
            citecolor=black,
            urlcolor=black,
            bookmarksnumbered=true,
            pdfstartpage=1,
            pdfpagemode=UseOutlines]{hyperref}

\usepackage[acronym]{glossaries}
\setacronymstyle{long-short}
\glsdisablehyper 

\newacronym{ac:1D}{1D}{one-dimensional}
\newacronym{ac:2D}{2D}{two-dimensional}
\newacronym{ac:3D}{3D}{three-dimensional}

\newacronym{ac:ALE}{ALE}{Arbitrary Lagrangean Eulerian}

\newacronym{ac:AMG}{AMG}{algebraic {\multigrid}}

\newacronym{ac:CSM}{CSM}{computational solid mechanics}

\newacronym{ac:CutFEM}{CutFEM}{cut finite element method}

\newacronym{ac:DD}{DD}{domain decomposition}

\newglossaryentry{ac:DOF}{ 
  name={DOF},
  description={degree of freedom},
  plural={DOFs},
  first={degree of freedom (DOF)},
  firstplural={degrees of freedom (DOFs)},
  text={DOF}
}

\newacronym{ac:FBI}{FBI}{fluid\hyp{}beam interaction}

\newacronym{ac:FCM}{FCM}{finite cell method}

\newacronym{ac:FEM}{FEM}{finite element method}

\newacronym{ac:FSI}{FSI}{fluid\hyp{}solid interaction}

\newacronym{ac:FVM}{FVM}{finite volume method}

\newacronym{ac:GMG}{GMG}{geometric {\multigrid}}

\newacronym{ac:GPTS}{GPTS}{Gauss\hyp{}point\hyp{}to\hyp{}segment}

\newacronym{ac:GPU}{GPU}{graphical processing unit}

\newacronym{ac:HPC}{HPC}{high\hyp{}performance computing}

\newacronym{ac:IBM}{IBM}{immersed boundary method}

\newacronym{ac:IGA}{IGA}{isogeometric analysis}

\newacronym{ac:ILU}{ILU}{incomplete LU}

\newacronym{ac:LagMult}{LM}{Lagrange multiplier}

\newacronym{ac:MG}{MG}{\multigrid}

\newacronym{ac:MPI}{MPI}{message passing interface}

\newacronym{ac:NURBS}{NURBS}{Non-Uniform Rational B-Splines}

\newacronym{ac:PDE}{PDE}{partial differential equation}

\newacronym{ac:PINN}{PINN}{physics-informed neural network}

\newacronym{ac:PI}{PI}{principal investigator}

\newacronym{ac:PAAMG}{PA\hyp{}AMG}{plain\hyp{}aggregation algebraic {\multigrid}}

\newacronym{ac:SAAMG}{SA\hyp{}AMG}{smoothed\hyp{}aggregation algebraic {\multigrid}}

\newacronym{ac:SMG}{SMG}{structured {\multigrid}}

\newacronym{ac:SPAI}{SPAI}{sparse approximate inverse}

\newacronym{ac:VEM}{VEM}{virtual element method}

\newacronym{ac:WP}{WP}{work package}

\newacronym{ac:XFEM}{XFEM}{extended finite element method}

\makeatletter
\def\ps@pprintTitle{%
  \let\@oddhead\@empty
  \let\@evenhead\@empty
  \def\@oddfoot{\reset@font\hfil\thepage\hfil}
  \let\@evenfoot\@oddfoot
}
\makeatother

\begin{document}

\graphicspath{{figures/}{../figures/}}

\begin{frontmatter}

\title{An approximate block factorization preconditioner for {\mixeddimensional} {\beamsolid} interaction}

\author[imcs]{Max Firmbach}\ead{max.firmbach@unibw.de}
\author[imcs]{Ivo Steinbrecher}\ead{ivo.steinbrecher@unibw.de}
\author[imcs]{Alexander Popp}\ead{alexander.popp@unibw.de}
\author[imcs,dsc]{Matthias Mayr\corref{cor1}}\ead{matthias.mayr@unibw.de}
\address[imcs]{Institute for Mathematics and Computer-Based Simulation, Universit\"{a}t der Bundeswehr M\"{u}nchen,\\Werner-Heisenberg-Weg 39, D-85577 Neubiberg, Germany}
\address[dsc]{Data Science \& Computing Lab, Universit\"{a}t der Bundeswehr M\"{u}nchen,\\Werner-Heisenberg-Weg 39, D-85577 Neubiberg, Germany}

\cortext[cor1]{corresponding author}

\begin{abstract}
This paper presents a scalable approximate block factorization preconditioner for {\mixeddimensional} models in {\beamsolid} interaction and their application in engineering.
In particular, it studies the linear systems arising from a regularized mortar-type approach
for embedding geometrically exact beams into solid continua.
Due to the lack of block diagonal dominance of the arising $2\times 2$ block system,
an approximate {\blockLU} preconditioner is used.
It exploits the sparsity structure of the beam {\subblock} to construct a sparse approximate inverse,
which is then not only used to explicitly form an approximation of the Schur complement,
but also acts as a smoother within the prediction and correction step of the arising {\blockLU} preconditioner.
The Schur complement equation is tackled with an algebraic multigrid method.
Although, for now, the beam {\subblock} is tackled by a one-level method only,
the {\multilevel} nature of the computationally demanding Schur equation delivers a scalable preconditioner in practice.
In numerical test cases,
the influence of different algorithmic parameters on the quality of the sparse approximate inverse is studied
and the weak scaling behavior of the proposed preconditioner on up to 1000 MPI ranks is demonstrated.
In addition, the robustness of the proposed method regarding
material parameters and geometric properties is shown,
before the preconditioner is finally applied for the analysis of steel-reinforced concrete structures in civil engineering.
\end{abstract}

\begin{keyword}
Approximate block factorization \sep physics-based block preconditioning \sep algebraic multigrid \sep sparse approximate inverse \sep {\mixeddimensional} modeling \sep {\beamsolid} interaction
\end{keyword}
\end{frontmatter}


\section{Introduction}
\label{sec:Introduction}

Originating from evolution in nature or human design processes,
thin fibers embedded into solid continua can enhance the constitutive or functional properties of systems in science, engineering, and biomedicine.
Applications can be found in different fields:
In civil engineering for example, steel-reinforced concrete is used to amplify the load bearing capacity of concrete structures such as bridges.
In aerospace engineering, fiber-reinforced composite materials are often used due to their unique combination of a high stiffness, but low specific weight.
In biological tissues, collagen fibers are distributed throughout the arterial walls of the circulatory system.
For all these application areas,
finite element simulations can provide detailed insight in the system's behavior and potentially assist in improving or optimizing the system's design.
While various mathematical models of fiber-enhanced continua are available in literature,
their efficient solution on parallel computing clusters has not been studied in detail.
To this end,
this paper sets out to develop an efficient and scalable {\multilevel} block preconditioning framework
for a penalty\hyp{}regularized {\mixeddimensional} approach recently proposed by Steinbrecher {\etal}~\cite{Steinbrecher2020a,Steinbrecher2021a}.

\begin{figure}
\centering
\subfigure[Homogenization]{\label{fig:ModelingHomogenization}\includegraphics[width=0.27\textwidth]{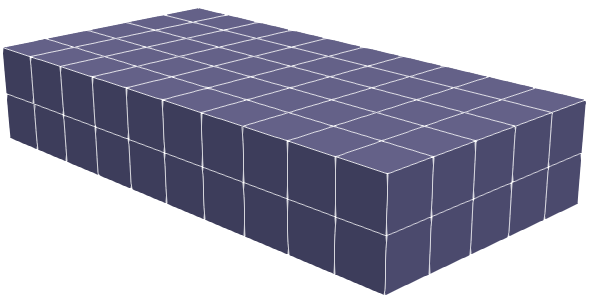}}
\hfill
\subfigure[Embedded fibers in continuum]{\label{fig:ModelingBeamToVolume}\includegraphics[width=0.27\textwidth]{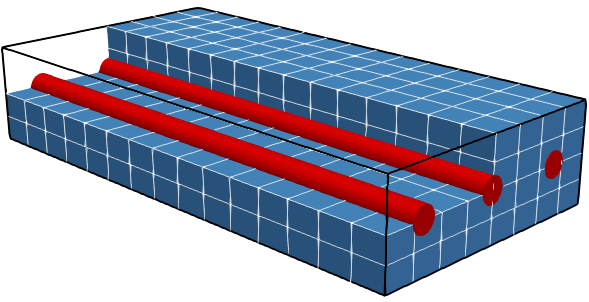}}
\hfill
\subfigure[Fully resolved 3D model]{\label{fig:ModelingFullyResolved}\includegraphics[width=0.27\textwidth]{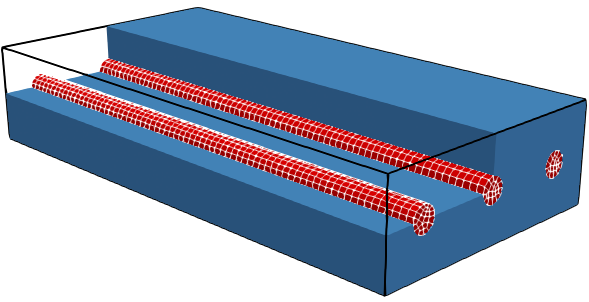}}
\caption{Spectrum of modeling techniques for fibers embedded into three-dimensional solids~\cite{Steinbrecher2020a}}
\label{fig:DifferentModelingApproaches}
\vspace{-1.5em plus .2em minus 0.5em}
\end{figure}

\Figref{fig:DifferentModelingApproaches} illustrates the range of modeling techniques for fully embedded fibers in solid bulk volumes.
On the one hand, homogenized formulations as depicted in \figref{fig:ModelingHomogenization}
incorporate all fiber information into the bulk constitutive law,
usually leading to anisotropic formulations with preferential directions along the fiber orientation~\cite{Agarwal2017a,Raghavan2005a}.
On the other end of the spectrum,
both bulk field and fibers are resolved as \gls{ac:3D} continua, {\cf} \figref{fig:ModelingFullyResolved},
which allows the reuse of existing constitutive models and finite element technology from classical \gls{ac:3D} computational solid dynamics.
This approach enables the analysis of very detailed micro-mechanical features of individual fibers
and the incorporation of advanced physical effects at the {\fibersolid} interface,
though it comes at significant computational cost.
Finally, to unify the high model quality of fully resolved models with the efficiency of homogenized models,
fibers can be represented by dimensionally reduced structural models such as beams or trusses,
which are embedded at arbitrary positions into a \gls{ac:3D} solid domain ({\cf} \figref{fig:ModelingBeamToVolume}).
In such {\mixeddimensional} {\oneDthreeD} models,
the bulk field still portrays the same effects as in the fully resolved 3D model,
but fibers are now reduced to a computationally efficient \gls{ac:1D} representation,
making such models good candidates to study fiber-enhanced continua \emph{at large scale}.

For the enforcement of coupling conditions in {\mixeddimensional} models,
embedded mesh techniques are required.
Although the imposition of constraints through Lagrange multipliers is well established in computational solid mechanics
for both boundary-fitted meshes~\cite{Puso2004c,Flemisch2007a} and embedded meshes~\cite{Bechet2009a,Hautefeuille2012a}
as well as in computational contact mechanics~\cite{Puso2004a,Puso2004b,Popp2010a,Wiesner2021a},
the construction of stable Lagrange multiplier spaces for {\mixeddimensional} {\beamsolid} coupling still poses an open research question.
Consequently, {\mixeddimensional} models for solid problems so far
either use a penalty regularization to enforce the {\fibersolid} coupling constraints~\cite{Durville2007a,Steinbrecher2020a,Steinbrecher2021a,Steinbrecher2022b,Kakaletsis2023},
employ a variationally consistent overlapping domain decomposition approach~\cite{Khristenko2021a},
or directly link embedded fibers to the surrounding volume discretization via the \gls{ac:XFEM}~\cite{Le2017a,Ao2022a}.
Similarly, {\mixeddimensional} {\oneDthreeD} models are also available for other types of physics,
{\eg} in \gls{ac:FBI} \cite{Hagmeyer2022a,Hagmeyer-TwoWay,Lespagnol2023a}
to counteract {\threeDthreeD} \gls{ac:FSI} models such as~\cite{Mayr2015a,Mayr2020a}.

Naturally, computational benefits of {\mixeddimensional} {\oneDthreeD} models are expected,
since beam models require much fewer \glspl{ac:DOF} than solid models to represent the embedded fibers.
In~\cite{Hagmeyer-TwoWay},
the {\mixeddimensional} approach reduces the number of DOFs for fiber modeling by 96\% while keeping the {\LTwo} error of the bulk field below 1.5\%.
Yet, the solution process of {\mixeddimensional} models at large scale is not widely studied in literature,
but will be tackled in this contribution.
In particular, when using Krylov methods to iteratively solve the arising linear systems,
suitable preconditioners are required to sufficiently improve the spectral properties of the linear system \cite{Saad2003a}.
A few approaches can be found in literature:
Block diagonal preconditioners for saddle point systems arising from {\oneDtwoD} coupling are investigated in~\cite{Kuchta2016a}.
A simplified model problem of {\oneDthreeD} coupling yielding a $3\times 3$ saddle point system is studied in~\cite{Kuchta2019a},
where a fractional Laplacian is used to approximate the Schur complement and a block diagonal preconditioner is employed for the coupled problem.
An additive multigrid preconditioner for the arising fractional Laplacian is proposed in~\cite{Baerland2019a}.
With {\mixeddimensional} {\oneDthreeD} couplings as one application area of interface-driven {\multiphysics} problems,
uniform convergence, parameter robustness and scalability have been achieved through a suitable subspace splitting and custom smoothers in \cite{Budisa2024a}.
Based on the framework of operator preconditioning~\cite{Mardal2011a},
robust preconditioners have been proposed for {\oneDthreeD} domains coupled with Lagrange multipliers
for applications in micro-circulation~\cite{Dimola2023a}.
For $2\times2$ systems, the use of {\RugeStueben} \gls{ac:AMG} methods to solve a Schur complement equation for the \gls{ac:3D} bulk domain,
while tackling the \gls{ac:1D} domain by a direct solver,
is briefly discussed in~\cite{Cerroni2019a}.
For penalty-based {\oneDthreeD} models and in particular for beam models serving as \gls{ac:1D} models,
preconditioners are yet to be developed.

In this paper,
we will focus on our prior work on a regularized {\mortartype} embedding of geometrically exact {\nonlinear} beams
into \gls{ac:3D} solids \cite{Steinbrecher2020a,Steinbrecher2021a}
and will devise scalable preconditioners for the arising systems of linear equations.
We will study key properties of the linear systems,
in particular their loss of both diagonal dominance and block diagonal dominance due to the penalty contributions,
and design a preconditioner that is mostly agnostic to these challenges.
To this end,
we will interpret the system matrix containing solid and beam contributions as a $2\times2$ block matrix
and employ approximate block factorizations to arrive at a {\blockLU} preconditioner.
For the approximate inversion of the beam {\subblock} of the coupled system matrix,
we will construct a \gls{ac:SPAI} \cite{Grote1997a},
which will not only allow us to explicitly form an approximate Schur complement,
but will also serve as a smoother within the application of the {\blockLU} preconditioner.
The original \gls{ac:SPAI} algorithm will be equipped with filtering and static enrichment algorithms to amplify its performance and robustness.
To achieve scalability on parallel computing clusters,
we will tackle the Schur complement equation itself with \gls{ac:AMG} methods from {\trilinos}\slash{}{\muelu} \cite{BergerVergiat2023a}.
We will then study the computational performance and demonstrate weak scalability, robustness under changes of physical parameters,
and applicability to practical use cases in a series of numerical experiments
and investigate savings in wall clock time due to the reuse of the preconditioner throughout the entire load step.
In sum,
this contribution builds upon our prior work \cite{Steinbrecher2020a,Steinbrecher2021a}
and equips these models with scalable iterative solvers to facilitate their efficient application to large models and systems with thousands of embedded fibers.
Furthermore, this contribution constitutes the first presentation of a scalable, preconditioned iterative solver for truly {\oneDthreeD} models applied to {\beamsolid} coupling
with weak scalability demonstrated on a distributed memory cluster using the \gls{ac:MPI} on up to 1000 \gls{ac:MPI} ranks.

The remainder of this manuscript is organized as follows:
\secref{sec:Equations} introduces the underlying mechanical problem of {\mixeddimensional} couplings of slender fibers embedded into solid continua
and outlines the finite element discretization and the resulting linear systems.
Relevant properties of these linear systems are then discussed in \secref{sec:LinearSystems}.
The design of the preconditioner and its building blocks, in particular the \gls{ac:SPAI} algorithm, will be detailed in \secref{sec:Preconditioner}
along with a brief comparison to existing methods from literature.
In \secref{sec:Experiments}, we will study the numerical properties of individual components of the preconditioner
and assess its performance and scalability when applied to academic and engineering test cases.
\secref{sec:Conclusion} will summarize our findings and hint at future research directions.

\section{Mixed-dimensional modeling of fiber/solid systems}
\label{sec:Equations}

Since this manuscript concerns itself with preconditioner development for the {\mixeddimensional} modeling of the interaction of solid continua with slender fibers,
we only give a brief introduction into the governing equations, discretization and coupling approach presented in \cite{Steinbrecher2020a,Steinbrecher2021a}.
For a broader overview of the fundamentals of {\beamsolid} interaction, the interested reader is referred to \cite{Steinbrecher2022a}.

\subsection{Pure solid problem}

The \gls{ac:3D} solid bodies considered in this work are modeled as hyperelastic Boltzmann continua.
The weak form of the equation of elastostatics describing the deformation of the solid body is given as
\begin{equation*}
\delta \indexedSolid{W}
= \int_{\indexedSolid{\Omega}} \tet{S}:\delta\tet{E} \; \mathrm{d}V
- \int_{\indexedSolid{\Omega}} \teo{b}\cdot \delta \indexedSolid{\teo{u}} \; \mathrm{d}V
- \int_{\indexedSolid{\Gamma}} \teo{t} \cdot \delta \indexedSolid{\teo{u}} \; \mathrm{d}A
= 0
\end{equation*}
with $\indexedSolid{\teo{u}}$ being the solid displacement field,
$\tet{S}$ and $\tet{E}$ representing the second {\PiolaKirchhoff} stress tensor and the {\GreenLagrange} strain tensor,
$\teo{b}$ denoting the body forces,
$\teo{t}$ standing for the external traction field,
and $\delta$ indicating virtual, but kinematically admissible quantities in line with the concept of virtual work.
Furthermore, $\indexedSolid{\Omega}$ is the solid domain and $\indexedSolid{\Gamma}$ is the Neumann boundary of the solid domain.
In this work, all quantities denoted with an $\indexedSolid{(\cdot)}$ are associated with the solid continuum.
For the spatial discretization of the solid domain, we employ displacement-based isoparametric finite
elements interpolated by Lagrange polynomials resulting in the following discretized linear system
\begin{equation*}
			\indexedSolid{\mat{K}_{ss}}
			\Delta \indexedSolid{\mao{\disp}}\\
		=
		-
			\indexedSolid{\mao{\residualNonlinear}_{s}}
\end{equation*}
to be solved in every iteration of the Newton solver.
Therein, $\indexedSolid{\mao{\residualNonlinear}_{s}}$ represents the {\nonlinear} residual vector
and~$\indexedSolid{\mat{K}_{ss}}$ denotes its linearization, {\ie} the solid stiffness matrix.
The discrete displacement vector and its increment are given by~$\indexedSolid{\mao{\disp}}$ and~$\Delta \indexedSolid{\mao{\disp}}$, respectively.

\subsection{Coupled {\beamsolid} system with {\SimoReissner} beam formulation}
\label{sec:SRBeams}

In the scope of this publication, {\SimoReissner} (SR) and {\torsionfree} {\KirchhoffLove} (TF) beam theories are considered.
With respect to the coupled {\beamsolid} problem, the SR beam theory results in the most general coupling formulation.
We first describe the coupled problem based on a SR beam formulation and defer the use of TF beam formulations to \secref{sec:TFBeams}.

The general weak form for a \gls{ac:1D} Cosserat continuum (applicable to SR and TF beam theory) is given by
\begin{equation*}
\delta \indexedBeam{W}
= \delta \indexedBeam{\Pi_{int,(\cdot)}}
- \delta \indexedBeam{W_{ext}}
= 0,
\end{equation*}
where~$\delta\indexedBeam{\Pi_{int,(\cdot)}}$ denotes the variation of the internal elastic energy function
and~$\delta \indexedBeam{W_{ext}}$ is the external virtual work acting on the beam.
All quantities denoted with the superscript~$\indexedBeam{(\cdot)}$ are associated with beam contributions.
The internal elastic energy for a SR beam, {\ie} a shear-deformable beam with six local modes of deformation, reads
\begin{equation*}
\delta \indexedBeam{\Pi_{int,\indexSimoReissner}}
= \frac{1}{2} \int_{\indexedBeam{\Omega}} \trans{\Gamma} \tet{C}_F \Gamma
+ \trans{\Omega} \tet{C}_M \Omega \; \mathrm{d}s.
\end{equation*}
Here, $\Gamma$ represents the material deformation,
$\Omega$ the material curvature,
and~$\textbf{C}_F$ as well as~$\textbf{C}_M$ are the respective constitutive matrices for the translational and rotational deformation modes.
We refer the interested reader to~\cite{Meier2015a,Meier2019a} for more details on the SR beam theory and its finite element discretization.

The weak form of the beam is defined on the undeformed \gls{ac:1D} centerline domain $\indexedBeam{\Omega}$.
The spatial interpolation of the beam finite elements is tailored to the particular beam model in order to ensure objectivity of the discrete formulation.
The interpolation of the beam centerline position employs $C^{1}$-continuous third-order Hermite shape functions~\cite{Meier2019a}.
An objective interpolation of the finite cross section orientations is a non-trivial task and results in a nonlinear and deformation-dependent interpolation strategy.
For a more detailed description of this topic, the reader is referred to~\cite{Meier2019a} and the references therein.
The resulting linearized system of a pure SR beam problem reads
\begin{equation*}
			\left(
		\begin{matrix}
			\indexedBeam{\mat{K}_{rr}} & 			\indexedBeam{\mat{K}_{r\theta}} \\
			\indexedBeam{\mat{K}_{\theta r}} & 			\indexedBeam{\mat{K}_{\theta\theta}}
		\end{matrix}
		\right)
		\left(
		\begin{matrix}
			\Delta \indexedBeam{\mao{\disp}}\\
			\Delta \indexedBeam{\bm{\theta}}\\
		\end{matrix}
		\right)
		=
		-
		\left(
		\begin{matrix}
			\indexedBeam{\mao{\residualNonlinear}_{r}}\\
			\indexedBeam{\mao{\residualNonlinear}_{\theta}}
		\end{matrix}
		\right).
\end{equation*}
To clarify the following equations and simplifications in the TF case,
the global discrete beam centerline degrees of freedom~$\indexedBeam{\mao{\disp}}$ are gathered together
as are the global beam orientational degrees of freedom~$\indexedBeam{\bm{\theta}}$.
Therefore, the beam stiffness matrix is written as a $2\times2$ block system
with the individual blocks~$\indexedBeam{\mat{K}_{\indexArbitraryOne\indexArbitraryTwo}}$ with indices~$\indexArbitraryOne,\indexArbitraryTwo\in\{r,\theta\}$.
Similarly, the residual is split into positional and rotational contributions~$\indexedBeam{\mao{\residualNonlinear}_{r}}$
and~$\indexedBeam{\mao{\residualNonlinear}_{\theta}}$, respectively.

To fully embed the beam in the solid matrix, the following six local coupling constraints are defined:
\begin{align}
\teo{0} &= \indexedBeam{\teo{r}} - \indexedSolid{\teo{x}} \  \text{on\ } \indexedBeam{\Omega} \label{eq:CouplingFullPos}, \\
\teo{0} &= \indexedSolidBeam{\teo{\psi}} \  \text{on\ } \indexedBeam{\Omega}.
\label{eq:CouplingFullRot}
\end{align}
Here, $\indexedBeam{\teo{r}}$ is the beam centerline position vector,
$\indexedSolid{\teo{x}}$ is the solid position vector,
and~$\indexedSolidBeam{\teo{\psi}}$ denotes the \mbox{(pseudo-)}rotation vector
describing the relative rotation between a suitable orthonormal triad field in the solid domain and the beam cross section orientation.
The constraints given in~\eqref{eq:CouplingFullPos} are referred to as the \emph{positional} coupling constraints,
since they enforce the position of the beam cross section centroid to be coupled to the underlying solid.
In a similar manner,
\eqref{eq:CouplingFullRot} is referred to as \emph{rotational} coupling
since a vanishing relative rotation enforces the beam cross section orientation to be coupled to the solid domain.
For a more elaborate discussion on the coupling constraints,
the interested reader is referred to~\cite{Steinbrecher2020a,Steinbrecher2021a}.
The coupling constraints~\eqref{eq:CouplingFullPos} and~\eqref{eq:CouplingFullRot} are enforced via a Lagrange multiplier method.
The total Lagrange multiplier potential reads
\begin{equation}
\Pi_{\lambda,\indexSimoReissner}
= \int_{\indexedBeam{\Omega}} (\trans{\indexedDispCoupling{\teo{\lambda}})} (\indexedBeam{\teo{r}}
- \indexedSolid{\teo{x}}) \; \mathrm{d}s
+ \int_{\indexedBeam{\Omega}} \trans{(\indexedRotCoupling{\teo{\lambda}})} \teo{\psi} \; \mathrm{d}s,
\label{eq:CouplingFullPot}
\end{equation}
where~$\indexedDispCoupling{\teo{\lambda}}$ and~$\indexedRotCoupling{\teo{\lambda}}$ are the Lagrange multiplier fields
introduced to enforce the positional and rotational coupling, respectively.
Accordingly, quantities denoted with~$\indexedDispCoupling{(\cdot)}$ and~$\indexedRotCoupling{(\cdot)}$ refer to positional and rotational coupling, respectively.

For the spatial interpolation of the Lagrange multiplier fields, we resort to the mortar-type approach and software implementation of~\cite{Steinbrecher2020a,Steinbrecher2021a}.
There it is shown that a linear interpolation of the Lagrange multiplier field with a subsequent penalty regularization of the saddle point system results in a stable coupling formulation for our envisioned application range.
The resulting global discrete Lagrange multiplier vectors for positional and rotational coupling are denoted with~$\indexedDispCoupling{\lagMultDiscVec}$ and~$\indexedRotCoupling{\lagMultDiscVec}$, respectively.
Variation of the coupling potential~\eqref{eq:CouplingFullPot} and insertion of the spatially discretized quantities results in the weak form of the coupling formulation.
With that we can now state the global discretized residual vector for the coupled problem:
\begin{equation*}
\left(
\begin{matrix}
	\indexedBeam{\mao{\residualNonlinear}_{r}} + \indexedDispCoupling{\mao{\residualNonlinear}}_{mt,r} \\
	\indexedBeam{\mao{\residualNonlinear}_{\theta}} + \indexedRotCoupling{\mao{\residualNonlinear}}_{mt,\theta} \\
	\indexedSolid{\mao{\residualNonlinear}}_{s} + \indexedDispCoupling{\mao{\residualNonlinear}}_{mt,s} + \indexedRotCoupling{\mao{\residualNonlinear}}_{mt,s} \\
	\indexedDispCoupling{\mao{\gap}}\\
	\indexedRotCoupling{\mao{\gap}}\\
\end{matrix}
\right) = \mao{0}.
\end{equation*}
Here, $\indexedDispCoupling{\mao{\gap}}$ and~$\indexedRotCoupling{\mao{\gap}}$ are the constraint residual vectors for positional and rotational coupling, respectively.
Moreover, ${\mao{\residualNonlinear}}_{mt,(\cdot)}^{(\cdot)}$ are the discrete coupling force contributions required to enforce the action of the Lagrange multipliers on the beam and solid degrees of freedom.
We resort to a penalty regularization with the penalty parameters
$\penaltyParamDisp>0$ and $\penaltyParamRot>0$ to express the Lagrange multiplier unknowns~$\indexedDispCoupling{\lagMultDiscVec}$
and~$\indexedRotCoupling{\lagMultDiscVec}$ in terms of beam and solid unknowns, {\ie}
\begin{align}
	\label{eq:penaltyPos}
	\indexedDispCoupling{\lagMultDiscVec} &\approx \penaltyParamDisp \inv{(\indexedDispCoupling{\mat{V}})} \indexedDispCoupling{\mao{\gap}}, \\
	\indexedRotCoupling{\lagMultDiscVec} &\approx \penaltyParamRot \inv{(\indexedRotCoupling{\mat{V}})} \indexedRotCoupling{\mao{\gap}}.
\end{align}
Therein, the diagonal matrices~$\indexedDispCoupling{\mao{V}}$ and~$\indexedRotCoupling{\mat{V}}$ are scaling matrices
to scale the regularized equations in order to pass patch-test like problems, cf.~\cite{Steinbrecher2020a}.
With the penalty regularization, the Lagrange multipliers are no longer unknowns.
Thus, we can state the final linearized system to be solved in every Newton iteration:
\begin{align}
\begin{split}
	\label{eq:RotationCouplingLinearSystem}
		& \left(
		\begin{matrix}
			\indexedBeam{\mat{K}_{rr}} + \penaltyParamDisp \trans{\mat{D}} \inv{(\indexedDispCoupling{\mat{V}})} \mat{D} &
			\indexedBeam{\mat{K}_{r \theta}} &
			- \penaltyParamDisp \trans{\mat{D}} \inv{(\indexedDispCoupling{\mat{V}})} \mat{M} \\
			\indexedBeam{\mat{K}_{\theta r}} &
			\indexedBeam{\mat{K}_{\theta\theta}} + \mat{Q}_{\theta\theta} + \penaltyParamRot \mat{Q}_{\theta\lambda} \inv{(\indexedRotCoupling{\mat{V}})} \mat{Q}_{\lambda\theta} &
			\mat{Q}_{\theta s} + \penaltyParamRot \mat{Q}_{\theta\lambda} \inv{\mat{V}} \mat{Q}_{\lambda s} \\
			- \penaltyParamDisp \trans{\mat{M}} \inv{(\indexedDispCoupling{\mat{V}})} \mat{D} &
			\mat{Q}_{s\theta} + \penaltyParamRot \mat{Q}_{s\lambda} \inv{(\indexedRotCoupling{\mat{V}})} \mat{Q}_{\lambda\theta} &
			\indexedSolid{\mat{K}_{ss}} + \mat{Q}_{ss} + \penaltyParamDisp \trans{\mat{M}} \inv{(\indexedDispCoupling{\mat{V}})} \mat{M} + \penaltyParamRot \mat{Q}_{s\lambda} \inv{(\indexedRotCoupling{\mat{V}})} \mat{Q}_{\lambda s} \\
		\end{matrix}
		\right)
		\left(
		\begin{matrix}
			\Delta \indexedBeam{\mao{\disp}}\\
			\Delta \indexedBeam{\bm{\theta}}\\
			\Delta \indexedSolid{\mao{\disp}}\\
		\end{matrix}
		\right)\\
		& =
		-
		\left(
		\begin{matrix}
			\indexedBeam{\mao{\residualNonlinear}_{r}}\\
			\indexedBeam{\mao{\residualNonlinear}_{\theta}}\\
			\indexedSolid{\mao{\residualNonlinear}_{s}}\\
		\end{matrix}
		\right)
\end{split}.
\end{align}
Here, $\mat{D}$ and~$\mat{M}$ are the so called mortar matrices for positional coupling, which only depend on the reference configuration, i.e., they are constant.
The matrices~$\mat{Q}_{\indexArbitraryOne\indexArbitraryTwo}$ with~$\indexArbitraryOne,\indexArbitraryTwo\in\{s,\theta,\lambda\}$
are the coupling matrices for rotational coupling, which depend on the current configuration.
We refer to the original publications~\cite{Steinbrecher2020a,Steinbrecher2021a} for details of the linearization procedure.

\subsection{Coupled {\beamsolid} system with {\torsionfree} (TF) beam formulation}
\label{sec:TFBeams}

The {\torsionfree} {\KirchhoffLove} (TF) beam formulation (see~\cite{Meier2015a,Meier2019a}) represents a special case of the {\SimoReissner} beam theory,
where the assumptions of vanishing shear and torsion deformations are incorporated in the beam model
and the resulting finite element formulation can be described solely by displacement degrees of freedom.
The assumptions are valid for fibers with high slenderness ratios, a double symmetric cross section and a straight centerline in the reference configuration.
For such fibers, the TF beam formulation results in an even more efficient numerical model than the SR formulation presented in the previous section.
The internal energy of a TF beam reads
\begin{equation*}
	\delta \indexedBeam{\Pi_{int,\indexTorsionFree}} = \frac{1}{2} \int_{\indexedBeam{\Omega}} EA\varepsilon^2 + EI\kappa^2 \; \mathrm{d}s
\end{equation*}
with $\youngs$ denoting the Young’s modulus, $A$ the cross section area, $I$ the moment of inertia,
$\varepsilon$ the axial tension and $\kappa$ the scalar curvature, respectively.
The TF beam formulation requires a $C^{1}$-continuous centerline interpolation of
centerline positions~$\teo{r}$, which is realized with third-order Hermite polynomials~\cite{Meier2015a}.

As the only field of unknowns along the TF beam centerline is a translation field, the total Lagrange multiplier potential of the coupling constraints in~\eqref{eq:CouplingFullPot} simplifies to
\begin{equation*}
	\Pi_{\lambda,\indexTorsionFree} = \int_{\Gamma} \trans{(\indexedDispCoupling{\teo{\lambda}})} (\indexedBeam{\teo{r}} - \indexedSolid{\teo{x}}) \; \mathrm{d}s
\end{equation*}
i.e., there is no rotational coupling for TF beams.
The same penalty regularization as stated in \eqref{eq:penaltyPos} is employed, resulting in the following global linearized system :
\begin{equation}
\label{eq:PositionCouplingLinearSystem}
		\left(
		\begin{matrix}
			\indexedBeam{\mat{K}_{rr}} + \penaltyParamDisp \trans{\mat{D}} \inv{(\indexedDispCoupling{\mat{V}})} \mat{D} &
			- \penaltyParamDisp \trans{\mat{D}} \inv{(\indexedDispCoupling{\mat{V}})} \mat{M} \\
			- \penaltyParamDisp \trans{\mat{M}} \inv{(\indexedDispCoupling{\mat{V}})} \mat{D} &
			\indexedSolid{\mat{K}_{ss}} + \penaltyParamDisp \trans{\mat{M}} \inv{(\indexedDispCoupling{\mat{V}})} \mat{M}
		\end{matrix}
		\right)
		\left(
		\begin{matrix}
			\Delta \indexedBeam{\mao{\disp}}\\
			\Delta \indexedSolid{\mao{\disp}}\\
		\end{matrix}
		\right)
		=
		-
		\left(
		\begin{matrix}
			\indexedBeam{\mao{\residualNonlinear}_{r}}\\
			\indexedSolid{\mao{\residualNonlinear}_{s}}\\
		\end{matrix}
		\right).
\end{equation}
Again, for a more detailed information on the derivations, the interested reader is referred to~\cite{Steinbrecher2020a}.
It should be noted, that the coupling contributions in~\eqref{eq:PositionCouplingLinearSystem}, i.e., $\mat{D}$ and~$\mat{M}$, are constant
due to the coupling taking place in the reference configuration.


\section{Characteristics of linear systems arising in {\beamsolid} coupling}
\label{sec:LinearSystems}

To be able to construct efficient algebraic block preconditioning techniques
to accelerate the convergence of the outer Krylov solver,
certain matrix properties of the underlying linear systems~$\linOp\mao{\sol}=\mao{\rhs}$ specified in~\eqref{eq:RotationCouplingLinearSystem}
and~\eqref{eq:PositionCouplingLinearSystem} are of particular interest.
A short explanation regarding conditioning, block diagonal dominance, symmetry and sparsity pattern is given below.
For the sake of simplicity, the block systems in~\eqref{eq:RotationCouplingLinearSystem} and~\eqref{eq:PositionCouplingLinearSystem}
are both abbreviated with the compact notation
\begin{equation}
\left(
\begin{matrix}
\matBeam & \trans{\matBeamSolid}\\
\matSolidBeam & \matSolid\\
\end{matrix}
\right)
\left(
\begin{matrix}
\indexedBeam{\mao{\sol}}\\
\indexedSolid{\mao{\sol}} \\
\end{matrix}
\right)
=
\left(
\begin{matrix}
\indexedBeam{\mao{\rhs}}\\
\indexedSolid{\mao{\rhs}}\\
\end{matrix}
\right)
\quad\text{ with }
\mao{\residualLinear}
=
\left(
\begin{matrix}
\indexedBeam{\mao{\residualLinear}}\\
\indexedSolid{\mao{\residualLinear}}\\
\end{matrix}
\right)
=
\left(
\begin{matrix}
\indexedBeam{\mao{\rhs}}\\
\indexedSolid{\mao{\rhs}}\\
\end{matrix}
\right)
-
\left(
\begin{matrix}
\matBeam & \trans{\matBeamSolid}\\
\matSolidBeam & \matSolid\\
\end{matrix}
\right)
\left(
\begin{matrix}
\indexedBeam{\mao{\sol}}\\
\indexedSolid{\mao{\sol}} \\
\end{matrix}
\right)
\label{eq:SimpleLinearSystem}
\end{equation}
throughout the remainder of this manuscript,
where we have grouped the unknowns based on their physical meaning, {\ie} being associated with the beams or the background solid.
The concrete identification of individual matrix blocks in~\eqref{eq:SimpleLinearSystem} with the linear systems from \secsref{sec:SRBeams}{sec:TFBeams}
depends on the employed beam theory.
In the case of a SR beam theory,
the individual blocks are defined such that~\eqref{eq:SimpleLinearSystem} represents~\eqref{eq:RotationCouplingLinearSystem}
with~$\indexedBeam{\mao{\rhs}} = \trans{(\trans{\Delta (\indexedBeam{\mao{\disp}})}, \trans{(\Delta \indexedBeam{\bm{\theta}})})}$
and~$\indexedSolid{\mao{\rhs}} = \Delta \indexedSolid{\mao{\disp}}$.
For the TF case, \eqref{eq:SimpleLinearSystem} represents~\eqref{eq:PositionCouplingLinearSystem}
with~$\indexedBeam{\mao{\rhs}} = \Delta \indexedBeam{\mao{\disp}}$
and~$\indexedSolid{\mao{\rhs}} = \Delta \indexedSolid{\mao{\disp}}$.
Generally speaking, $\matBeam$ denotes the matrix block containing the
beam stiffness matrices $\indexedBeam{\mat{K}_{\indexArbitraryOne\indexArbitraryTwo}}$ as well as stiffness and penalty contributions of the coupling constraints.
In similar fashion, the {\subblock} $\mat{C}$
refers to the sum of the solid stiffness matrix $\indexedSolid{\mat{K}_{\indexArbitraryOne\indexArbitraryTwo}}$ and the respective interaction terms.
The off-diagonal {\subblock}s~$\trans{\matBeamSolid}$ and~$\matSolidBeam$ represent the remaining coupling terms between both
fields, respectively.

\subsection{Ill-conditioning due to penalty regularization}
\label{sec:PenaltyRegularization}

Due to the discretization and coupling approach introduced in \secref{sec:Equations},
the linear system of equations suffers from ill-conditioning,
which directly originates from the penalty parameters~$\penaltyParamDisp$ and~$\penaltyParamRot$,
that are steering the strength of the interaction between solid and fibers.
Naturally, larger values of~$\penaltyParamDisp$ and~$\penaltyParamRot$ lead to a more accurate constraint enforcement,
however enlarge the eigenvalue spectrum of the matrix and, thus, worsen the conditioning problems.
To show this exemplarily,
a small eigenvalue study is done for test cases~I and~IV introduced later in \secref{sec:NumericalStudySpai}
with all parameters fixed despite~$\penaltyParamDisp$ and~$\penaltyParamRot$ as shown in \tabref{tab:EigenvalueAnalysis}.
\begin{table}
\centering
\caption{Minimum and maximum eigenvalues $\lambda_{min}$, $\lambda_{max}$ of the given block system with condition number estimates for different beam models and varying positional and rotational penalty parameters $\penaltyParamDisp$, $\penaltyParamRot$}
\label{tab:EigenvalueAnalysis}
\begin{tabular}{c c c | c c c c}
\hline
beam model & $\penaltyParamDisp$  & $\penaltyParamRot$ & $\lambda_{min}$ & $\lambda_{max}$ & $\lambda_{max}/\lambda_{min}$ & $\normOne{\linOp} \normOne{\inv{\linOp}}$ \\
\hline
TF & $1$    & - & $\qty{0.3920e-7}{}$ & $1.0000$ & $\qty{2.5510e7}{}$ & $\qty{4.0078e7}{}$ \\
TF & $10$   & - & $\qty{0.3920e-7}{}$ & $1.5042$ & $\qty{3.8372e7}{}$ & $\qty{4.7060e7}{}$ \\
TF & $100$  & - & $\qty{0.3920e-7}{}$ & $14.915$ & $\qty{3.8048e8}{}$ & $\qty{4.6951e8}{}$ \\
TF & $1000$ & - & $\qty{0.3920e-7}{}$ & $149.03$ & $\qty{3.8018e9}{}$ & $\qty{4.6868e9}{}$ \\
\hline
SR & $1$    & $0.1$ & $\qty{0.1047e-6}{}$ & $4.7644$ & $\qty{4.5505e7}{}$ & $\qty{1.9389e6}{}$ \\
SR & $10$   & $1$   & $\qty{0.1047e-6}{}$ & $47.199$ & $\qty{4.5081e8}{}$ & $\qty{1.9089e7}{}$ \\
SR & $100$  & $10$  & $\qty{0.1047e-6}{}$ & $471.55$ & $\qty{4.5039e9}{}$ & $\qty{1.8604e8}{}$ \\
SR & $1000$ & $100$ & $\qty{0.1047e-6}{}$ & $4715.1$ & $\qty{4.5034e10}{}$ & $\qty{1.9031e9}{}$ \\
\hline
\end{tabular}
\end{table}
For both beam models, an increase of the penalty parameters results in an increasing maximum eigenvalue~$\lambda_{max}$ of the overall system,
while the minimum eigenvalue~$\lambda_{min}$ remains constant.
This results in growing condition number estimates given by the well-established definitions~$\lambda_{max}/\lambda_{min}$ and~$\normOne{\linOp} \normOne{\inv{\linOp}}$.
Despite the bad conditioning, {\subblock}~$\matBeam$ is still nonsingular,
being an essential requirement for factorizations of the block matrix~$\linOp$.

\subsection{Loss of block diagonal dominance}
\label{sec:BlockDiagonalDominance}
Independent of the actual choice of the beam model within {\beamsolid} interaction,
the matrices in the arising linear systems in~\eqref{eq:RotationCouplingLinearSystem} and~\eqref{eq:PositionCouplingLinearSystem} exhibit a 2 $\times$ 2 block structure
based on a physically motivated grouping of unknowns into beam unknowns~$\indexedBeam{\mao{\sol}}$
and solid unknowns~$\indexedSolid{\mao{\sol}}$, respectively.
This becomes particularly evident in the unified notation of~\eqref{eq:SimpleLinearSystem}.
For a closer look at the arising matrices,
we adopt the concept of \emph{block diagonal dominance} for block matrices from \cite{feingold1962a}:

\begin{definition}[Block diagonal dominance]
\label{def:BlockDiagonalDominance}
Let $\mat{\anyMatrix} \in \mathbb{C}^{\numBlockRows \times \numBlockRows}$ be a square block matrix with~$\numBlockRows$ block rows and block columns, respectively.
With a given matrix norm~$\norm{\anyQuantity}$,
we assume~$\mat{\anyMatrix}$ to only contain nonsingular matrix blocks~$\mat{\anyMatrix}_{\indBlockRow\indBlockRow}$ on its main diagonal,
{\ie} $\det{\mat{\anyMatrix}_{\indBlockRow\indBlockRow}} \neq 0 \, \forall \indBlockRow \in 1,\hdots,\numBlockRows$.
Then, a matrix~$\mat{\anyMatrix}$ is referred to as \emph{block diagonally dominant},
if
\begin{align}
\label{eq:BlockDiagonalDominance}
\sum_{\substack{\indBlockCol=1 \\ \indBlockCol \neq \indBlockRow}}^{\numBlockRows}
\norm{\mat{\anyMatrix_{\indBlockRow\indBlockCol}}} \leq \inv{\norm{\inv{\mat{\anyMatrix}_{\indBlockRow\indBlockRow}}}}
\quad \text{ for } \indBlockRow=1,\dots,\numBlockRows.
\end{align}
\end{definition}
In general, the matrices in \eqref{eq:RotationCouplingLinearSystem} and~\eqref{eq:PositionCouplingLinearSystem}
do not satisfy the conditions for block diagonal dominance as outlined in \defref{def:BlockDiagonalDominance}.
For illustration purposes, we consider the {\mixeddimensional} modeling approach from \secref{sec:Equations}
in the practical case of fiber\hyp{}reinforced solids, where the stiffness of the fibers is much higher than the stiffness of the embedding solid, {\ie} $E^\indexBeam \gg E^\indexSolid$.
To this end, we assume a fixed geometry and mesh, constant material parameters, and fiber and solid constitutive properties satisfying $E^\indexBeam \gg E^\indexSolid$.
Since the projection operators~$\mat{D}$ and~$\mat{M}$ solely depend on the mesh,
the only variable parameters left are $\penaltyParamDisp \gg 0$ and $\penaltyParamRot \gg 0$.
With increasing penalty parameters~$\penaltyParamDisp$ and~$\penaltyParamRot$,
the norm of the off-diagonal matrix blocks increases, too.
In addition, the inversion of the diagonal matrix blocks results in denser matrices with a rapid growth of the norm
and, thus, decreasing values on the right-hand side of inequality~\eqref{eq:BlockDiagonalDominance}.
For positional coupling,
the block diagonal dominance property of the matrix becomes harder to achieve with an increasing penalty parameter~$\penaltyParamDisp$,
especially for~$\penaltyParamDisp \approx E^\indexBeam$ as recommended for practical computations.
The same holds true for the rotational coupling contributions for the recommended choice~$\penaltyParamRot \approx E^\indexBeam \beamRadius^2$
with $\beamRadius$ being the radius of the beam along the centerline (see~\cite{Steinbrecher2022b}).
To support this argument,
we assess the property of block diagonal dominance for the first block row in \eqref{eq:SimpleLinearSystem}, {\ie} specifically the contribution of~$\matBeam$,
by anticipating a small numerical example, in particular test case~I introduced later in \secref{sec:NumericalStudySpai}.
Therein, the off-diagonal matrix block~$\matBeamSolid$ exhibits a norm~$\norm{\matBeamSolid}=1.0153$,
while the main diagonal block~$\matBeam$'s contribution evaluates to $\inv{\norm{\inv{\matBeam}}}=3.9202\cdot 10^{-8}$,
hence violating the condition outlined in \eqref{eq:BlockDiagonalDominance}.

Due to this lack of block diagonal dominance,
conventional block preconditioning methods based on block Jacobi or block {\GaussSeidel} schemes are not applicable without major convergence problems (or even divergence)
as already evidenced in \cite{Chacon2008a,Cyr2016a}.
Independently, the individual matrix blocks on the main diagonal lose their diagonal dominance as well due to the penalty contributions.
Hence, conventional relaxation-based smoothers cannot be applied on individual blocks either.

\subsection{Potential loss of symmetry}
The symmetry of~$\matBeam$ is governed by the beam formulation at hand:
TF beam models always result in a symmetric~$\matBeam$,
while all other beam models yield a {\nonsymmetric} beam {\subblock}~$\matBeam$.
For pure positional coupling as applied for TF beam models,
the off-diagonal matrix block are symmetric, {\ie} $\matBeamSolid = \matSolidBeam$.
In contrast, the additional coupling of rotational degrees of freedom for SR beams introduces {\nonsymmetric} off-diagonal terms.

\subsection{Sparsity pattern}
Of special interest is the matrix graph~$\graphOf{\matBeam}$ of
the {\subblock}~$\matBeam$ related to the beam problem.
Since we restrict ourselves so far to cases where the embedded fibers do not interact with each other but only with the surrounding solid,
the matrix~$\matBeam$ features a block diagonal sparsity structure, where the size of the small blocks depends on the
number of beam finite elements used to discretize each individual fiber.
We will illustrate and study the sparsity pattern~$\graphOf{\matBeam}$ for different test cases in \secref{sec:NumericalStudySpai},
in particular in \figref{fig:SparsityTestCases}.

\section{Block preconditioning for {\beamsolid} interaction}
\label{sec:Preconditioner}

The construction of a preconditioner, that captures the coupling interactions properly and is tailored to the specific matrix properties,
is crucial for an efficient and scalable solution process. Preconditioners based on approximate block factorizations have been shown to be
suited for similar problem types such as contact problems~\cite{Wiesner2021a},
incompressible flow~\cite{Patankar1972a,Elman2008a},
\gls{ac:FSI}~\cite{Langer2016a,Jodlbauer2019a},
or magneto-hydro dynamics~\cite{Chacon2008a,Phillips2014a,Phillips2016a}.

\subsection{Preconditioning based on a block factorization of the system matrix}
\label{sec:SchurComplementBlockFactorization}

Due to the lack of block diagonal dominance and the ill-conditioning of the matrix discussed in \secref{sec:LinearSystems},
we resort to a block factorization preconditioner.
Specifically, we perform a block factorization into a lower triangular matrix~$\matFactL$, a diagonal matrix~$\matFactD$, and an upper triangular matrix~$\matFactU$.
The $\matFactL\matFactD\matFactU$ decomposition of the system matrix reads
\begin{equation*}
\left(
\begin{matrix}
\matBeam & \trans{\matBeamSolid}\\
\matSolidBeam & \matSolid\\
\end{matrix}
\right)
=
\left(
\begin{matrix}
\matIdentity & \matZero \\
\matSolidBeam \inv{\matBeam} & \matIdentity \\
\end{matrix}
\right)
\left(
\begin{matrix}
\matBeam & \matZero \\
\matZero & \schurMatrix \\
\end{matrix}
\right)
\left(
\begin{matrix}
\matIdentity & \inv{\matBeam}\trans{\matBeamSolid} \\
\matZero & \matIdentity \\
\end{matrix}
\right)
\end{equation*}
with the Schur complement $\schurMatrix := \matSolid - \matSolidBeam \inv{\matBeam} \trans{\matBeamSolid}$.
To this end, the preconditioning matrix~$\precMatrix$ based on the above factorization is given by
\begin{equation*}
\precMatrix
:=
\left(
\begin{matrix}
\matBeam & \matZero \\
\matSolidBeam & \schurMatrix \\
\end{matrix}
\right)
\left(
\begin{matrix}
\matIdentity & \inv{\matBeam}\trans{\matBeamSolid} \\
\matZero & \matIdentity \\
\end{matrix}
\right).
\end{equation*}
Expressing the application of the preconditioner as a {\fixedpoint} iteration over the index~$\indSweep$,
one application of the preconditioner yields
\begin{equation*}
\left(
\begin{matrix}
\indexedBeamSweep{\mao{\sol}}{\indSweep+1} \\
\indexedSolidSweep{\mao{\sol}}{\indSweep+1} \\
\end{matrix}
\right)
=
\left(
\begin{matrix}
\indexedBeamSweep{\mao{\sol}}{\indSweep} \\
\indexedSolidSweep{\mao{\sol}}{\indSweep} \\
\end{matrix}
\right)
+
\inv{\precMatrix}
\left(
\begin{matrix}
\indexedBeamSweep{\mao{\residualLinear}}{\indSweep}\\
\indexedBeamSweep{\mao{\residualLinear}}{\indSweep}\\
\end{matrix}
\right)
=
\left(
\begin{matrix}
\indexedBeamSweep{\mao{\sol}}{\indSweep} \\
\indexedSolidSweep{\mao{\sol}}{\indSweep} \\
\end{matrix}
\right)
+
\left(
\begin{matrix}
\matBeam & \matBeam\inv{\matBeam}\trans{\matBeamSolid}\\
\matSolidBeam & \matSolidBeam \inv{\matBeam} \trans{\matBeamSolid} + \schurMatrix \\
\end{matrix}
\right)^{-1}
\left[
\left(
\begin{matrix}
\indexedBeam{\mao{\rhs}}\\
\indexedSolid{\mao{\rhs}}\\
\end{matrix}
\right)
-
\left(
\begin{matrix}
\matBeam & \trans{\matBeamSolid} \\
\matSolidBeam & \matSolid \\
\end{matrix}
\right)
\left(
\begin{matrix}
\indexedBeamSweep{\mao{\sol}}{\indSweep}\\
\indexedSolidSweep{\mao{\sol}}{\indSweep}\\
\end{matrix}
\right)
\right].
\end{equation*}
It is usually executed via a predictor-corrector scheme by first solving an equation related to the beam contribution,
afterwards one related to the Schur complement, and lastly a correction step to the beam solution.
To this end, after having to form an explicit representation of the Schur complement,
a total of three linear systems have to be solved in every application of the preconditioner.
The overall algorithm is given as \algref{alg:PreconditionerRaw}.

Since an exact version of a block factorization preconditioner is hard to achieve due to its immense computational cost,
the following sections are devoted to the construction of an approximate block factorization preconditioner
and its application to {\mixeddimensional} {\beamsolid} interaction.

\begin{algorithm}
\caption{Full block factorization preconditioner for {\fibersolid} coupling}
\label{alg:PreconditionerRaw}
\SetKwProg{myproc}{Procedure}{}{}
\myproc{Preconditioner($\indSweep_{max}$)}
{
// Form explicit Schur complement \newline
$\schurMatrix = \matSolid - \matSolidBeam \inv{\matBeam} \trans{\matBeamSolid}$

\For{$\indSweep = 1, \dots, \indSweep_{max}$}
{
	$\left(
	\begin{matrix}
	\indexedBeamSweep{\mao{\residualLinear}}{\indSweep}\\
	\indexedSolidSweep{\mao{\residualLinear}}{\indSweep}\\
	\end{matrix}
	\right)
	=
	\left(
	\begin{matrix}
	\indexedBeam{\mao{\rhs}}\\
	\indexedSolid{\mao{\rhs}}\\
	\end{matrix}
	\right)
	-
	\left(
	\begin{matrix}
	\matBeam & \trans{\matBeamSolid} \\
	\matSolidBeam & \matSolid \\
	\end{matrix}
	\right)
	\left(
	\begin{matrix}
	\indexedBeamSweep{\mao{\sol}}{\indSweep}\\
	\indexedSolidSweep{\mao{\sol}}{\indSweep}\\
	\end{matrix}
	\right)$

    // Prediction step: solve for $\indexedBeamSweep{\mao{\sol}}{\indSweep+\frac{1}{2}}$ \newline
    $\matBeam \indexedBeamSweep{\mao{\sol}}{\indSweep+\frac{1}{2}} = \indexedBeamSweep{\mao{\residualLinear}}{\indSweep}$

    // Schur complement step: solve for $\indexedSolidSweep{\mao{\sol}}{\indSweep+1}$ \newline
    $\schurMatrix \indexedSolidSweep{\mao{\sol}}{\indSweep+1} = \indexedSolidSweep{\mao{\residualLinear}}{\indSweep}
    -\matSolidBeam \indexedBeamSweep{\mao{\sol}}{\indSweep+\frac{1}{2}}$

	// Correction step: solve for $\indexedBeamSweep{\mao{\sol}}{\indSweep+1}$ \newline
	$\matBeam \indexedBeamSweep{\mao{\sol}}{\indSweep+1} = \indexedBeamSweep{\mao{\residualLinear}}{\indSweep}
	-\trans{\matBeamSolid}\indexedSolidSweep{\mao{\sol}}{\indSweep+1}$
}
\Return $\left(\begin{matrix} \indexedBeamSweep{\mao{\sol}}{\indSweep_{max}} \\ \indexedSolidSweep{\mao{\sol}}{\indSweep_{max}}\end{matrix} \right)$
}
\end{algorithm}

\subsection{Explicit sparse inverse approximation}
\label{sec:SPAI}

The first major step of the preconditioner calculation consists of finding an explicit approximation of the Schur complement $\approxMat{\schurMatrix} := \matSolid - \matSolidBeam \inv{\approxMat{\matBeam}} \trans{\matBeamSolid}$ with~$\approxMat{\matBeam}$ denoting an easy-to-invert approximation of $\matBeam$. Both quality and computational cost of the preconditioner are mainly governed by the choice~$\approxMat{\matBeam}$ and~$\approxMat{\schurMatrix}$ to approximate~$\matBeam$ and~$\schurMatrix$, respectively.
In traditional Schur complement based block preconditioners,
the inverse~$\inv{\approxMat{\matBeam}} \approx \inv{\matBeam}$ appearing due to the block factorization and also in the Schur complement calculation itself
is often approximated by some diagonal matrix,
since the inversion of a diagonal matrix comes at a negligible computational cost.
The most simple approach is to base the inverse approximation on the diagonal part of the $\nRows\times\nRows$ matrix~$\matBeam$
resulting in $\approxMat{\matBeam} = \diag{\matBeamEntry{\indRow}{\indRow}}, \indRow = 1,\hdots,\nRows$.
Another well-known approach takes the row sums of $\matBeam$~\cite{VanDoormaal1984a}, reading
\begin{equation*}
\approxMat{\matBeam} = \diag{\sum_{\indCol=1}^\nRows \abs{\matBeamEntry{\indRow}{\indCol}}}, \quad \indRow = 1,\hdots,\nRows.
\end{equation*}
However, such simple diagonal approximations cannot be used in the present scenario
since~$\matBeam$ lacks diagonal dominance, {\cf} \secref{sec:BlockDiagonalDominance}.
Since $\matBeam$ resembles the {\subblock} related to the beam equations, which themselves yield a block diagonal structure of~$\matBeam$,
a more sophisticated explicit approximation scheme of the inverse can be applied taking this particular sparsity structure into account.

Although in general the inverse of the sparse matrix~$\matBeam$ cannot be expected to be sparse as well,
explicit \glspl{ac:SPAI} aim at creating an explicit matrix representation~$\spaiOf{\matBeam}$ of the approximation of the exact inverse~$\inv{\matBeam}$,
that itself is still a sparse matrix.
Ideally, $\nnzOf{\spaiOf{\matBeam}}$, the number of {\nonzeros} of~$\spaiOf{\matBeam}$, does not exceed~$\nnzOf{\matBeam}$,
since a matrix-vector product with~$\spaiOf{\matBeam}$ must be performed at each Krylov iteration~\cite{Grote1997a}.
Although incomplete LU factorizations fall into that category,
they require considerable effort to parallelize~\cite{Chow2015a}.


To take advantage of the block diagonal structure of~$\matBeam$,
we pursue a fully parallelizable approach to construct an explicit sparse approximate inverse~$\spaiOf{\matBeam}$ of the matrix~$\matBeam$
based on the minimization of the Frobenius norm of the residual matrix~$\matBeam\spaiOf{\matBeam} - \idmat$,
see \cite{Grote1997a,Sedlacek2012a}.
By choosing an appropriate sparsity pattern~$\graphOf{\spaiOf{\matBeam}}$ for the \gls{ac:SPAI}~$\spaiOf{\matBeam}$ from the set~$\graph$ describing all known patterns,
the following least-squares problem needs to be solved:
\begin{equation}
\min_{\graphOf{\spaiOf{\matBeam}} \; \in \; \graph} \normFrobenius{\matBeam\spaiOf{\matBeam} - \idmat}
\label{eq:SparseApproximateInverseMinimizationProblem}
\end{equation}
We represent the minimization procedure to compute~$\spaiOf{\matBeam}$ by the operation~$\spaiOf{\matBeam} \leftarrow \spaiMinimizeOp{\matBeam}{\graphOf{\spaiOf{\matBeam}}}$.
To this end, \eqref{eq:SparseApproximateInverseMinimizationProblem} requires to select an appropriate sparsity pattern~$\graphOf{\spaiOf{\matBeam}}$
and a practical approach to the minimization of the Frobenius norm in a distributed memory environment,
which we will address in \secsref{sec:SPAIGraphSelection}{sec:SPAIFrobeniusNorm}, respectively.

\subsubsection{Selection of a sparsity pattern for the sparse approximate inverse calculation}
\label{sec:SPAIGraphSelection}

The main challenge in~\eqref{eq:SparseApproximateInverseMinimizationProblem} is the selection of a sparsity pattern~$\graphOf{\spaiOf{\matBeam}}$
to be used as input into the minimization procedure.
An appropriate pattern needs to contain enough information of the inverse by retaining high values,
but should also act as a filter to remove small entries in order to reduce fill-in.
Straightforward approaches are based on a static sparsity pattern selection and include the choices~$\graphOf{\spaiOf{\matBeam}} := \graphOf{\matBeam}$
and~$\graphOf{\spaiOf{\matBeam}} := \graphOf{\trans{\matBeam}}$,
which are easy to obtain, but do not guarantee a good approximation quality.
Especially in cases with a partially known sparsity structure of the inverse, {\eg} for block-diagonal matrices such as~$\matBeam$ in~\eqref{eq:SimpleLinearSystem},
more advanced static selections are able to deliver a satisfying approximation,
yet not requiring dynamic pattern selection approaches as proposed in~\cite{Grote1997a}.

In this work, we follow the static pattern selection proposed in~\cite{Chow2001a}
and use powers of a sparsified version~$\threshMat{\graphOf{\matBeam}}$ of the graph~$\graphOf{\matBeam}$ of the original matrix~$\matBeam$
to obtain an enriched sparsity pattern to be used for the minimization in~\eqref{eq:SparseApproximateInverseMinimizationProblem}.
First, $\threshMat{\graphOf{\matBeam}}$ is obtained through a thresholding of~$\graphOf{\matBeam}$ based on the entries in~$\matBeam$
and using a drop-off tolerance~$\spaiDropTol$.
We represent this thresholding by the filter operation
\begin{align*}
\threshMat{\graphOf{\matBeam}} \leftarrow \thresholdingOf{{\matBeam}}{\spaiDropTol}{\graph}
\end{align*}
delivering individual entries~$\threshMat{\graphElementOf{\matBeam}{\indRow}{\indCol}}$
of the filtered graph via
\begin{equation}
\threshMat{\graphElementOf{\matBeam}{\indRow}{\indCol}} :=
\begin{cases}
1 \quad \text{if} \; \indRow=\indCol \; \text{or} \; \abs{\diagEntry{\indRow}^{-\frac{1}{2}} \matBeamEntry{\indRow}{\indCol} \diagEntry{\indRow}^{-\frac{1}{2}}} > \spaiDropTol, \\
0 \quad \text{otherwise},
\end{cases}
\quad \text{where} \quad
\diagEntry{\indRow} :=
\begin{cases}
\abs{\matBeamEntry{\indRow}{\indRow}} \quad \text{if} \; \abs{\matBeamEntry{\indRow}{\indRow}} > 0, \\
1 \quad \text{otherwise}.
\end{cases}
\label{eq:SparseApproximateInverseSparsification}
\end{equation}
The additional Jacobi scaling in~\eqref{eq:SparseApproximateInverseSparsification} simplifies the thresholding and choosing of~$\spaiDropTol$
if~$\matBeam$ is poorly scaled.
In a second step, a refined sparsity pattern~$\graphOf{\matBeam^\spaiRefinementLevel}$
is calculated by taking powers~$\spaiRefinementLevel$ of the sparsified graph~$\threshMat{\graphOf{\matBeam}}$,
reading
\begin{align*}
\graphOf{\spaiOf{\matBeam}} \leftarrow \spaiRefinementOp{\threshMat{\graphOf{\matBeam}}}{\spaiRefinementLevel}.
\end{align*}
The matrix $\matBeam^\spaiRefinementLevel$ is never calculated explicitly, but the powers are directly computed on its sparsified graph~$\threshMat{\graphOf{\matBeam}}$.

\subsubsection{Evaluation of the Frobenius norm on a parallel computer}
\label{sec:SPAIFrobeniusNorm}

Due to the 2-norm compatibility of the Frobenius norm, the problem can be decoupled into a sum of Euclidian norms, reading
\begin{equation*}
\min_{\graphOf{\spaiOf{\matBeam}} \; \in \; \graph} \normFrobenius{\matBeam\spaiOf{\matBeam} - \idmat}
= \sum_{\spaiIndRow=1}^\nRows \min_{\graphOf{\spaiOf{\matBeam}} \; \in \; \graph}
\normTwo{\left(\matBeam\spaiOf{\matBeam} - \idmat\right) \indexedRow{\mao{e}}{\spaiIndRow}}^2
= \sum_{\spaiIndRow=1}^\nRows \min_{\graphOf{\indexedRow{\spaiOf{\matBeam}}{\spaiIndRow}} \; \in \; \indexedRow{\graph}{\spaiIndRow}}
\normTwo{\matBeam\indexedRow{\spaiOf{\matBeam}}{\spaiIndRow} - \indexedRow{\mao{e}}{\spaiIndRow}}^2,
\end{equation*}
which can be solved independently for each row~$\spaiIndRow=1,\hdots,\nRows$ of $\spaiOf{\matBeam}$.
Since the matrix~$\matBeam$ is usually stored in a row-wise distribution,
where each parallel process stores a subset of all rows of~$\matBeam$,
this decomposition renders the method, besides an initial communication step, inherently parallel.
To this end, calculating a row of the \gls{ac:SPAI} means solving a small least-squares problem by applying a dense QR factorization.

\subsubsection{Practical algorithm}
\label{sec:SpaiPracticalAlgorithm}

In practice, one usually does not solve~\eqref{eq:SparseApproximateInverseMinimizationProblem} directly,
but rather combines it with some pre- and post-operations,
foremost the graph computation from \secref{sec:SPAIGraphSelection} and the handling of the Frobenius norm outlined in \secref{sec:SPAIFrobeniusNorm}.
In addition, it appears beneficial to perform a post-filtering of $\spaiOf{\matBeam}$ by dropping all entries
with~$\abs{\indexedRowCol{\spaiOf{\matBeamElement}}{\indRow}{\indCol}} < \spaiDropTol \quad \forall \indRow,\indCol=1,\hdots,\nRows$
to further reduce fill-in of~$\inv{\approxMat{\matBeam}}$ and, thus, the cost of applying the preconditioner~\cite{Chow2001a}.
The post-filtering will be denoted by the operator~$\thresholdingOf{\spaiOf{\matBeam}}{\spaiDropTol}{\matBeam}$.

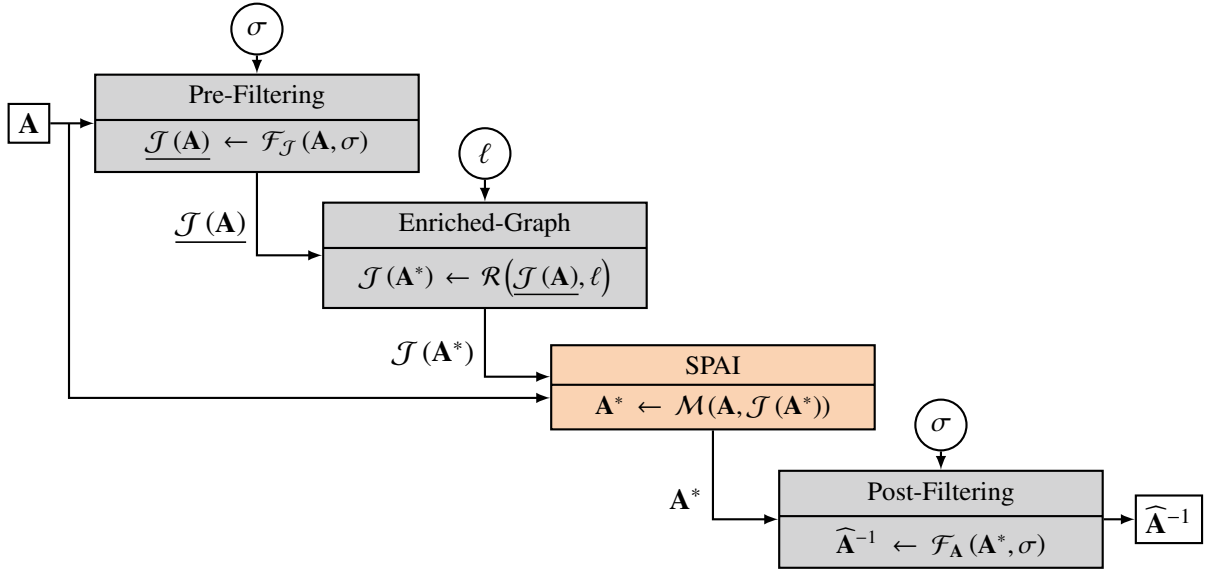
\begin{figure}
\centering

\tikzstyle{box} = [draw, rectangle, thick, node distance=7em, text width=6em, text centered, minimum height=3.5em]
\tikzstyle{line} = [draw, thick, -latex]
\tikzstyle{block} = [thick, rectangle split, draw, rectangle split parts=2, text width=10em, text centered, minimum height=4em]
\tikzstyle{circ} = [thick, draw, circle]

\begin{tikzpicture}
\node [block, node distance=7cm, font=\fontsize{10}{0}\selectfont, fill=colUniBwGr!30] (Pre) at (0,0) {Pre-Filtering
\nodepart[text width=4cm]{two} $\threshMat{\graphOf{\matBeam}} \leftarrow \thresholdingOf{{\matBeam}}{\spaiDropTol}{\graph}$};

\node [block, node distance=7cm, font=\fontsize{10}{0}\selectfont, fill=colUniBwGr!30] (Refine) at (3,-1.75) {Enriched-Graph
\nodepart[text width=4cm]{two} $\graphOf{\spaiOf{\matBeam}} \leftarrow \spaiRefinementOp{\threshMat{\graphOf{\matBeam}}}{\spaiRefinementLevel}$};

\node [block, node distance=7cm, font=\fontsize{10}{0}\selectfont, fill=colUniBwOr!30] (Spai) at (6,-3.5) {SPAI
\nodepart[text width=4cm]{two} $\spaiOf{\matBeam} \leftarrow \spaiMinimizeOp{\matBeam}{\graphOf{\spaiOf{\matBeam}}}$};

\node [block, node distance=7cm, font=\fontsize{10}{0}\selectfont, fill=colUniBwGr!30] (Post) at (9,-5.25) {Post-Filtering
\nodepart[text width=4cm]{two} $\inv{\approxMat{\matBeam}} \leftarrow \thresholdingOf{\spaiOf{\matBeam}}{\spaiDropTol}{\matBeam}$};

\node [draw, thick] (A) at (-3, 0) {$\matBeam$};
\node [draw, thick] (Ainv) at (12, -5.25) {$\inv{\approxMat{\matBeam}}$};
\node [circ] (Taupre) at (0, 1.25) {$\spaiDropTol$};
\node [circ] (Taupost) at (9, -4) {$\spaiDropTol$};
\node [circ] (Refinelevel) at (3, -0.4) {$\spaiRefinementLevel$};

\draw [line] (A.east) -- (Pre.west);
\draw [line] (Pre.south) |- node [above left, midway] {$\threshMat{\graphOf{\matBeam}}$} (Refine.west);
\draw [line] (Refine.south) |- node [above left, midway] {$\graphOf{\spaiOf{\matBeam}}$} ([yshift=+4pt]Spai.west);
\draw [line] (Spai.south) |- node [above left, midway] {$\spaiOf{\matBeam}$} (Post.west);
\draw [line] (Post.east) -- (Ainv.west);	

\draw [line] ([xshift=7pt]A.east) |- ([yshift=-4pt]Spai.west);

\draw [line] (Taupre) -- (Pre.north);
\draw [line] (Taupost) -- (Post.north);
\draw [line] (Refinelevel) -- (Refine.north);

\end{tikzpicture}
\caption{Optional (gray) and mandatory (orange) steps of the \gls{ac:SPAI} computation
and its flow of information with computed data (in boxes) and user parameters (in circles)}
\label{fig:SpaiAlgorithm}
\end{figure}

We summarize all the steps to compute the sparse approximate inverse~$\spaiOf{\matBeam}$ in \figref{fig:SpaiAlgorithm}.
Therein, optional steps are marked in gray, while the mandatory computation of~$\spaiOf{\matBeam}$ is highlighted in orange.
User-given data to configure the individual steps is depicted in circles,
while computed input and output data is put into rectangular boxes.
The arrows indicate the flow of data between the individual steps.

Even though the method can be steered quite effectively by the refinement level~$\spaiRefinementLevel$ and threshold tolerance~$\spaiDropTol$,
there are still several problems to avoid.
The method can still produce rather dense matrices with poor approximation quality.
Depending on the threshold parameter~$\spaiDropTol$,
an aggressive dropping of values might result in a loss of information,
which again results in a poor result and the approximation of the inverse might even be singular.

\subsection{One-level approach for the predictor and corrector step}
\label{sec:SpaiSmoother}

As outlined in \algref{alg:PreconditionerRaw},
the first and last linear equation to be solved during the application of the block preconditioner resemble the predictor step
\begin{equation}
\matBeam \indexedBeamSweep{\mao{\sol}}{\indSweep+\frac{1}{2}} = \indexedBeamSweep{\mao{\residualLinear}}{\indSweep}
\label{eq:PredictorStep}
\end{equation}
and the corrector step
\begin{equation}
\matBeam \indexedBeamSweep{\mao{\sol}}{\indSweep+1} = \indexedBeamSweep{\mao{\residualLinear}}{\indSweep}
-\trans{\matBeamSolid}\indexedSolidSweep{\mao{\sol}}{\indSweep+1},
\label{eq:CorrectorStep}
\end{equation}
respectively.
Both use the beam matrix~$\matBeam$ to update the beam unknowns.
Following the assumptions made in \secref{sec:BlockDiagonalDominance},
traditional smoothing approaches are not applicable to approximate the solutions of the linear systems in~\eqref{eq:PredictorStep} and~\eqref{eq:CorrectorStep}
without a major loss in convergence properties.

Based on the explicit \gls{ac:SPAI} calculation for the approximation of the Schur complement,
a rather good approximation $\inv{\approxMat{\matBeam}}$ is already available.
Following~\cite{Broeker2002a}, the sparse approximate inverse is reused as a smoother by applying
the {\fixedpoint} iteration over index~$\indSpaiSweep$ to \eqref{eq:PredictorStep}, reading
\begin{equation}
\indexedBeamMultiSweep{\mao{\sol}}{\indSweep+\frac{1}{2}}{\indSpaiSweep + 1}
= \indexedBeamMultiSweep{\mao{\sol}}{\indSweep+\frac{1}{2}}{\indSpaiSweep} + \inv{\approxMat{\matBeam}}
\indexedBeamSweep{\mao{\residualLinear}}{\indSweep},
\label{eq:PredictorSparseApproximateInverseSmoother}
\end{equation}
and to \eqref{eq:CorrectorStep}, reading
\begin{equation}
\indexedBeamMultiSweep{\mao{\sol}}{\indSweep+1}{\indSpaiSweep+1}
= \indexedBeamMultiSweep{\mao{\sol}}{\indSweep+1}{\indSpaiSweep} + \inv{\approxMat{\matBeam}}
\left(\indexedBeamSweep{\mao{\residualLinear}}{\indSweep}
-\trans{\matBeamSolid}\indexedSolidSweep{\mao{\sol}}{\indSweep+1}\right).
\label{eq:CorrectorSparseApproximateInverseSmoother}
\end{equation}
A comparison to more traditional smoothers, {\eg} Jacobi and {\GaussSeidel} methods, can also be found in the original publication~\cite{Broeker2002a}.

\begin{remark}[Challenges for multilevel methods]
Expanding the solution procedure of the predictor and corrector steps to an \gls{ac:AMG} approach to solve the beam-related
equation is challenging, as there are still several open questions around the construction of \gls{ac:AMG} hierarchies for beam models:
The beams are represented as \gls{ac:1D} elements, for which a suitable coarsening scheme to form aggregates
for coarser \gls{ac:MG} levels is still part of active research.
Even with appropriate coarsening, fibers discretized with only a few elements would quickly form single-node
aggregates being suboptimal for the restrictor and prolongator construction as well as effectively stalling the coarsening process.
The addition of rotational components into the beam equation introduces another difficulty,
as nodes with a different number of \glspl{ac:DOF} exist, hindering the application of conventional aggregation strategies.
To project and, thus, treat important error modes correctly on coarser levels,
a proper near nullspace has to be constructed, which strongly depends on the underlying beam formulation.
For the use cases presented in this manuscript, which mostly feature a small number of beam elements used to discretize a fiber,
the \gls{ac:SPAI} smoother from~\eqref{eq:PredictorSparseApproximateInverseSmoother} and~\eqref{eq:CorrectorSparseApproximateInverseSmoother} proved to be very competitive
in terms of approximation quality of the inverse, computational efficiency, robustness, and weak scaling behavior,
see \secref{sec:Experiments}.
\end{remark}

\subsection{Multilevel approach for the Schur complement step}
\label{sec:AmgForSchur}

The computationally most demanding part of the preconditioning algorithm involves the solution procedure for the
Schur complement equation given as
\begin{equation*}
\schurMatrix \indexedSolidSweep{\mao{\sol}}{\indSweep+1} = \indexedSolidSweep{\mao{\residualLinear}}{\indSweep}
-\matSolidBeam \indexedBeamSweep{\mao{\sol}}{\indSweep+\frac{1}{2}}.
\end{equation*}
Due to the explicit sparse approximation of the inverse of~$\matBeam$
used to form the approximate Schur complement~$\approxMat{\schurMatrix} := \matSolid - \matSolidBeam \inv{\approxMat{\matBeam}} \trans{\matBeamSolid}$,
the resulting matrix~$\approxMat{\schurMatrix}$ is rather
dense compared to calculations using one of the diagonal approximation approaches for~$\matBeam$
given in \secref{sec:SchurComplementBlockFactorization}.
The inverse of the Schur complement is approximated by a standard aggregation-based \gls{ac:AMG} method.
As level smoother, a one-level domain decomposition with overlap~$\iluOverlap$ and an incomplete LU factorization with
fill-in~$\iluLevelFill$ and thresholding~$\iluDropTol$ of small entries  is applied, often abbreviated as ILUT~\cite{Saad1994a}.
To accelerate convergence, a smoothing of the tentative prolongation operator basis functions can be done,
also known as \gls{ac:SAAMG}~\cite{Vanek1996a}.
This however leads to a higher fill-in of the coarse system matrix representations increasing the computational cost
of the preconditioner setup. Especially the Galerkin product for the calculation of the coarse level operator and
the incomplete LU factorization smoother are negatively influenced by the additional fill-in. Furthermore,
the operator smoothing is classically based on a Jacobi method, which heavily relies on diagonal dominance.
Thus, in some cases it makes sense to skip the prolongator smoothing and take advantage of the robustness of
\gls{ac:PAAMG} by applying aggregate-wise constant basis functions~\cite{Thomas2019a}.
Caused by the additional {\mixeddimensional} coupling terms in the bulk field's matrix,
a similar increase in the fill-in of the coarse level operators of a {\RugeStueben} \gls{ac:AMG} method has been observed in~\cite{Cerroni2019a}.

\subsection{Approximate block factorization preconditioner for {\beamsolid} interaction}
\label{sec:FinalPreconditioner}
The components described in \secsrangeref{sec:SchurComplementBlockFactorization}{sec:AmgForSchur} are now put together
to tailor the block preconditioner from \algref{alg:PreconditionerRaw} to the specifics of {\mixeddimensional} {\beamsolid} interaction,
yielding the final preconditioner summarized in \algref{alg:Preconditioner}.
\begin{algorithm}
\caption{Approximate block factorization preconditioner for {\fibersolid} coupling}
\label{alg:Preconditioner}
\SetKwProg{myproc}{Procedure}{}{}
\myproc{Preconditioner($\indSweep_{max}$)}
{
// Pre-compute SPAI of $\matBeam$ \newline
$\inv{\approxMat{\matBeam}} \leftarrow \thresholdingOf{\spaiMinimizeOp{\matBeam}{\spaiRefinementOp{\thresholdingOf{{\matBeam}}{\spaiDropTol}{\graph}}{\spaiRefinementLevel}}}{\spaiDropTol}{\matBeam}$

// Form explicit, approximate Schur complement \newline
$\approxMat{\schurMatrix} = \matSolid - \matSolidBeam \inv{\approxMat{\matBeam}} \trans{\matBeamSolid}$

\For{$\indSweep = 1, \dots, \indSweep_{max}$}
{
	$\left(
	\begin{matrix}
	\indexedBeamSweep{\mao{\residualLinear}}{\indSweep}\\
	\indexedSolidSweep{\mao{\residualLinear}}{\indSweep}\\
	\end{matrix}
	\right)
	=
	\left(
	\begin{matrix}
	\indexedBeam{\mao{\rhs}}\\
	\indexedSolid{\mao{\rhs}}\\
	\end{matrix}
	\right)
	-
	\left(
	\begin{matrix}
	\matBeam & \trans{\matBeamSolid} \\
	\matSolidBeam & \matSolid \\
	\end{matrix}
	\right)
	\left(
	\begin{matrix}
	\indexedBeamSweep{\mao{\sol}}{\indSweep}\\
	\indexedSolidSweep{\mao{\sol}}{\indSweep}\\
	\end{matrix}
	\right)$
	
    // Prediction step: solve for $\indexedBeamSweep{\mao{\sol}}{\indSweep+\frac{1}{2}}$ with SPAI smoother
    
    \For{$\indSpaiSweep = 1 \dots \indSpaiSweep_{max}$}
    {
	$\indexedBeamMultiSweep{\mao{\sol}}{\indSweep+\frac{1}{2}}{\indSpaiSweep + 1}
	= \indexedBeamMultiSweep{\mao{\sol}}{\indSweep+\frac{1}{2}}{\indSpaiSweep} + \inv{\approxMat{\matBeam}}
	\indexedBeamSweep{\mao{\residualLinear}}{\indSweep}$
	}

    // Schur complement step: solve for $\indexedSolidSweep{\mao{\sol}}{\indSweep+1}$ with AMG
    
    $\approxMat{\schurMatrix} \indexedSolidSweep{\mao{\sol}}{\indSweep+1} = \indexedSolidSweep{\mao{\residualLinear}}{\indSweep}
    -\matSolidBeam \indexedBeamSweep{\mao{\sol}}{\indSweep+\frac{1}{2}}$
    
    // Correction step: solve for $\indexedBeamSweep{\mao{\sol}}{\indSweep+1}$ with SPAI smoother
    
    \For{$\indSpaiSweep = 1 \dots \indSpaiSweep_{max}$}
    {
    $\indexedBeamMultiSweep{\mao{\sol}}{\indSweep+1}{\indSpaiSweep+1}
	= \indexedBeamMultiSweep{\mao{\sol}}{\indSweep+1}{\indSpaiSweep} + \inv{\approxMat{\matBeam}}
	\left(\indexedBeamSweep{\mao{\residualLinear}}{\indSweep}
	-\trans{\matBeamSolid}\indexedSolidSweep{\mao{\sol}}{\indSweep+1}\right)$
    }

}
\Return $\left(\begin{matrix} \indexedBeamSweep{\mao{\sol}}{\indSweep_{max}} \\ \indexedSolidSweep{\mao{\sol}}{\indSweep_{max}}\end{matrix} \right)$
}
\end{algorithm}
In a pre-computation step, an explicit representation~$\inv{\approxMat{\matBeam}}$ of the \gls{ac:SPAI} of the beam matrix~$\matBeam$ is formed
and also used to compute an approximation~$\approxMat{\schurMatrix}$ to the Schur complement.
Based on the sweep index~$\indSweep$ of the preconditioner,
the main computation loop consists of the three steps:
First, we predict the beam unknowns~$\indexedBeamSweep{\mao{\sol}}{\indSweep+\frac{1}{2}}$ by using the \gls{ac:SPAI} as a smoother.
Then, we solve the Schur complement equation for the solid unknowns~$\indexedSolidSweep{\mao{\sol}}{\indSweep+1}$ using an \gls{ac:AMG} method.
Finally, we again use the \gls{ac:SPAI} as a smoother to correct the beam unknowns to their final values~$\indexedBeamSweep{\mao{\sol}}{\indSweep+1}$.

In terms of computational effort,
the computation of the \gls{ac:SPAI} as outlined in \secref{sec:SPAI} comes at a certain cost,
however is perfectly parallelizable
and is used at three steps in \algref{alg:Preconditioner}:
once in the approximation of the Schur complement and twice to update the beam solution in the predictor and corrector step.
A single application of \gls{ac:SPAI} as a smoother for the predictor or corrector step boils down to a sparse matrix-vector multiplication.

\subsection{Comparison to existing methods in literature}
\label{sec:ComparisonWithLiterature}

Having discussed all the details of our proposed preconditioner,
we now want to outline its commonalities and differences compared to preconditioners available in the literature.
In particular, we will discuss the work in~\cite{Kuchta2016a,Kuchta2019a,Baerland2019a,Cerroni2019a,Dimola2023a}.

In contrast to our work,
the preconditioners from~\cite{Kuchta2016a,Kuchta2019a,Baerland2019a,Dimola2023a} are all tailored to saddle point systems.
In the construction of a block diagonal preconditioner, they use only the diagonal part of an $\matFactL\matFactD\matFactU$ factorization of the original matrix.
In~\cite{Kuchta2019a,Baerland2019a,Dimola2023a},
the arising Schur complement is approximated by a spectrally equivalent fractional Laplacian.
\Gls{ac:AMG} is used to tackle matrices arising from the \gls{ac:3D} bulk field,
while the embedded \gls{ac:1D} domains are always handled by a direct solver.

The case of $2\times 2$ systems as they arise for example from a penalty regularization
is covered in \cite{Cerroni2019a}.
Due to the low cost of inverting the matrix of the \gls{ac:1D} domain,
they use an exact factorization to represent the Schur complement.

Similar to~\cite{Cerroni2019a},
we base the construction of the Schur complement on the original block matrix,
however avoid the exact inversion of the beam {\subblock}~$\matBeam$ and rather use its \gls{ac:SPAI} $\inv{\approxMat{\matBeam}}$.
In line with all available preconditioners,
the matrix block associated with the \gls{ac:3D} bulk discretization is tackled by an \gls{ac:AMG} method.
In our work though,
the matrix arising from the \gls{ac:1D} discretization is never explicitly inverted or factorized,
but also treated in an approximate fashion ({\ie} using the \gls{ac:SPAI} as a smoother, {\cf} \secref{sec:SpaiSmoother})
to facilitate large numbers of embedded fibers as well as finely resolved fiber discretizations.
Regarding the underlying physical problems and applications,
existing work is concerned with transport problems on the \gls{ac:1D} domain,
while our work is the first preconditioner tailored to {\multidimensional} \glspl{ac:PDE} on the \gls{ac:1D} domain,
in particular geometrically exact beam models with up to nine \glspl{ac:DOF} per mesh node depending on the actual beam model at hand.

\section{Numerical experiments}
\label{sec:Experiments}

We present numerical examples to illustrate the influence of the different algorithmic parameters of the \gls{ac:SPAI} computation
proposed in \secref{sec:SPAI}, to study and demonstrate the weak scaling behavior of the proposed preconditioner, to investigate
the robustness of the proposed method regarding material parameters and geometric properties and to showcase its applicability
to practical problems in civil engineering.
All computations are done with our in-house {\multiphysics} code {\baci}~\cite{BaciURL},
which is built upon the {\trilinos} project~\cite{Heroux2005a,TrilinosURL}.
All preconditioning operations are done through the multigrid package {\muelu}~\cite{BergerVergiat2023a} and its dependencies within the {\trilinos} project.
For the generation of the beam geometries, we rely on {\meshpy}~\cite{MeshPyWebsite}.

\subsection{Numerical study of the sparse approximate inverse calculation for the beam {\subblock}}
\label{sec:NumericalStudySpai}

A major component of a Schur complement based preconditioner is the approximation of the
inverse appearing in the Schur complement calculation itself.
In the presented approach, the combination of the drop-off tolerance $\spaiDropTol$ of small values and the allowed fill-in (indirectly described by the refinement level~$\spaiRefinementLevel$)
during the sparse approximate inverse calculation have a great influence on the quality of the sparse approximate inverse~$\spaiOf{\matBeam}$
and, thus, on the convergence behavior of the block preconditioning method.
In the following, we investigate test cases with different sparsity patterns of the sub-matrix $\matBeam$
and study the impact of different choices of both~$\spaiDropTol$ and~$\spaiRefinementLevel$
on the quality of the sparse approximate inverse as well as on the convergence behavior of the preconditioned linear solver.

\begin{figure}
\centering
\begin{tikzpicture}
\node at (0,0) {\includegraphics[width=0.5\textwidth]{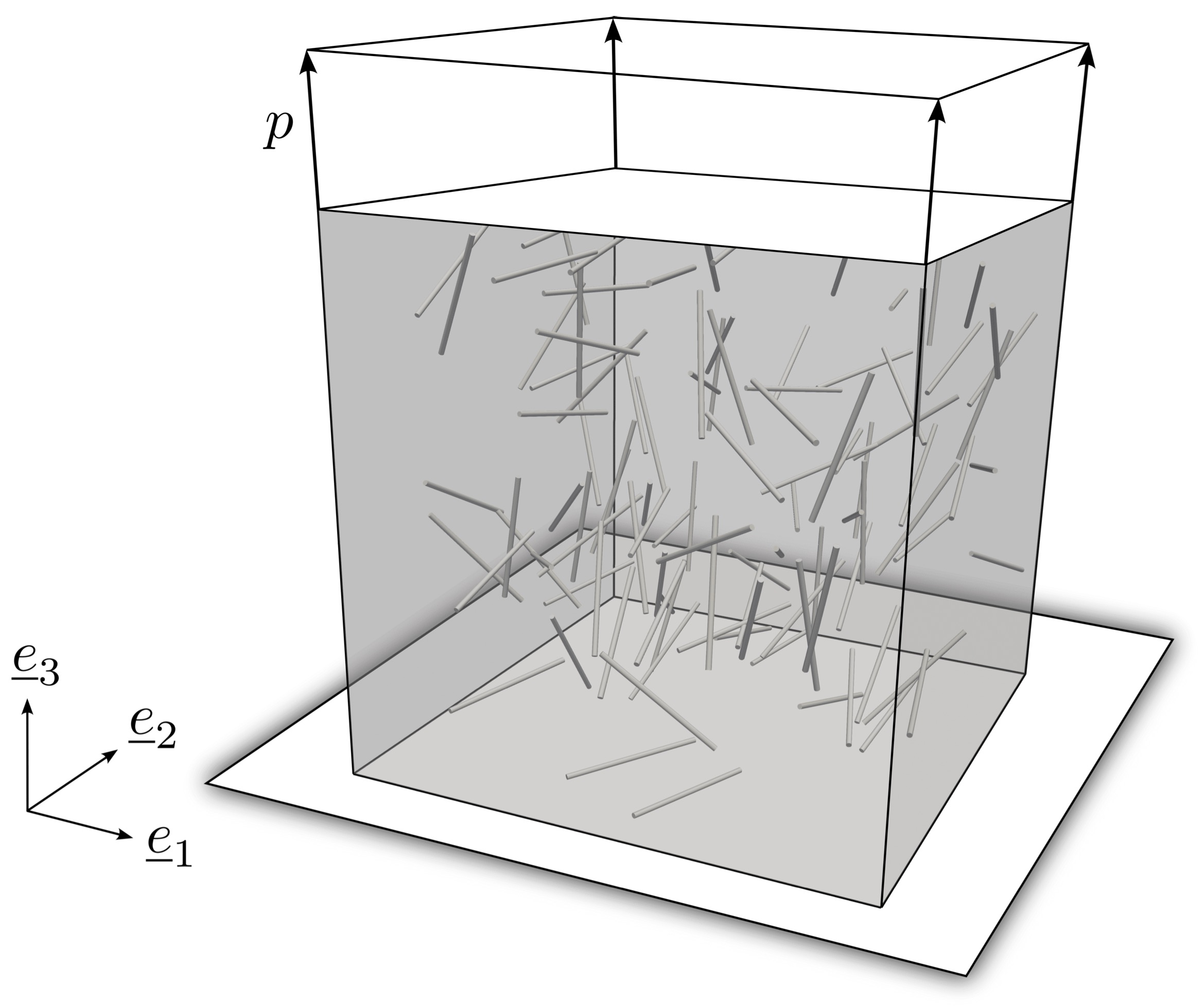}};
\node [fill=white] at (-2.3, 2.5) {$\distributedLoad$};
\node [fill=white] at (-2.9,-2.4) {$\teo{e}_1$};
\node [fill=white] at (-3.0,-1.5) {$\teo{e}_2$};
\node [fill=white] at (-3.9,-1.08) {$\teo{e}_3$};
\end{tikzpicture}
\caption{Geometry and setup for the numerical study of the sparse approximate inverse calculation: a solid cube with edge length~$\indexedSolid{l} = \qty{1}{\meter}$
is randomly filled with fibers of the same length~$\indexedBeam{l} = \qty{0.25}{\meter}$,
clamped at its bottom and loaded with a distributed external load~$\distributedLoad=\qty{1}{\newton/\square\meter}$.}
\label{fig:SolidBlockRandomFibers}
\end{figure}
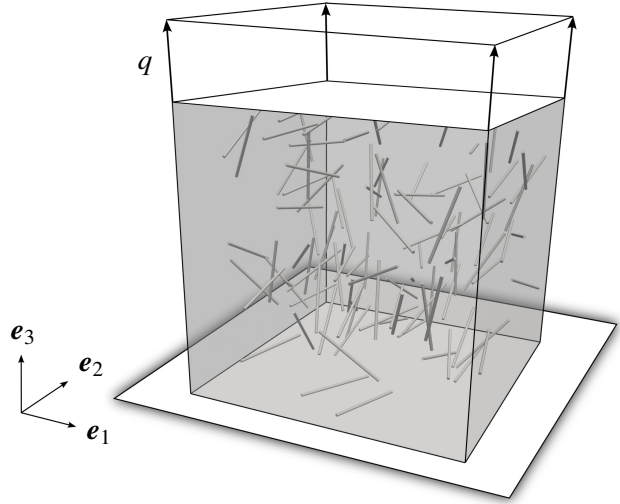

To this end, we consider a simple \gls{ac:3D} {\beamsolid} interaction problem as shown in \figref{fig:SolidBlockRandomFibers}:
A solid cube with edge length~$\indexedSolid{l} = \qty{1}{\meter}$ is filled with randomly placed straight fibers,
such that the {\beamsolid} volume ratio is $\textfrac{\indexedBeam{\volume}}{\indexedSolid{\volume}}\approx 0.2\%$
and that the fibers do not stick out of the solid volume.
The cube is clamped at its bottom surface and loaded with a constant tensile load of $\distributedLoad=\qty{1}{\newton/\square\meter}$ at its top face.
The solid is modeled by a {\StVenantKirchhoff} material
(Young's modulus~$\indexedSolid{\youngs} = \qty{1}{\newton/\square\meter}$, Poisson's ratio~$\indexedSolid{\poisson}=0.3$)
and discretized by first-order hexahedral finite elements.
The fibers are represented by either {\torsionfree} {\KirchhoffLove} beams (TF) or {\SimoReissner} beam elements (SR)
using the following parameters: Young's modulus~$\indexedBeam{\youngs}= \qty{10}{\newton/\square\meter}$,
radius~$\beamRadius = \qty{0.005}{\meter}$ and length~$\indexedBeam{l} = \qty{0.25}{\meter}$. In addition,
a Poisson's ratio of~$\indexedBeam{\poisson}=0.0$ is used in the cases with SR beam models.
The coupling conditions are enforced with penalty parameters~$\penaltyParamDisp = \qty{10}{\newton/\square\meter}$ and $\penaltyParamRot = \qty{10}{\newton\meter/\meter}$,
if applicable.

\begin{figure}
\centering
\subfigure[Test case I: one beam element per fiber]{\label{fig:SparsityTestCase1}\includegraphics[width=150pt]{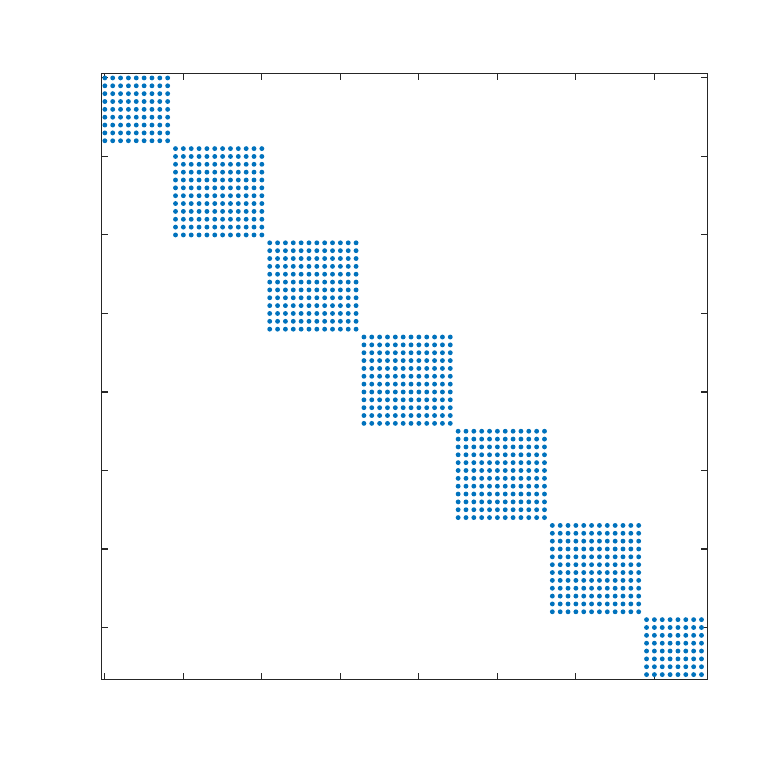}}
\subfigure[Test case II: four beam elements per fiber]{\label{fig:SparsityTestCase2}\includegraphics[width=150pt]{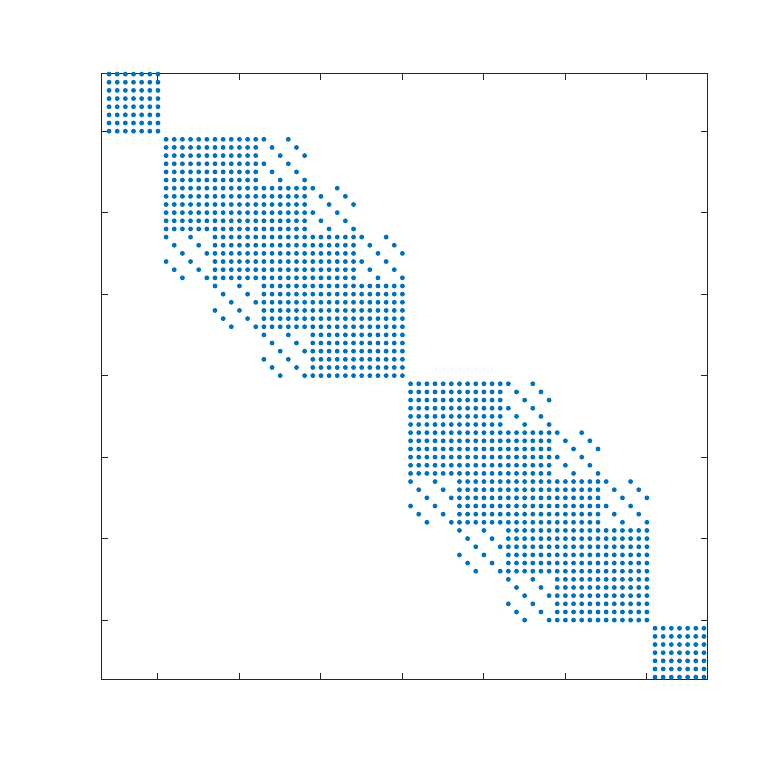}}
\subfigure[Test case III: random number (1, 2, 3 or 4) of beam elements per fiber]{\label{fig:SparsityTestCase3}\includegraphics[width=150pt]{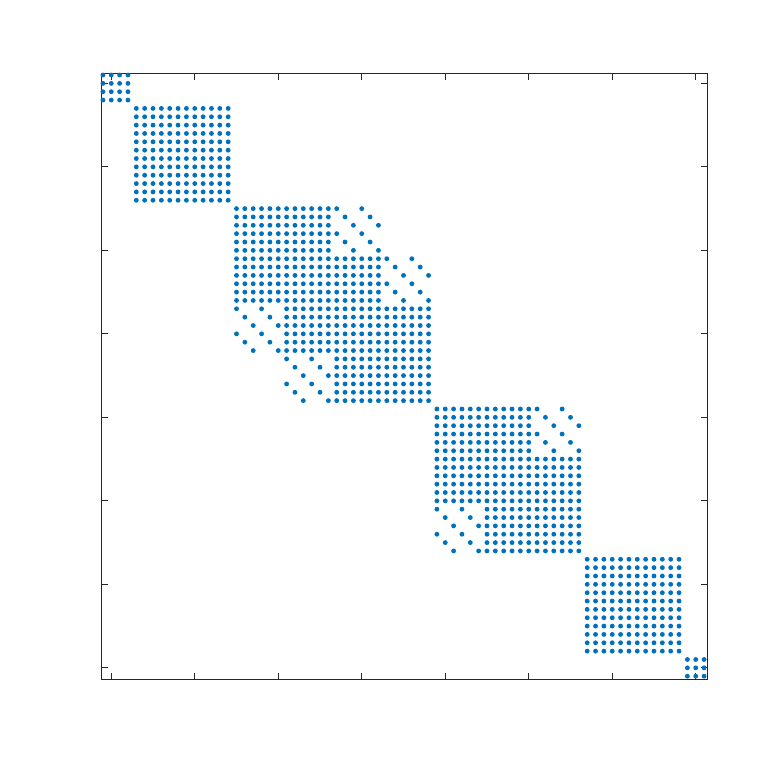}}
\caption{Partial visualization of the sparsity structure~$\graphOf{\matBeam}$ of matrix~$\matBeam$ for test cases~I--III}
\label{fig:SparsityTestCases}
\end{figure}

For the calculation of the sparse approximate inverse, the matrix block~$\matBeam$ describing the contribution of the fibers is of particular interest.
Therefore, six test cases with different beam formulations and varying number of beam finite elements per fiber are set up to trigger different sparsity patterns~$\graphOf{\matBeam}$.
In the test cases~I and~IV, the fibers are discretized by just one beam element, resulting in a block-diagonal matrix with fully populated {\subblocks}.
For test cases~II and~V, four beam elements are used per fiber, resulting in bigger and sparser {\subblock}s.
Test cases~III and~VI mix the other scenarios by randomly using between one and four beam elements per fiber.
All test cases and their respective matrix sizes and number of {\nonzeros} of $\graphOf{\matBeam}$ are summarized in \tabref{tab:SPAITestCaseSetup}.
The resulting sparsity patterns are illustrated exemplarily for the test cases~I--III in \figref{fig:SparsityTestCases}.

\begin{table}
\centering
\caption{Matrix size of $\matBeam$ and number of {\nonzeros} of the graph $\graphOf{A}$ for the six different test cases}
\label{tab:SPAITestCaseSetup}
\begin{tabular}{l c c c}
\hline
~ & beam model & size($\matBeam$) & $\nnzOf{\graphOf{\matBeam}}$ \\
\hline
Test case I & TF & 1224 $\times$ 1224 & \num{14688} \\
Test case II & TF & 3060 $\times$ 3060 & \num{55077} \\
Test case III & TF & 1854 $\times$ 1854 & \num{28546} \\
\hline
Test case IV & SR & 2142 $\times$ 2142 & \num{44982} \\
Test case V & SR & 5814 $\times$ 5814 & \num{184518} \\
Test case VI & SR & 3402 $\times$ 3402 & \num{92862} \\
\hline
\end{tabular}
\end{table}

The overall simulation is of quasi-static nature and imposes the total load over the course of two load steps,
which is sufficient for investigating the key features of the linear solver like the iteration count and setup/solve timings.
The {\nonlinear} solver converges,
if the {\nonlinear} residual $\normTwo{\mao{\residualNonlinear}}$ drops below $10^{-6}$
and if the full displacement increment $\normTwo{\Delta \mao{\disp}}$ is smaller than $10^{-8}$.
In each {\nonlinear} iteration, a linear system is solved using a preconditioned GMRES method~\cite{Saad1986a}.
The preconditioner is configured as follows: the number of sweeps through the preconditioner is set to~$\indSweep=1$,
and the number of iterations for the SPAI smoother is chosen as~$\indSpaiSweep=1$.
The Schur complement equation is solved with a \gls{ac:SAAMG} scheme with an ILUT
smoother with overlap~$\iluOverlap = 1$, fill-in level~$\iluLevelFill=1$ and
drop-off tolerance~$\iluDropTol=10^{-4}$.
The linear solver is assumed to be converged
if the relative residual~$\normTwo{\indexedIter{\mao{\residualLinear}}{\indLinIter}} / \normTwo{\indexedIter{\mao{\residualLinear}}{0}}$
falls below $10^{-8}$.
All simulations are done in serial on a single processor.

Using \glspl{ac:SPAI},
the convergence of the linear solver is tightly related to the parameters chosen for the \gls{ac:SPAI} calculation.
In \tabref{tab:SPAINumericalStudy}, different combinations of the drop-off tolerance~$\spaiDropTol$ and refinement level~$\spaiRefinementLevel$ are given for each of the six test cases.
Each value pair represents the largest possible drop-off value~$\spaiDropTol$ possible for a fixed refinement level~$\spaiRefinementLevel$,
such that the linear solver still converges.
For certain values of~$\spaiRefinementLevel$,
no convergence could be a achieved at all for some test cases,
even with a very small drop-off tolerance~$\spaiDropTol$.
In these situations, an appropriate value of~$\spaiDropTol$ is chosen,
such that only explicit zero values are dropped from the matrix graph
to still be able to make a fair comparison with the other test cases regarding the error norm and number of {\nonzero} entries in the approximate inverse.
In addition, the number of {\nonzero} entries of the filtered graph~$\threshMat{\graphOf{\matBeam}}$ used as starting point for the sparsity pattern construction
as well as the number of {\nonzero} entries of the graph of the inverse approximation~$\graphOf{\spaiOf{\matBeam}}$ are given.
The number of {\nonzeros} for the unfiltered graph, $\nnzOf{\graphOf{\matBeam}}$, is illustrated in \tabref{tab:SPAITestCaseSetup} for comparison.
The behavior of the iterative linear solver is assessed by the averaged number of iterations per nonlinear solver step and three timings
concerning the preconditioner setup time~$\tSetup$,
the time spent for solving the linear system~$\tSolve$, and the total time~$\tTotal = \tSetup + \tSolve$.
The overall quality of the inverse approximation~$\spaiOf{\matBeam}$ is quantified by the error norm relative to the exact inverse~$\inv{\matBeam}$.

\begin{table}
\centering
\caption{Comparison of the influence of different values of $\spaiDropTol$ and $\spaiRefinementLevel$ on the number of linear solver iterations and timings}
\label{tab:SPAINumericalStudy}
\begin{tabular}{l l l l l l l l l l}
	\hline
    ~ & $\spaiDropTol$ & $\spaiRefinementLevel$ & $\nnzOf{\threshMat{\graphOf{\matBeam}}}$ & $\nnzOf{\graphOf{\spaiOf{\matBeam}}}$ & $\#$iter & \multicolumn{3}{c}{CPU time ($\unit{\second}$)} & rel. error norm\\
    ~ & ~      & ~   & ~   &  ~         & ~              & $\tSetup$ & $\tSolve$ & $\tTotal$ & $\frac{\normFrobenius{\spaiOf{\matBeam} - \inv{\matBeam}}}{\normFrobenius{\inv{\matBeam}}}$ \\
	\hline
    Test case I & $10^{-10}$ & $1$ & $\num{14572}$ & $\num{14572}$ & 10 & $3.390$ & $0.755$ & $4.145$ & $2.100\cdot 10^{-3}$ \\
    ~           & $10^{-8}$ & $2$ & $\num{11576}$ & $\num{14688}$ & 10 & $3.338$ & $0.754$ & $4.092$ & $1.248\cdot 10^{-12}$ \\
    ~           & $10^{-6}$ & $3$ & $\num{9552}$ & $\num{14688}$ & 10 & $3.464$ & $0.750$ & $4.214$ & $1.248\cdot 10^{-12}$ \\
    \hline
    Test case II & $10^{-11}$ & $1$ & $\num{44335}$ & $\num{44335}$ & - & - & - & - & $7.653\cdot 10^{-1}$ \\
    ~            & $10^{-10}$ & $2$ & $\num{37222}$ & $\num{86861}$ & 10 & $3.863$ & $0.781$ & $4.644$ & $6.000\cdot 10^{-12}$ \\
    ~            & $10^{-9}$ & $3$ & $\num{32436}$ & $\num{91592}$ & 10 & $3.732$ & $0.800$ & $4.532$ & $1.981\cdot 10^{-14}$ \\
    \hline
    Test case III & $10^{-12}$ & $1$ & $\num{26966}$ & $\num{26966}$ & 20 & $3.559$ & $1.463$ & $5.022$ & $1.415\cdot 10^{-1}$ \\
    ~             & $10^{-10}$ & $2$ & $\num{23919}$ & $\num{35961}$ & 10 & $3.859$ & $0.819$ & $4.678$ & $3.433\cdot 10^{-13}$ \\
    ~             & $10^{-8}$ & $3$ & $\num{17544}$ & $\num{35814}$ & 10 & $3.830$ & $0.839$ & $4.669$ & $3.433\cdot 10^{-13}$ \\
    \hline
    Test case IV & $10^{-12 }$ & $1$ & $\num{26662}$ & $\num{26662}$ & - & - & - & - & $9.277\cdot 10^{-1}$ \\
    ~           & $10^{-11}$ & $2$ & $\num{26614}$ & $\num{44978}$ & 9 & $3.901$ & $0.719$ & $4.620$ & $1.082\cdot 10^{-12}$ \\
    ~           & $10^{-7}$ & $3$ & $\num{16748}$ & $\num{42641}$ & 9 & $4.100$ & $0.761$ & $4.861$ & $1.081\cdot 10^{-12}$ \\
    \hline
    Test case V & $10^{-12}$ & $1$ & $\num{92731}$ & $\num{92731}$ & - & - & - & - & $9.519\cdot 10^{-1}$ \\
    ~            & $10^{-12}$ & $2$ & $\num{92731}$ & $\num{249034}$ & - & - & - & - & $1.528\cdot 10^{-1}$ \\
    ~            & $10^{-11}$ & $3$ & $\num{87453}$ & $\num{331152}$ & 9 & $5.172$ & $0.834$ & $6.006$ & $1.954\cdot 10^{-7}$ \\
    \hline
    Test case VI & $10^{-12}$ & $1$ & $\num{49614}$ & $\num{49614}$ & - & - & - & - & $9.530\cdot 10^{-1}$ \\
    ~             & $10^{-12}$ & $2$ & $\num{49614}$ & $\num{109705}$ & - & - & - & - & $5.970\cdot 10^{-2}$ \\
    ~             & $10^{-9}$ & $3$ & $\num{38420}$ & $\num{121400}$ & 10 & $4.187$ & $0.794$ & $4.981$ & $3.829\cdot 10^{-12}$ \\
    \hline
\end{tabular}
\end{table}

In test case~I, the dense {\subblock}s of the block-diagonal sparsity pattern of~$\matBeam$ already lead to the exact graph of the inverse, which makes the approximation rather simple.
For the given parameter combinations the upper bound of number of {\nonzeros} to exactly compute~$\inv{\matBeam}$ is quickly reached resulting in low iteration counts and error norms.
For the given problem, the setup timings are nearly identical, with only the solver timings for the first parameter combination taking a bit longer due to a not fully populated sparsity pattern.

The second test case is based on rather sparse {\subblock}s of bigger size compared to test case~I.
Without a pattern refinement, the linear solver did not converge as the approximation quality of the inverse is not sufficient.
For higher refinement levels, the number of {\nonzeros} in~$\graphOf{\spaiOf{\matBeam}}$ quickly increases resulting in convergence of the method with still acceptable timings.
Yet, there is not a lot of flexibility in choosing the drop-off tolerance~$\spaiDropTol$ to still retain convergence.

As test case~III is a combination of the first two problems, the resulting behavior is a mix of these.
Using the initial sparsity graph for the inverse approximation results in $20$ linear solver iterations until convergence,
which explains the high solving time~$\tSolve$.
On the other hand,
the setup time~$\tSetup$ is comparable to the other tests.
For increased refinement levels~$\spaiRefinementLevel$, a similar behavior as in test cases~I and~II is observed.

For test case~IV, the beam elements are switched to a {\SimoReissner} formulation,
which contributes additional rotational degrees of freedom into~$\matBeam$. In contrast to test case~I,
this results in already sparse {\subblock}s for using one beam element per fiber. Therefore,
using just the graph~$\graphOf{\matBeam}$
results in a poor approximation of the sparsity pattern of the inverse and, thus, leads to no convergence.
Using higher refinement levels~$\spaiRefinementLevel$ to enrich the input graph for the \gls{ac:SPAI} computation quickly
heals this problem and even allows one to work with a more aggressive dropping scheme, {\ie} using larger values for~$\spaiDropTol$.

Test cases~V and~VI show a similar behavior and only converge with~$\spaiRefinementLevel=3$. The additional beam
elements used per fiber increase the block size of each {\subblock} and thus leave more room for possible sparse
approximations for the inverse. The static approach for choosing an appropriate sparsity pattern for the inverse presented
in \secref{sec:SpaiPracticalAlgorithm} is still able to produce good approximations, thus leading to convergence of the
linear solver, but only for rather dense representations of~$\spaiOf{\matBeam}$.

In conclusion, a robust parameter combination is highly problem dependent and necessary to achieve convergence of the linear solver.
In the presented cases, a refinement level of~$\spaiRefinementLevel=3$ and a drop-off tolerance of~$\spaiDropTol=10^{-9}$
to get rid of small values polluting the sparsity pattern showed to be sufficient to enable convergence
even for challenging cases.

\subsection{Weak scaling behavior}
\label{sec:WeakScalingStudy}

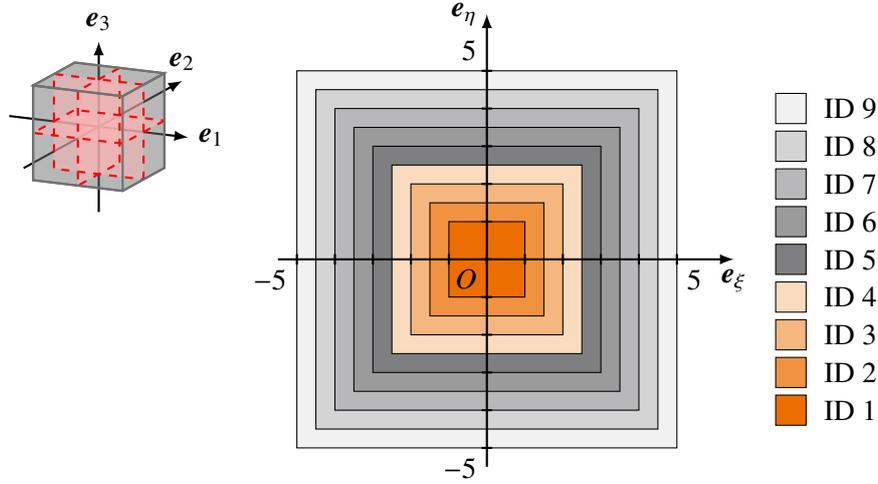
\begin{figure}
\centering

\begin{tikzpicture}

\def\h{0.5}
\def\dist{0.07}
\tikzstyle{axisline}=[thick,-{latex}]

\draw [fill=colUniBwGr!10] (-5*\h,-5*\h) rectangle (5*\h,5*\h);
\draw [fill=colUniBwGr!30] (-4.5*\h,-4.5*\h) rectangle (4.5*\h,4.5*\h);
\draw [fill=colUniBwGr!50] (-4*\h,-4*\h) rectangle (4*\h,4*\h);
\draw [fill=colUniBwGr!70] (-3.5*\h,-3.5*\h) rectangle (3.5*\h,3.5*\h);
\draw [fill=colUniBwGr!90] (-3*\h,-3*\h) rectangle (3*\h,3*\h);
\draw [fill=colUniBwOr!25] (-2.5*\h,-2.5*\h) rectangle (2.5*\h,2.5*\h);
\draw [fill=colUniBwOr!50] (-2*\h,-2*\h) rectangle (2*\h,2*\h);
\draw [fill=colUniBwOr!75] (-1.5*\h,-1.5*\h) rectangle (1.5*\h,1.5*\h);
\draw [fill=colUniBwOr] (-1*\h,-1*\h) rectangle (1*\h,1*\h);

\draw [axisline] (-5.5*\h,0) -- (6.5*\h,0);
\draw [axisline] (0,-5.5*\h) -- (0,6.5*\h);

\foreach \i in {-5,-4,...,5}{
  \draw [thick] (-\dist, \i*\h) -- (\dist, \i*\h);
  \draw [thick] (\i*\h, -\dist) -- (\i*\h, \dist);
}

\node [below left] at (0,0) {$O$};
\node [below] at (6.5*\h,0) {$\teo{e}_{\indBaseVecOne}$};
\node [left] at (0,6.5*\h) {$\teo{e}_{\indBaseVecTwo}$};

\node [below right] at (5*\h,0) {$5$};
\node [below left] at (-5*\h,0) {$-5$};
\node [above left] at (0,5*\h) {$5$};
\node [below left] at (0,-5*\h) {$-5$};


\node [draw, rectangle, fill=colUniBwGr!10, minimum width=0.4cm, minimum height=0.4cm] at (8*\h, 4*\h) {};
\node [right=0.3cm,align=left] at (8*\h, 4*\h) {ID 9};

\node [draw, rectangle, fill=colUniBwGr!30, minimum width=0.4cm, minimum height=0.4cm] at (8*\h, 3*\h) {};
\node [right=0.3cm,align=left] at (8*\h, 3*\h) {ID 8};

\node [draw, rectangle, fill=colUniBwGr!50, minimum width=0.4cm, minimum height=0.4cm] at (8*\h, 2*\h) {};
\node [right=0.3cm,align=left] at (8*\h, 2*\h) {ID 7};

\node [draw, rectangle, fill=colUniBwGr!70, minimum width=0.4cm, minimum height=0.4cm] at (8*\h, 1*\h) {};
\node [right=0.3cm,align=left] at (8*\h, 1*\h) {ID 6};

\node [draw, rectangle, fill=colUniBwGr!90, minimum width=0.4cm, minimum height=0.4cm] at (8*\h, 0*\h) {};
\node [right=0.3cm,align=left] at (8*\h, 0*\h) {ID 5};

\node [draw, rectangle, fill=colUniBwOr!25, minimum width=0.4cm, minimum height=0.4cm] at (8*\h, -1*\h) {};
\node [right=0.3cm,align=left] at (8*\h, -1*\h) {ID 4};

\node [draw, rectangle, fill=colUniBwOr!50, minimum width=0.4cm, minimum height=0.4cm] at (8*\h, -2*\h) {};
\node [right=0.3cm,align=left] at (8*\h, -2*\h) {ID 3};

\node [draw, rectangle, fill=colUniBwOr!75, minimum width=0.4cm, minimum height=0.4cm] at (8*\h, -3*\h) {};
\node [right=0.3cm,align=left] at (8*\h, -3*\h) {ID 2};

\node [draw, rectangle, fill=colUniBwOr, minimum width=0.4cm, minimum height=0.4cm] at (8*\h, -4*\h) {};
\node [right=0.3cm,align=left] at (8*\h, -4*\h) {ID 1};

\tdplotsetmaincoords{-75}{25}
\begin{scope}[tdplot_main_coords,shift={(-7.5,4)},scale=1.3]

\draw[axisline] (-1.0,0,0) -- (1.0,0,0) node[right] {$\teo{e}_1$};
\draw[axisline] (0,-1.8,0) -- (0,2.0,0) node[above] {$\teo{e}_2$};
\draw[axisline] (0,0,0.9) -- (0,0,-0.9) node[above] {$\teo{e}_3$};

\fill[fill=colUniBwGr,fill opacity=0.5] (-0.5,-0.5,-0.5) -- (0.5,-0.5,-0.5) -- (0.5,0.5,-0.5) -- (-0.5,0.5,-0.5) -- cycle;
\fill[thick,draw=colUniBwGr,fill=colUniBwGr,fill opacity=0.5] (-0.5,-0.5,-0.5) -- (0.5,-0.5,-0.5) -- (0.5,0.5,-0.5) -- (0.5,0.5,0.5) -- (0.5,-0.5,0.5) -- (-0.5,-0.5,0.5) -- cycle;

\fill [fill=red!30,fill opacity=0.5] (0.5,0.5,0) -- (0.5,-0.5,0) -- (-0.5,-0.5,0) -- (-0.5,0.5,0) -- cycle;
\fill [fill=red!30,fill opacity=0.5] (0.5,0,0.5) -- (0.5,0,-0.5) -- (-0.5,0,-0.5) -- (-0.5,0,0.5) -- cycle;
\fill [fill=red!30,fill opacity=0.5] (0,0.5,0.5) -- (0,0.5,-0.5) -- (0,-0.5,-0.5) -- (0,-0.5,0.5) -- cycle;

\draw [thick,dashed,draw=red] (0.5,0.5,0) -- (0.5,-0.5,0) -- (-0.5,-0.5,0) -- (-0.5,0.5,0) -- cycle;
\draw [thick,dashed,draw=red] (0.5,0,0.5) -- (0.5,0,-0.5) -- (-0.5,0,-0.5) -- (-0.5,0,0.5) -- cycle;
\draw [thick,dashed,draw=red] (0,0.5,0.5) -- (0,0.5,-0.5) -- (0,-0.5,-0.5) -- (0,-0.5,0.5) -- cycle;

\draw[thick,draw=colUniBwGr] (-0.5,-0.5,-0.5) -- (0.5,-0.5,-0.5) -- (0.5,0.5,-0.5) -- (-0.5,0.5,-0.5) -- cycle;
\draw[thick,draw=colUniBwGr] (-0.5,-0.5,-0.5) -- (0.5,-0.5,-0.5) -- (0.5,0.5,-0.5) -- (0.5,0.5,0.5) -- (0.5,-0.5,0.5) -- (-0.5,-0.5,0.5) -- cycle;
\draw[thick,draw=colUniBwGr] (-0.5,-0.5,-0.5) -- (-0.5,0.5,-0.5) -- (0.5,0.5,-0.5);
\draw[thick,draw=colUniBwGr] (0.5,-0.5,-0.5) -- (0.5,-0.5,0.5);

\end{scope}

\end{tikzpicture}
\caption{Intersections of all cubes (IDs 1--9) of the weak scaling studies with planes spanned
by basis vectors~$\teo{e}_{\indBaseVecOne}$ and~$\teo{e}_{\indBaseVecTwo}$, $\indBaseVecOne,\indBaseVecTwo\in\{1,2,3\}, \indBaseVecOne\neq \indBaseVecTwo$
in the cartesian frame of reference.
Orientation of the cutting planes is sketched in the top left.}
\label{fig:WeakScalingGeometry}
\end{figure}

To study the performance of the proposed block preconditioner also for large-scale examples and on parallel
computing clusters, we now conduct a weak scaling study.
The problem setup is similar to test case~I from \secref{sec:NumericalStudySpai}.
To guarantee that large problems exhibit the same fiber distribution at least in parts of the domain,
the meshes are setup as follows:
We first create the geometry and mesh for the largest problem
by placing a cube with edge length $\qty{10}{\meter}$ inside a cartesian frame of reference,
such that the cube's center of mass coincides with the origin~$O$ and its edges are oriented along the cartesian axes.
Then, the cube is filled with randomly positioned and oriented straight fibers with length~$\indexedBeam{l} = \qty{0.25}{\meter}$ and radius~$\beamRadius = \qty{0.005}{\meter}$.
Only fibers, which are fully contained in the cube, are considered.
This problem will be solved on 1000 MPI ranks.
For smaller problems,
the geometry is cut out of this initial cube.
Also for the cut out problem, only fully contained fibers are considered.
\figref{fig:WeakScalingGeometry} shows the intersections of the series of cubes with each coordinate plane spanned
by basis vectors~$\teo{e}_{\indBaseVecOne}$ and~$\teo{e}_{\indBaseVecTwo}$, $\indBaseVecOne,\indBaseVecTwo\in\{1,2,3\}, \indBaseVecOne\neq\indBaseVecTwo$.
This process not only yields an almost constant {\beamsolid} volume ratio for all problem sizes,
but also guarantees that larger meshes are just extensions of the smaller meshes.
The load per processor is kept constant at around $50k$ degrees of freedom.
Meshing details are given in \tabref{tab:WeakScalingHierarchy}.
\begin{table}
\centering
\caption{Mesh refinement schedule for the weak scaling study}
\label{tab:WeakScalingHierarchy}
\begin{tabular}{c c c c c c }
	\hline
	ID & $\nproc$ & $n^{S}_{DOF}$ & $n^{B}_{DOF}$ & $n^{total}_{DOF}$ & $n^{total}_{DOF/proc}$ \\
	\hline
	1  & 8    &   \num{397953} &    \num{9132} &   \num{407085} & \num{50885.6} \\
	2  & 27   &  \num{1316928} &   \num{31824} &  \num{1348752} & \num{49953.8} \\
	3  & 64   &  \num{3090903} &   \num{74916} &  \num{3165819} & \num{49465.9} \\
	4  & 125  &  \num{6001128} &  \num{146724} &  \num{6147852} & \num{49182.8} \\
	5  & 216  & \num{10328853} &  \num{257448} & \num{10586301} & \num{49010.7} \\
	6  & 343  & \num{16355328} &  \num{413976} & \num{16769304} & \num{48890.1} \\
	7  & 512  & \num{24361803} &  \num{621876} & \num{24983679} & \num{48796.2} \\
	8  & 729  & \num{34629528} &  \num{892932} & \num{35522460} & \num{48727.7} \\
	9  & 1000 & \num{47439753} & \num{1222320} & \num{48662073} & \num{48662.1} \\
	\hline
\end{tabular}
\end{table}

The boundary conditions are identical to test case~I from \secref{sec:NumericalStudySpai},
{\ie} the bottom surface is clamped and the top surface loaded with a constant tensile load of $\distributedLoad=\qty{1}{\newton/\square\meter}$.
The solid is again modeled by a {\StVenantKirchhoff} material
(Young's modulus~$\indexedSolid{\youngs} = \qty{1}{\newton/\square\meter}$, Poisson's ratio~$\indexedSolid{\poisson}=0.3$)
and discretized with first-order hexahedral finite elements.
The fibers are modeled using {\torsionfree} {\KirchhoffLove} beam elements with Young's modulus~$\indexedBeam{\youngs}= \qty{10}{\newton/\square\meter}$.
The positional coupling condition is enforced with a penalty parameter of~$\penaltyParamDisp = \qty{10}{\newton/\square\meter}$.

The overall simulation is again of quasi-static nature and imposes the total load over the course of two load steps.
The {\nonlinear} solver converges, if the {\nonlinear} residual~$\normTwo{\mao{\residualNonlinear}}$ drops below~$10^{-6}$
and the full displacement increment~$\normTwo{\Delta \mao{\disp}}$ is smaller than~$10^{-8}$.
For the solution of the linear system arising in
each {\nonlinear} iteration, a preconditioned GMRES method is applied with the proposed preconditioner.
Hereby, the parameters are set as follows: to increase the robustness of the smoothers in the parallel setting,
the number of sweeps through the preconditioner is changed to~$\indSweep=3$
as well as the number of iterations for the SPAI smoother to~$\indSpaiSweep=3$ (in contrast to the parameter choice in \secref{sec:NumericalStudySpai}).
The \gls{ac:SPAI} computation for the beam sub-matrix~$\matBeam$ uses a drop tolerance~$\spaiDropTol=10^{-8}$
and a refinement level~$\spaiRefinementLevel=2$ to enrich the sparsity pattern.
This choice is inspired by the results of the test case~I in \secref{sec:NumericalStudySpai}.
Due to the inherently parallel nature of the \gls{ac:SPAI} computation ({\cf} \secref{sec:SPAIFrobeniusNorm}),
it has a marginal influence on the weak scalability compared to the other components of the
overall preconditioner, especially the \gls{ac:AMG} components. Therefore this part of the
algorithm is kept constant without parameter variation.
The Schur complement equation is solved with an aggregation-based \gls{ac:AMG} method.
Coarsening is performed until the number of unknowns on the coarsest level drops below 6500.
The \gls{ac:AMG} hierarchy is traversed using a V-cycle with level transfer operators arising from
either \gls{ac:PAAMG} or \gls{ac:SAAMG}. On all but the coarsest level, the level smoother is chosen as ILUT
with overlap~$\iluOverlap = 1$, fill-in level~$\iluLevelFill=1$ and drop-off tolerance~$\iluDropTol=10^{-4}$.
The coarse level is solved with a direct solver using the distributed memory version of
{\superlu}~\cite{Li2003a}. The outer GMRES solver is assumed to be converged
if the relative residual~$\normTwo{\indexedIter{\mao{\residualLinear}}{\indLinIter}} / \normTwo{\indexedIter{\mao{\residualLinear}}{0}}$
falls below $10^{-8}$.

The scaling study is run on our in-house cluster
(16 nodes with 2x Intel Xeon Cascade Lake CPUs with 26 cores, 20 nodes with 2x Intel Xeon Skylake CPUs with 24 cores, 1312 cores in total, Mellanox Infiniband Interconnect).
The overall weak scaling performance is quantified by the averaged number of iterations per nonlinear Newton iteration
and the timings for setting up the preconditioner~$\tSetup$, for solving the linear system~$\tSolve$, and the total solver time~$\tTotal = \tSetup + \tSolve$.
Since the setup of the preconditioner is expected to be expensive,
we also examine the option of reusing the preconditioner throughout all Newton steps of a load step
with the aim to reduce~$\tSetup$ and, thus, also reduce~$\tTotal$.

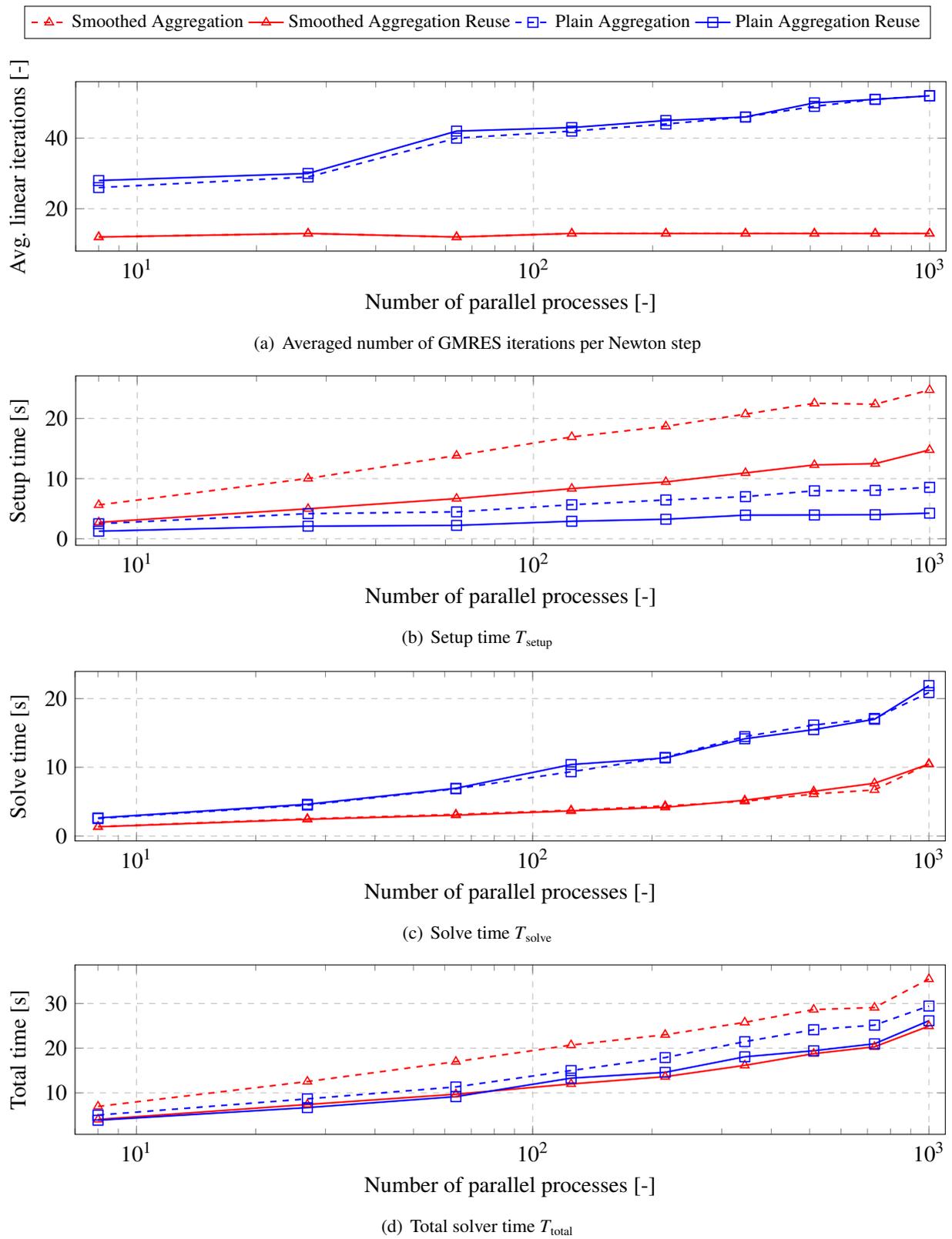
\begin{figure}
\centering

\tikzstyle{SaAmgDirect}=[thick, dashed, red, mark=triangle, mark options={scale=1.2, solid, semithick}]
\tikzstyle{SaAmgDirectReuse}=[thick, solid, red, mark=triangle, mark options={scale=1.2, solid, semithick}]
\tikzstyle{PaAmgDirect}=[thick, dashed, blue, mark=square, mark options={scale=1.2, solid, semithick}]
\tikzstyle{PaAmgDirectReuse}=[thick, solid, blue, mark=square, mark options={scale=1.2, solid, semithick}]

\begin{tikzpicture}
\begin{axis}[%
  hide axis,
  xmin=0,
  xmax=1,
  ymin=0,
  ymax=1,
  legend cell align=left,
  legend style={font=\footnotesize, at={(0.5,1.2)}, anchor=center},
  legend columns = 4
  ]

\addlegendimage{SaAmgDirect}
\addlegendentry{Smoothed Aggregation} 
\addlegendimage{SaAmgDirectReuse}
\addlegendentry{Smoothed Aggregation Reuse}

\addlegendimage{PaAmgDirect}
\addlegendentry{Plain Aggregation} 
\addlegendimage{PaAmgDirectReuse}
\addlegendentry{Plain Aggregation Reuse}

\end{axis}
\end{tikzpicture}

\subfigure[Averaged number of GMRES iterations per Newton step]{
\label{fig:WeakScalingIterationComparison}

\begin{tikzpicture}
\pgfplotstableread{data/block_weak_scaling/block_weak_eb_coarse_direct_smoothed.data}\ebCoarseDirectSmoothed
\pgfplotstableread{data/block_weak_scaling/block_weak_eb_coarse_direct_plain.data}\ebCoarseDirectPlain

\begin{axis}[%
 width = \textwidth,
 height = 4.5cm,
 xlabel={Number of parallel processes [-]},
 ylabel={Avg. linear iterations [-]},
 xmin=7,
 xmax=1100,
 xmode=log,
 ymajorgrids=true,
 xmajorgrids=true,
 grid style=dashed
]
\addplot[SaAmgDirect] table[x = num_procs, y = iterations] from \ebCoarseDirectSmoothed;
\addplot[SaAmgDirectReuse] table[x = num_procs, y = iterations_reuse] from \ebCoarseDirectSmoothed;
\addplot[PaAmgDirect] table[x = num_procs, y = iterations] from \ebCoarseDirectPlain;
\addplot[PaAmgDirectReuse] table[x = num_procs, y = iterations_reuse] from \ebCoarseDirectPlain;
\end{axis}
\end{tikzpicture}
} 

\subfigure[Setup time~$\tSetup$]{
\label{fig:WeakScalingSetupTimeComparison}

\begin{tikzpicture}
\pgfplotstableread{data/block_weak_scaling/block_weak_eb_coarse_direct_smoothed.data}\ebCoarseDirectSmoothed
\pgfplotstableread{data/block_weak_scaling/block_weak_eb_coarse_direct_plain.data}\ebCoarseDirectPlain

\begin{axis}[%
 width = \textwidth,
 height = 4.5cm,
 xlabel={Number of parallel processes [-]},
 ylabel={Setup time [$\unit{\second}$]},
 xmin=7,
 xmax=1100,
 xmode=log,
 ymajorgrids=true,
 xmajorgrids=true,
 grid style=dashed
]
\addplot[SaAmgDirect] table[x = num_procs, y = setup_time] from \ebCoarseDirectSmoothed;
\addplot[SaAmgDirectReuse] table[x = num_procs, y = setup_time_reuse] from \ebCoarseDirectSmoothed;
\addplot[PaAmgDirect] table[x = num_procs, y = setup_time] from \ebCoarseDirectPlain;
\addplot[PaAmgDirectReuse] table[x = num_procs, y = setup_time_reuse] from \ebCoarseDirectPlain;
\end{axis}
\end{tikzpicture}
} 

\subfigure[Solve time~$\tSolve$]{
\label{fig:WeakScalingSolveTimeComparison}

\begin{tikzpicture}
\pgfplotstableread{data/block_weak_scaling/block_weak_eb_coarse_direct_smoothed.data}\ebCoarseDirectSmoothed
\pgfplotstableread{data/block_weak_scaling/block_weak_eb_coarse_direct_plain.data}\ebCoarseDirectPlain

\begin{axis}[%
 width = \textwidth,
 height = 4.5cm,
 xlabel={Number of parallel processes [-]},
 ylabel={Solve time [$\unit{\second}$]},
 xmin=7,
 xmax=1100,
 xmode=log,
 ymajorgrids=true,
 xmajorgrids=true,
 grid style=dashed
]
\addplot[SaAmgDirect] table[x = num_procs, y = solve_time] from \ebCoarseDirectSmoothed;
\addplot[SaAmgDirectReuse] table[x = num_procs, y = solve_time_reuse] from \ebCoarseDirectSmoothed;
\addplot[PaAmgDirect] table[x = num_procs, y = solve_time] from \ebCoarseDirectPlain;
\addplot[PaAmgDirectReuse] table[x = num_procs, y = solve_time_reuse] from \ebCoarseDirectPlain;
\end{axis}
\end{tikzpicture}
} 

\subfigure[Total solver time~$\tTotal$]{
\label{fig:WeakScalingTotalTimeComparison}

\begin{tikzpicture}
\pgfplotstableread{data/block_weak_scaling/block_weak_eb_coarse_direct_smoothed.data}\ebCoarseDirectSmoothed
\pgfplotstableread{data/block_weak_scaling/block_weak_eb_coarse_direct_plain.data}\ebCoarseDirectPlain

\begin{axis}[%
 width = \textwidth,
 height = 4.5cm,
 xlabel={Number of parallel processes [-]},
 ylabel={Total time [$\unit{\second}$]},
 xmin=7,
 xmax=1100,
 xmode=log,
 ymajorgrids=true,
 xmajorgrids=true,
 grid style=dashed
]
\addplot[SaAmgDirect] table[x = num_procs, y = total_time] from \ebCoarseDirectSmoothed;
\addplot[SaAmgDirectReuse] table[x = num_procs, y = total_time_reuse] from \ebCoarseDirectSmoothed;
\addplot[PaAmgDirect] table[x = num_procs, y = total_time] from \ebCoarseDirectPlain;
\addplot[PaAmgDirectReuse] table[x = num_procs, y = total_time_reuse] from \ebCoarseDirectPlain;
\end{axis}
\end{tikzpicture}
} 

\caption{Weak scaling behavior}
\label{fig:WeakScalingResults}
\end{figure}
\Figref{fig:WeakScalingResults} summarizes the results of the weak scaling study.
We first discuss the case where the preconditioner is built in every Newton step.
Looking at the iteration counts in \figref{fig:WeakScalingIterationComparison},
the \gls{ac:SAAMG} method delivers iteration counts independent of the problem size (with mostly 13 iterations per solve),
while \gls{ac:PAAMG} exhibits an increase in iterations by a factor of $2\times$ (from 26 to 52 iterations) when increasing the problem size by $120\times$,
{\ie} from mesh ID 1 to mesh ID 9, {\cf} \tabref{tab:WeakScalingHierarchy}.
Regarding the setup time~$\tSetup$ required to build the preconditioner shown in \figref{fig:WeakScalingSetupTimeComparison},
\gls{ac:PAAMG} is more than twice as fast as \gls{ac:SAAMG}
due to the smaller support of \gls{ac:PAAMG} interpolation functions and, thus, less fill-in in coarse level operators.
In contrast,
the time to solve the linear system is more than $2\times$ smaller for \gls{ac:SAAMG}
due to the better approximation properties of smoothed interpolation functions in \gls{ac:SAAMG},
{\cf} \figref{fig:WeakScalingSolveTimeComparison}.
When looking at the combined time~$\tTotal = \tSetup + \tSolve$ as shown in \figref{fig:WeakScalingTotalTimeComparison},
both types of transfer operators result in very similar timings.

With a moderate increase of~$\tSetup$ for \gls{ac:PAAMG} for an increasing number of parallel processes,
the increase in solver time~$\tSolve$ as well as total time~$\tTotal$ of the \gls{ac:PAAMG} scheme
appears to be directly linked to the number of iterations required to achieve the desired tolerance of the iterative linear solver.
In contrast,
\gls{ac:SAAMG} requires a rather constant number of iterations for all problem sizes and spends most of its time in the preconditioner setup,
thus resulting in total solver timings~$\tTotal$ that are dominated by the preconditioner setup time.
This hints at potential savings when setting up the preconditioner once and then reusing it to solve multiple subsequent linear systems,
{\eg} through the course of a Newton scheme. 

We now look at the option of building the preconditioner only in the first Newton step and then reusing it throughout an entire load step.
In the present study, each load step requires two Newton steps,
hence we expect to save 50\% of~$\tSetup$ and hope to not worsen in terms of iteration numbers and solver time~$\tSolve$.
When looking at the setup time in \figref{fig:WeakScalingSetupTimeComparison},
setup costs for both \gls{ac:SAAMG} and \gls{ac:PAAMG} are reduced by $\approx 50\%$ as expected.
Furthermore, the iteration numbers as well as the solver time~$\tSolve$ stay nearly the same for both \gls{ac:SAAMG} and \gls{ac:PAAMG},
{\cf} \figsref{fig:WeakScalingIterationComparison}{fig:WeakScalingSolveTimeComparison}.
Overall, the reuse of the preconditioner positively impacts the total solver time~$\tTotal$
with the best option being \gls{ac:SAAMG} with reuse of the preconditioner,
which appears to be roughly~$30\%$ faster than the variant without reuse of the preconditioner.
We note that the actual benefit of reusing the preconditioner depends on the number of nonlinear solver iterations per load step:
The more Newton steps are required, the greater savings are to be expected from reusing the preconditioner.

In conclusion,
this example has shown weak scalability of the proposed preconditioner.
The iteration numbers remain perfectly constant for \gls{ac:SAAMG}.
Due to the overlap~$\iluOverlap = 1$ of the ILUT smoother in the \gls{ac:MG} hierarchy,
the setup time~$\tSetup$ increases by a factor of $\approx 5\times$ when increasing the problem size by~$120\times$,
{\ie} from mesh ID 1 to mesh ID 9, {\cf} \tabref{tab:WeakScalingHierarchy}.
We stress that the choice of transfer operators has a
great influence on the total weak scalability of the preconditioner. In the presented test cases, the best scalability in terms of the iteration count
were obtained by \gls{ac:SAAMG} transfer operators and when reusing the preconditioner though all Newton steps of a load step.
In sum, \gls{ac:SAAMG} appears as the method of choice to demonstrate weak scalability and keep the iteration count low.
For more complex application scenarios however, it might be beneficial to fall back to \gls{ac:PAAMG}
due to its reduced fill-in during coarsening.
Given the intricate nature of {\beamsolid} applications and their arising systems of linear equations,
we deem the present weak scaling behavior acceptable and adequate.

\subsection{Robustness of the preconditioner under varying physical parameters}
\label{sec:RobustnessStudy}

To assess the preconditioner's robustness, we now study its behavior {\wrt} iteration numbers of the linear solver under changes of critical physical parameters.
Therefore, a composite plate with four fiber layers is considered, where the {\beamsolid} stiffness ratio as well as the  beam radius are varied.
This allows to cover a wide range of possible parameter combinations for the coupled problem.

\begin{figure}
	\hfill
	\subfigure[Problem setup of the fiber-reinforced composite plate]{
	\begin{tikzpicture}
	\node at (0,0) {\includegraphics[scale=1]{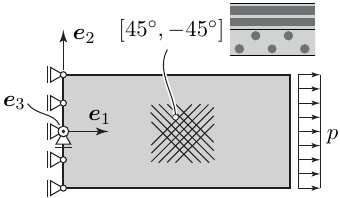}};
	\node [fill=white] at (2.8, -0.65) {$\distributedLoad$};
	\end{tikzpicture}
	}
	\hfill
	\subfigure[Deformed configuration of the plate for configuration for~$\indexedBeam{\youngs}=\qty{8}{\giga\pascal}$ and~$\beamRadius=\qty{0.004}{\meter}$]{
	\begin{tikzpicture}
	\node at (0,0) {\includegraphics[scale=0.07]{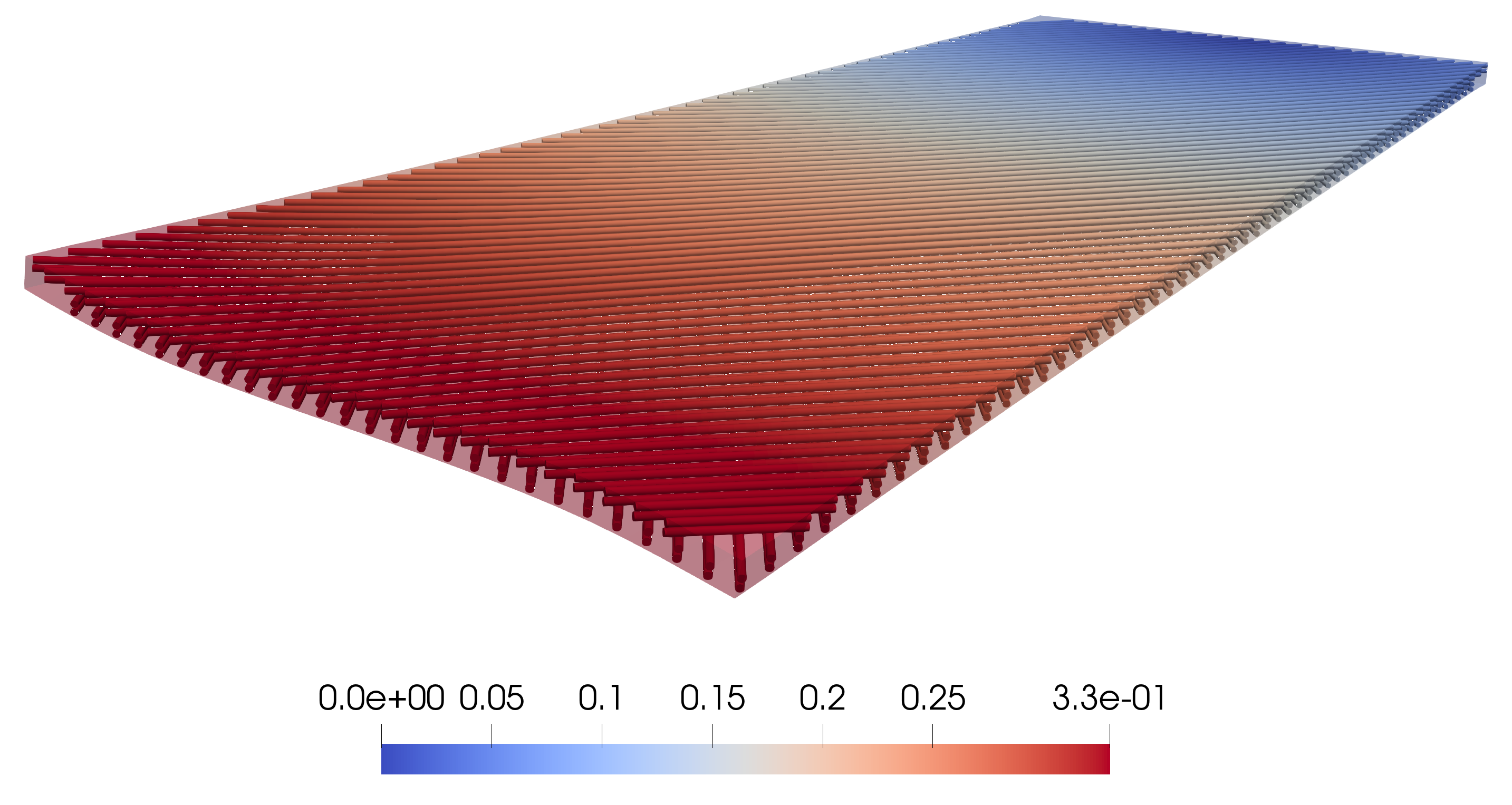}};
	\node at (0, -2.0) {\footnotesize $||d||$};
	\end{tikzpicture}
	}
    \hfill~
	\caption{Model problem for the robustness study.
	The left figure is adapted from our previous work~\cite{Steinbrecher2020a}, permissions granted under the Creative Commons~\mbox{(CC BY)} license.}
	\label{fig:RobustnessStudyProblem}
\end{figure}

The geometrical setup is identical to the composite plate presented in our prior work \cite{Steinbrecher2020a} with a length of $\qty{2}{\metre}$,
a width of $\qty{1}{\metre}$ and a total thickness of $\thickness=\qty{0.04}{\metre}$, where two layers are oriented in
$\qty{45}{\degree}$ and $\qty{-45}{\degree}$ angles, respectively.
The solid bulk domain is modeled as {\StVenantKirchhoff} material with fixed constitutive properties
($\indexedSolid{\youngs}=\qty{1}{\giga\pascal}$, $\indexedSolid{\poisson}=0.3$).
The embedded fibers are modeled as {\torsionfree} {\KirchhoffLove} beams,
where we vary the beam radius~$\beamRadius\in\{\qty{0.001}{\meter}, \qty{0.002}{\meter}, \qty{0.004}{\meter}, \qty{0.008}{\meter}\}$
and the beam's Young's modulus~$\indexedBeam{\youngs}\in\{\qty{2}{\giga\pascal}, \qty{8}{\giga\pascal}, \qty{32}{\giga\pascal}, \qty{128}{\giga\pascal}, \qty{256}{\giga\pascal}, \qty{512}{\giga\pascal}\}$.
The examined {\beamsolid} stiffness ratios~$\indexedBeam{\youngs}/\indexedSolid{\youngs}$ span a wide array of practical applications, ranging from natural fiber composites with low ratios to steel-reinforced concrete and carbon fiber composites with high ratios.
The penalty parameter is chosen as~$\penaltyParamDisp=\indexedBeam{\youngs}$ to properly enforce the positional coupling constraints.
Considering boundary conditions, the left side of the plate is fixed and a distributed tensile load~$\distributedLoad$ is applied to the right side.
The problem setup and the deformed configuration (exemplary for~$\indexedBeam{\youngs}=\qty{8}{\giga\pascal}$ and~$\beamRadius=\qty{0.004}{\meter}$)
are depicted in \figref{fig:RobustnessStudyProblem}.
For a more detailed description of the problem, the reader is referred to \cite{Steinbrecher2020a}.
Since the overall compound plate stiffness changes with the different parameter combinations described above,
the load~$\distributedLoad$ is adapted such that the axial deformations of the plate are the same for each parameter combination and match the example shown in \cite{Steinbrecher2020a}.
The overall simulation is of quasi-static nature and the load is applied incrementally over the course of 10 load steps.

The nonlinear solver converges for all test cases if the nonlinear residual~$\normTwo{\mao{\residualNonlinear}}$ drops
below~$10^{-6}$ and the displacement increment~$\normTwo{\Delta \mao{\disp}}$ falls below~$10^{-6}$.
For the solution of the linear system arising in
each {\nonlinear} iteration, a preconditioned GMRES method is applied with the proposed preconditioner.
Hereby, the parameters for the block method are similar to test case~I from \secref{sec:NumericalStudySpai}:
We set the number of applications of the preconditioner per linear solver iteration to~$\indSweep=1$,
build the preconditioner once per load step and then reuse it in every nonlinear iteration of this load step.
Due to the rather long fibers appearing in this example, the settings for the SPAI smoother are changed to~$\indSpaiSweep=3$.
In addition, the \gls{ac:SPAI} computation for the beam sub-matrix~$\matBeam$ uses a drop tolerance~$\spaiDropTol=10^{-8}$
and an increased refinement level~$\spaiRefinementLevel=4$ to handle the larger individual beam sub-blocks properly.
Similar to before, the Schur complement equation is solved with an aggregation-based \gls{ac:AMG} method.
Coarsening is performed until the number of unknowns on the coarsest level drops below 6500.
The \gls{ac:AMG} hierarchy is traversed using a V-cycle with level transfer operators arising from \gls{ac:SAAMG}.
On all but the coarsest level, the level smoother is chosen as ILUT
with overlap~$\iluOverlap = 1$, an increased fill-in level~$\iluLevelFill=2.5$ (again necessary due to the length of the fibers
and their discretization with sometimes more than ten beam elements, resulting in a denser Schur complement matrix) and
drop-off tolerance~$\iluDropTol=10^{-4}$. The coarse level is solved with a direct solver using the distributed memory
version of {\superlu}~\cite{Li2003a}. The linear solver is assumed to be converged
if the relative residual~$\normTwo{\indexedIter{\mao{\residualLinear}}{\indLinIter}} / \normTwo{\indexedIter{\mao{\residualLinear}}{0}}$
falls below $10^{-6}$. All simulations are done in serial on a single processor.
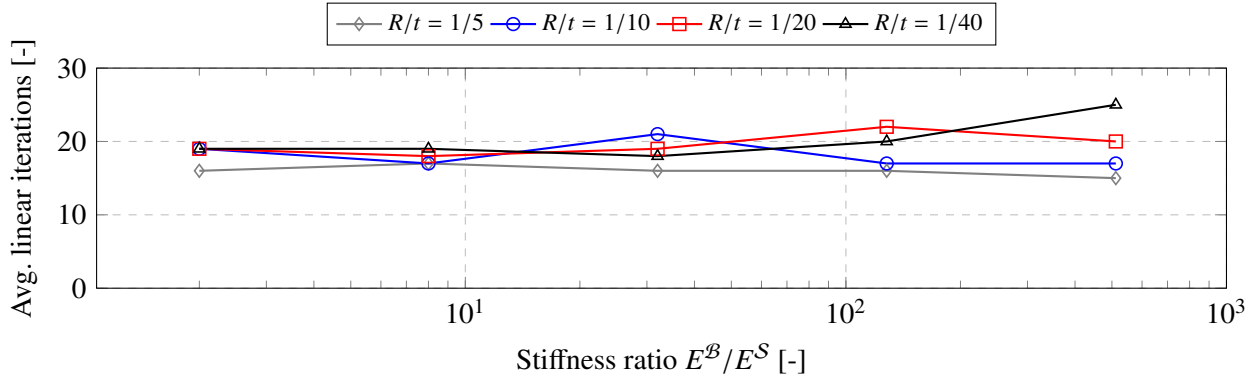
\begin{figure}
\tikzstyle{thicknessToRadiusFive}=[thick, solid, gray, mark=diamond, mark options={scale=1.2, solid, semithick}]
\tikzstyle{thicknessToRadiusTen}=[thick, solid, blue, mark=o, mark options={scale=1.2, solid, semithick}]
\tikzstyle{thicknessToRadiusTwenty}=[thick, solid, red, mark=square, mark options={scale=1.2, solid, semithick}]
\tikzstyle{thicknessToRadiusFourty}=[thick, solid, black, mark=triangle, mark options={scale=1.2, solid, semithick}]
\begin{tikzpicture}
\begin{axis}[%
 width = \textwidth,
 height = 4.5cm,
 xlabel={Stiffness ratio $\stiffnessRatio$ [-]},
 ylabel={Avg. linear iterations [-]},
 xmin=0,
 xmax=1000,
 ymin=0,
 ymax=30,
 xmode=log,
 ymajorgrids=true,
 xmajorgrids=true,
 grid style=dashed,
 legend cell align=left,
 legend style={font=\footnotesize, at={(0.5,1.2)}, anchor=center},
 legend columns = 4
]
\legend{$\geometryRatio=1/5$, $\geometryRatio=1/10$, $\geometryRatio=1/20$, $\geometryRatio=1/40$}
\addplot [thicknessToRadiusFive] table {
2   16
8   17
32  16
128 16
512	15
}; 
\addplot [thicknessToRadiusTen] table {
2   19
8   17
32  21
128 17
512	17
}; 
\addplot [thicknessToRadiusTwenty] table {
2   19
8   18
32  19
128 22
512	20
}; 
\addplot [thicknessToRadiusFourty] table {
2   19
8   19
32  18	
128 20
512	25
}; 
\end{axis}
\end{tikzpicture}
\caption{Robustness study of the preconditioning regarding the beam to solid stiffness ratio and plate thickness to beam radius ratio:
Iteration numbers are not affected by changes in physical parameters.}
\label{fig:RobustnessStudy}
\end{figure}

Averaged iteration counts of the linear solver are shown in \figref{fig:RobustnessStudy}.
At global scope, the number of iterations appears to be independent of the stiffness ratio~$\stiffnessRatio$
as well as the geometric ratio~$\geometryRatio$.
This is particularly true for a larger beam radius, {\ie} $\geometryRatio = 1/5$.
While smaller beam radii are expected to be more challenging to handle,
their impact on the solver performance is very limited:
the increase in iterations is very small compared to the case of~$\geometryRatio = 1/5$.
Intermediate geometric ratios~$\geometryRatio = \{1/10, 1/20\}$ exhibit slight changes in the iteration number
under an increasing stiffness ratio~$\stiffnessRatio$,
though neither outliers nor a trend towards increasing iteration numbers has been observed.
Considering the smallest beam radius, {\ie} $\geometryRatio = 1/40$,
the iteration count remains rather constant for all but the largest stiffness ratio.
Only the largest stiffness ratio~$\stiffnessRatio = \qty{512}{\giga\pascal}$
in combination with the smallest radius results in an outliner in iteration numbers
that is slightly above all other cases.
The challenges of small radii for the iterative solvers have also been reported in~\cite{Dimola2023a}.
Overall, this study reveals a satisfying robustness of the proposed preconditioner in relevant application scenarios.

\subsection{Application: Concrete wall with steel reinforcements}
\label{sec:ApplicationExample}

\begin{figure}
\centering
\subfigure[Measurements (unit: $\unit{\meter}$) of the concrete wall with a thickness of $\qty{0.24}{\meter}$]{

\begin{tikzpicture}

\def\dist{0.1}
\def\auxlen{0.3}

\tikzstyle{domainline}=[thick]
\tikzstyle{hiddenline}=[thick,dashed]

\tikzstyle{auxline}=[dashed]

\tikzstyle{measline}=[thick, {latex}-{latex}]
\tikzstyle{measauxline}=[thick]

\draw [domainline, fill=colUniBwGr!50] (0,0) -- (3.2,0) -- (3.2,2) -- (4.2,2) -- (4.2,0) -- (5,0) -- (5,2.4) -- (0,2.4) -- cycle;
\draw [domainline, fill=white] (1,1) -- (2,1) -- (2,2) -- (1,2) -- cycle;

\foreach \x in {0,1,2,3.2,4.2,5}{
  \draw [measauxline] (\x,-\dist) -- (\x,-\auxlen-2*\dist);
}

\draw [measline] (0,-\auxlen-\dist) -- (1,-\auxlen-\dist) node[below,pos=0.5] {1};
\draw [measline] (1,-\auxlen-\dist) -- (2,-\auxlen-\dist) node[below,pos=0.5] {1};
\draw [measline] (2,-\auxlen-\dist) -- (3.2,-\auxlen-\dist) node[below,pos=0.5] {1.2};
\draw [measline] (3.2,-\auxlen-\dist) -- (4.2,-\auxlen-\dist) node[below,pos=0.5] {1};
\draw [measline] (4.2,-\auxlen-\dist) -- (5,-\auxlen-\dist) node[below,pos=0.5] {0.8};

\draw [auxline] (1,\dist) -- (1,1-\dist);
\draw [auxline] (2,\dist) -- (2,1-\dist);

\foreach \y in {0,1,2,2.4}{
  \draw [measauxline] (-\dist,\y) -- (-\auxlen-2*\dist,\y);
}

\draw [measline] (-\auxlen-\dist,0) -- (-\auxlen-\dist,1) node[left,pos=0.5] {1};
\draw [measline] (-\auxlen-\dist,1) -- (-\auxlen-\dist,2) node[left,pos=0.5] {1};
\draw [measline] (-\auxlen-\dist,2) -- (-\auxlen-\dist,2.4) node[left,pos=0.5] {0.4};

\draw [auxline] (\dist,1) -- (1-\dist,1);
\draw [auxline] (\dist,2) -- (1-\dist,2);
\draw [auxline] (2+\dist,2) -- (3.2-\dist,2);

\end{tikzpicture}}
\hfill
\subfigure[Concrete wall with steel reinforcements]{\label{fig:ConcreteWallSetupB}
\begin{tikzpicture}
\node at (0,0) {\includegraphics[width=250pt]{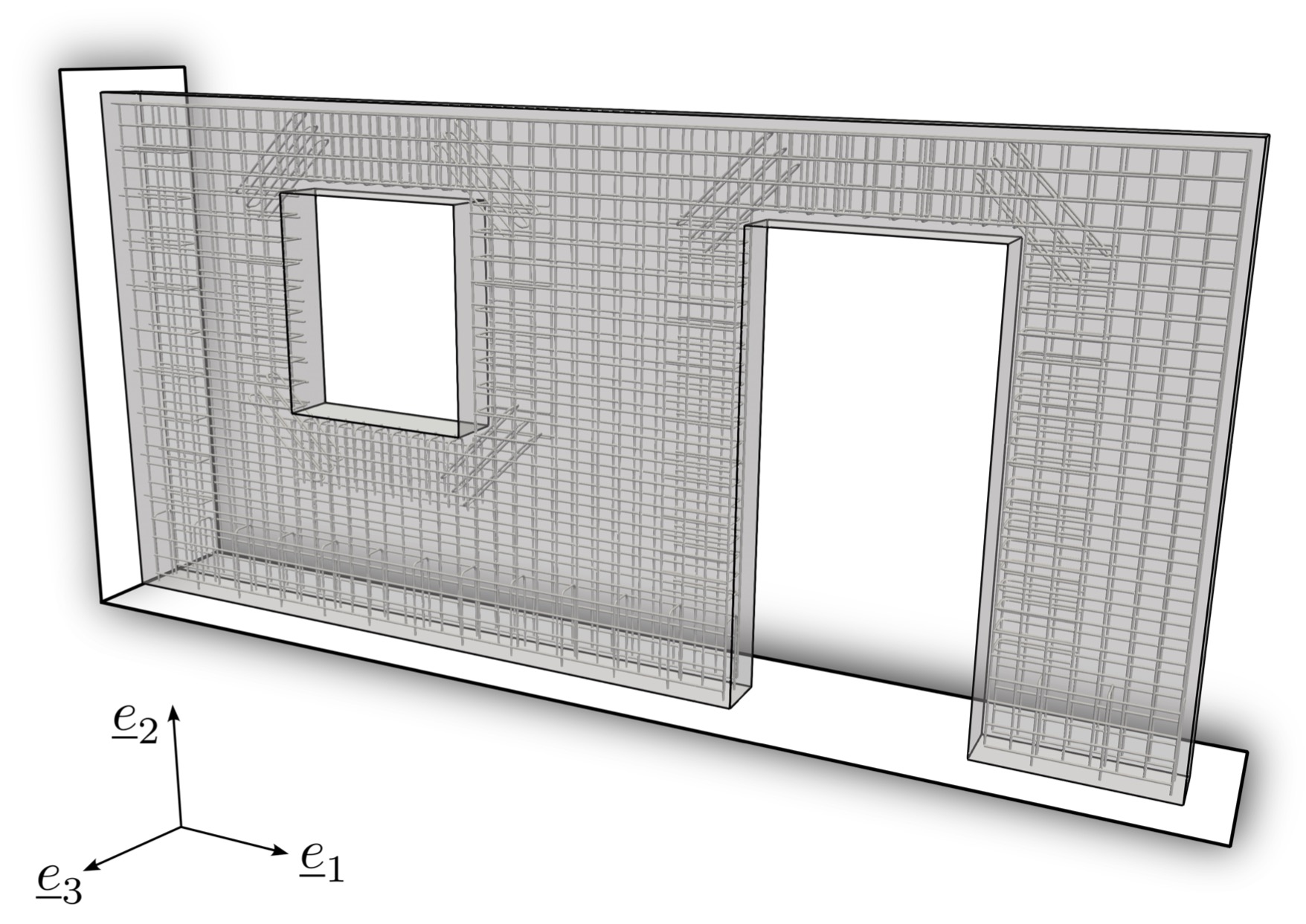}};
\node [fill=white] at (-2.1,-2.7) {$\teo{e}_1$};
\node [fill=white] at (-3.55,-1.8) {$\teo{e}_2$};
\node [fill=white] at (-4.1,-2.8) {$~$};
\node [fill=white] at (-3.9,-3.0) {$\teo{e}_3$};
\end{tikzpicture}}
\caption{Geometric configuration of a steel-reinforced concrete wall}
\label{fig:ConcreteWall}
\end{figure}
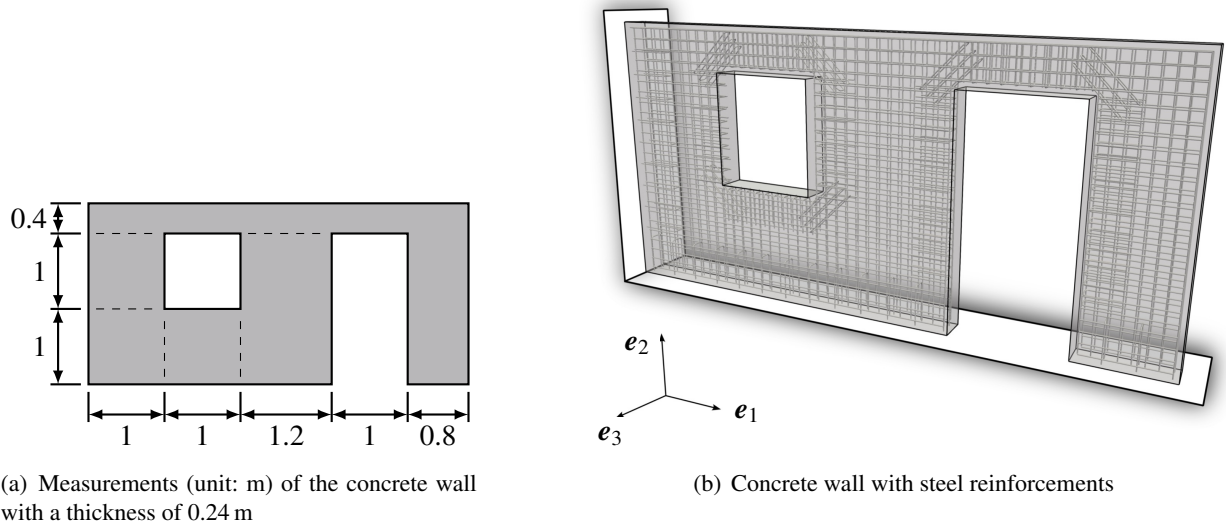

To show the applicability of the proposed block preconditioner to real-world problems,
we study an example from civil engineering,
in particular the loading of a steel-reinforced concrete wall.
The problem setup and its dimensions are shown in \figref{fig:ConcreteWall}.

Since we are focusing on the performance of the linear solver and the proposed preconditioner,
we restrict ourselves in this example to an idealized, fully elastic constitutive behavior of concrete.
Therefore, it suffices to model the solid with a {\StVenantKirchhoff} material
($\indexedSolid{\youngs}=\qty{30}{\giga\pascal}$, $\indexedSolid{\poisson}=0.3$).
In this study, we, thus, refrain from using more elaborate constitutive models,
that also cover inelastic effects, such as the Drucker--Prager model~\cite{Drucker1952a},
which could serve as a smeared, phenomenological model for concrete undergoing damage or crack initiation.
Yet, we study a complex reinforcement design:
The initially curved reinforcement bars are modeled as {\SimoReissner} beams
($\indexedBeam{\youngs}=\qty{210}{\giga\pascal}$, $\indexedBeam{\poisson}=0.0$)
with a fiber cross section radius of~$\beamRadius=\qty{0.005}{\meter}$.
The bottom and left side of the wall are clamped, restricting the displacement of the
solid and fibers. Additionally, a distributed load of~$\qty{-3e7}{\newton/\square\meter}$
is applied on the top surface in~$\teo{e}_2$ direction and~$\qty{1.5e5}{\newton/\square\meter}$
on the back surface in~$\teo{e}_3$ direction.
The reinforced concrete wall is discretized with first-order hexahedral finite elements for the solid domain
and {\SimoReissner} beam elements for the reinforcement fibers, respectively.
We study this problem for three different mesh sizes (coarse, medium, fine),
where each refinement quadruples the total number of unknowns.
Details on these meshes including the resulting number of degrees of freedom for each field are given in \tabref{tab:ConceteWallSolverTimings}
along with the computational resources, {\ie} number of MPI ranks~$\nproc$, used for each mesh.
The penalty parameters for positional and rotational couplings are set
to~$\penaltyParamDisp=\qty{2.1e11}{\newton/\square\meter}$
and~$\penaltyParamRot=\qty{5.25e6}{\newton\meter/\meter}$, respectively.
We perform a quasi-static simulation and impose the total load over the course of four load steps.

Exemplarily, \figref{fig:ConcreteWallSparsityPattern} depicts the sparsity pattern of the system matrix for the coarse mesh.
\begin{figure}
\centering
\input{figures/sparsity_pattern_concrete_wall_I.tex}
\caption{Sparsity pattern of the linear system arising from the coarse mesh of the concrete wall ({\cf} \tabref{tab:ConceteWallSolverTimings})
with the block-diagonal beam {\subblock}~$\matBeam$ (orange),
the solid {\subblock}~$\matSolid$ (white),
and the off-diagonal coupling blocks~$\trans{\matBeamSolid}$ and~$\matSolidBeam$ (gray)}
\label{fig:ConcreteWallSparsityPattern}
\end{figure}
The $2\times 2$ blocking as introduced in \eqref{eq:SimpleLinearSystem} is highlighted through colors.
The beam {\subblock}~$\matBeam$ (orange) internally exhibits block diagonal structure as illustrated in \figref{fig:SparsityTestCases},
which allows for an efficient construction of the \gls{ac:SPAI}, since the sparsity pattern of the inverse can be estimated very well based on the block-diagonal structure of~$\matBeam$.
Arising from the penalty contributions assembled into~$\matSolid$, the solid {\subblock}~$\matSolid$ (white) contains many entries far away from its diagonal,
exemplifying the attested lack of diagonal dominance, {\cf} \secref{sec:BlockDiagonalDominance}.
Both beams and solid are connected through the coupling blocks~$\trans{\matBeamSolid}$ and~$\matSolidBeam$ (gray),
which are the main reasons for the loss of block diagonal dominance, again see \secref{sec:BlockDiagonalDominance}.

For all simulations, the nonlinear solver converges
if the nonlinear residual~$\normTwo{\mao{\residualNonlinear}}$ drops below~$10^{-8}$
and the displacement increment~$\normTwo{\Delta \mao{\disp}}$ falls below~$10^{-8}$.
To get an idea of the effectiveness of the proposed preconditioner in an application scenario,
we compare different methods to solve the arising linear system in each Newton step,
in particular a direct solver (by applying the distributed memory version of {\superlu}~\cite{Li2003a}),
a naive approach represented by a GMRES solver with an \gls{ac:ILU} preconditioner,
and a GMRES solver with the proposed block preconditioner from \secref{sec:Preconditioner}.
Where applicable, we also study the effect of reusing the preconditioner over multiple invocations of the linear solver within a single load step
to better amortize the potentially expensive setup of the block preconditioner.
In case of GMRES, we assume convergence of the linear solver,
if the full relative residual norm~$\normTwo{\indexedIter{\mao{\residualLinear}}{\indLinIter}} / \normTwo{\indexedIter{\mao{\residualLinear}}{0}}$
drops to at least $10^{-6}$.

In case of the block preconditioner,
we apply $\indSweep = 3$ sweeps of the proposed preconditioner within each GMRES iteration.
The \gls{ac:SPAI} computation for the beam sub-matrix~$\matBeam$ uses a drop tolerance~$\spaiDropTol=10^{-8}$
and a refinement level~$\spaiRefinementLevel=4$ to enrich the sparsity pattern.
For the predictor and corrector step, the described SPAI smoother is applied with $\indSpaiSweep=3$ sweeps.
The Schur complement equation is tackled with a
\gls{ac:PAAMG} hierarchy with a maximum size of the coarse level problem of 6500 unknowns,
which results in three \gls{ac:MG} levels for all meshes.
Prolongator smoothing (as proposed by~\cite{Vanek1996a} and usually beneficial for problems in solid mechanics) is explicitly disabled
to reduce the fill-in of the coarse level matrices; see~\cite{Thomas2019a} for a detailed comparison.
An ILUT method with~$\iluOverlap=1$, $\iluLevelFill=2.5$, and~$\iluDropTol=10^{-4}$ is used as level smoother.
The coarse level equations are solved directly using the distributed memory version of {\superlu}~\cite{Li2003a}.

The solver options and their iteration counts and timings are summarized in \tabref{tab:ConceteWallSolverTimings}.
\begingroup
\setlength{\tabcolsep}{0.3em}
\begin{table}
\centering
\caption{Comparison of averaged linear solver timings per nonlinear iteration for a steel-reinforced concrete wall
(methods: \textit{Direct} --- direct solver based on an LU factorization,
\textit{Naive} --- a naive-preconditioned GMRES solver using an \gls{ac:ILU} factorization as preconditioner,
 \textit{Block} --- a GMRES solver using the block preconditioner proposed in \secref{sec:Preconditioner})}
\label{tab:ConceteWallSolverTimings}
\begin{tabular}{c c c c c c c c c c c}
\hline
$\nproc$ & $n^{\indexSolid}_{DOF}$ & $n^{\indexBeam}_{DOF}$ & $n^{total}_{DOF}$ & Method & Reuse & $\#$iter & \multicolumn{3}{c}{CPU time ($\unit{\second}$)} & Speedup \\
~ & ~ & ~ & ~ & ~ & ~ & ~ & $\tSetup$ & $\tSolve$ & $\tTotal$ & $\frac{\tTotal^{direct}}{\tTotal^{iterative}} $ \\
\hline
\multirow{4}{*}{8} & \multirow{4}{*}{\num{180888}} & \multirow{4}{*}{\num{45894}} & \multirow{4}{*}{\num{226782}} & Direct   & no & - & 1070 & 2.8 & 1072.8 & ~ \\
~ & ~ & ~ & ~ & Naive    & no & \multicolumn{5}{c}{no convergence} \\
~ & ~ & ~ & ~ & Block & no  & 25 & 100.9 & 13.9 & 114.8 & $\approx$~9 \\
~ & ~ & ~ & ~ & Block & yes & 26 & 19.4 & 13.3 & 32.7 & $\approx$~33 \\
\hline
\multirow{4}{*}{32} & \multirow{4}{*}{\num{844545}} & \multirow{4}{*}{\num{45894}} & \multirow{4}{*}{\num{890439}} & Direct & no  & - & 5358 & 8.3 & 5366.3 & ~ \\
~ & ~ & ~ & ~ & Naive & no  & \multicolumn{5}{c}{no convergence} \\
~ & ~ & ~ & ~ & Block & no  & 25 & 145.9 & 17.6 & 163.5 & $\approx$~33 \\
~ & ~ & ~ & ~ & Block & yes & 26 & 29.2 & 18.9 & 48.1 & $\approx$~112 \\
\hline
\multirow{4}{*}{128} & \multirow{4}{*}{\num{3323565}} & \multirow{4}{*}{\num{45894}} & \multirow{4}{*}{\num{3369459}} & Direct & no & \multicolumn{5}{c}{not feasible} \\
~ & ~ & ~ & ~ & Naive & no  & \multicolumn{5}{c}{no convergence} \\
~ & ~ & ~ & ~ & Block & no  & 22 & 198.3 & 20.4 & 218.7 & n/a \\
~ & ~ & ~ & ~ & Block & yes & 23 & 38.5 & 21.1 & 59.6 & n/a \\
\end{tabular}
\end{table}
\endgroup
Each of the four load steps requires five Newton iterations to reach convergence of the {\nonlinear} solver.
The reported iteration counts and timings have been averaged over all load steps and Newton steps,
such that the numbers now give a good estimate for the cost of a single invocation of the linear solver.
For the direct solver, the coarse and medium mesh could be solved,
while the fine mesh was infeasible, {\ie} one load step taking more than three days of wall clock time on a cluster, such that we do not report the final result.
For the {\outofthebox} iterative solver using a GMRES method preconditioned with an incomplete LU factorization,
the iterative solver was not able to reach convergence within 1000 iterations.
Only GMRES with the proposed block preconditioner from \secref{sec:Preconditioner} was able to solve all three levels of mesh refinement.
Moreover, the iterative method
outperforms the direct solver in the sense that it is roughly~$9\times$
faster than the direct method for the coarse discretization.
When moving to the medium-sized mesh,
the discrepancy in solver timings increases even further:
the direct method requires a total solver time~$\tTotal = \qty{5366.3}{\second}$ for a single solve,
while the total solver time~$\tTotal$ of the preconditioned GMRES only takes~$\qty{163.5}{\second}$,
resulting in a {\speedup} of approximately~$33\times$. Increasing the problem size even
further makes the application of a direct method infeasible, as memory consumption and
computing time become prohibitively high, leaving the iterative approach as the sole viable option.
In all working cases, the linear iterations for the preconditioned iterative method per Newton step
remain almost constant over the whole simulation for each problem and in addition also stay
almost constant for each mesh size.
For the coarse and medium meshes, approximately~$25$ iterations are necessary to achieve the desired tolerance,
whereas the large setup requires~$22$ iterations on average.
Still, the discrepancy between~$\tSetup$ and~$\tSolve$
is rather large for the iterative method,
as most of the computation time is spent in the construction of
the factorization of the ILUT level smoother.
To better amortize the expensive setup cost,
one can build the preconditioner only once per load step and then reuse it for each Newton step with the goal of decreasing the overall simulation time.
This shows to be an effective option, as it reduces the setup time~$\tSetup$ by a factor of $5\times$ in each case,
which is perfectly in line with using the preconditioner five times, but only building it once per load step.
Due to the reuse, the preconditioner is not perfectly fitting the system matrix anymore,
occasionally resulting in a slight increase in iteration numbers (\#iter) and solver time~$\tSolve$.
Overall though,
the total solver time~$\tTotal$ is reduced, which also manifests itself in the {\speedup} factors of~$33\times$ and~$112\times$ in \tabref{tab:ConceteWallSolverTimings}
for the coarse and medium mesh.
The infeasibility of a direct solver for the fine mesh prevents the calculation of a {\speedup} factor (n/a),
however clearly testifies to the beneficial impact of the proposed preconditioner on the solvability of large-scale application examples.
For both rebuilding or reusing the preconditioner,
the iteration counts appear to be independent of the mesh size also in this example from engineering practice.

Overall, this example demonstrates not only the applicability of the proposed
preconditioner in engineering problems,
but also its benefits in terms of efficiency and {\speedup},
ultimately enabling the analysis of large and complex {\fibersolid} systems,
which have not been accessible with existing linear solvers so far.

\section{Concluding remarks}
\label{sec:Conclusion}

In this paper,
we have proposed a physics-based {\multilevel} block preconditioner for the scalable solution
of {\mixeddimensional} models in {\beamsolid} interaction,
specifically tailored to systems with many independent fibers being embedded into a solid domain.
The regularized mortar-type coupling approach leads to $2\times 2$ block systems
exhibiting particular properties,
most prominently the lack of block diagonal dominance stemming from the penalty terms on the off-diagonal coupling blocks,
which render classical block relaxation preconditioners inapplicable.
To precondition an outer Krylov solver,
we utilize an approximate block factorization to enable the tackling of the individual blocks and their coupling within the preconditioner.
To this end, we exploit the beam-related {\subblock}'s sparsity structure resembling a block diagonal matrix
to explicitly construct a \acrfull{ac:SPAI} by solving a minimization problem over the Frobenius norm on a given sparsity pattern,
which in practice is decomposed into row-wise minimizations to be solved in parallel.
To increase its robustness and approximation quality,
the \gls{ac:SPAI} computation is equipped with pre-processing steps such as a filtering of small entries and a static enrichment of the sparsity pattern
as well as a post-filtering of small entries.
This approximation has then not only been used for the explicit formation of an approximation to the Schur complement,
but also as a smoother in the {\blockLU}'s prediction and correction steps.
To solve the Schur complement equation,
we have employed an \gls{ac:AMG} hierarchy.
Due to the Schur complement's fill-in stemming from the \gls{ac:SPAI} matrix as well as the penalty contributions,
an ILUT factorization with fill-in~$\iluLevelFill$, threshold~$\iluDropTol$, and overlap~$\iluOverlap$ serves as level smoother on all levels
except for the direct solver on the coarse level.
All building blocks of the proposed preconditioner have been implemented in {\trilinos}
and are available as open-source software to the entire scientific community.

We have studied the influence of the \gls{ac:SPAI} algorithm's parameters and have found,
that a static enrichment of the graph~$\graphOf{\matBeam}$ of at least~$\spaiRefinementLevel=2$ greatly improves the quality of the approximation
as well as the performance of the iterative solver,
while more enrichment might be required for particularly challenging problems.
Regarding the scaling behavior,
we were able to demonstrate weak scalability up to 1000 \gls{ac:MPI} ranks.
While the iteration count is completely independent of the problem size and number of \gls{ac:MPI} ranks for \gls{ac:SAAMG},
the setup and solver time exhibit a minor increase with an increasing problem size.
This is mainly attributed to the use of an ILUT level smoother within the {\blockLU}'s Schur complement step.
We have demonstrated the robustness of the preconditioner by showing that the iteration counts remain constant when changing critical physical parameters
such as the stiffness ratio between fibers and bulk field or the fiber radii.
Finally, we have investigated an application example from civil engineering, in particular a steel-reinforced concrete wall,
and have compared the performance of the proposed {\multilevel} preconditioner to established one-level preconditioners and direct solvers.
Even for a coarse mesh,
the one-level preconditioner failed to converge.
The proposed {\multilevel} preconditioner delivered a {\speedup} by a factor up to $112\times$ compared to the direct solver on small and medium sized meshes,
whereas the application of a direct method on the finest mesh was not feasible anymore, leaving the proposed preconditioned iterative method as the only working option.
Again, the iteration count is independent of the mesh size.
While intractable for existing solvers,
the proposed preconditioner enables the analysis of {\mixeddimensional} {\fibersolid} systems with complex reinforcement structures for the first time.
Considering computational performance, the option of building the preconditioner only once per load step and then reusing it in every iteration of the {\nonlinear} solver
has shown to cut down the setup time~$\tSetup$ at the expected rate,
while still delivering a strong preconditioning effect,
such that the convergence of the linear solver is not impeded
and the total solver time~$\tTotal$ is reduced significantly.

In future work,
the proposed preconditioner and its building blocks can be extended to other types of {\beamsolid} interaction phenomena
such as the coupling of beams onto a solid's surface~\cite{Steinbrecher2022b} or the contact between beams and solid bodies~\cite{Steinbrecher2022a}.
Similarly, other {\mixeddimensional} {\multiphysics} systems such as {\fiberfluid} interaction are likely to be amenable to such a methodology.
It seems worthwhile to use semi-structured grids to discretize the solid domain, if applicable,
which in turn allow one to further enhance its computational performance in many application scenarios~\cite{Mayr2022a}.
Performance and scalability bottlenecks associated with the evaluation of the {\beamsolid} coupling terms on a distributed-memory parallel computing cluster need to be addressed,
{\eg} following ideas from mortar methods for contact mechanics to improve data locality and load balancing~\cite{Mayr2023a}.

\section*{Acknowledgements}
The work described in this contribution has been funded by \emph{dtec.bw - Digitalization and Technology Research Center of the Bundeswehr}
under the project ``hpc.bw - Competence Platform for High Performance Computing''.
dtec.bw is funded by the European Union – NextGenerationEU.
Partial support has been provided by the \emph{Deutsche Forschungsgemeinschaft (DFG, German Research Foundation)} 
within the project
``Stable discretization methods and scalable solvers for embedded fiber\slash{}solid coupling'' (project number 528397555)
and by the \emph{Bayerisches Verbundforschungsprogramm (BayVFP) -- Digitalisierung} within the project ``IFB-Individuelle Fließfertigung für Betonfertigteile'' (project number DIK-2201-0016 / DIK0411/03).
The authors gratefully acknowledge the computing resources provided by the Data Science \& Computing Lab at the University of the Bundeswehr Munich.

Some sketches in this work have been created using the Adobe Illustrator plug-in LaTeX2AI (\url{https://github.com/isteinbrecher/latex2ai}).

\section*{Data Availability Statement}
Building blocks of the AMG preconditioner developed and applied in this study are openly available in Trilinos at \url{https://github.com/trilinos/Trilinos} \cite{TrilinosURL,Heroux2005a}.
All other data that support the findings of this study are available from the corresponding author upon reasonable request.

\section*{Declaration of Competing Interest}

The authors declare that they have no known competing financial interests or personal relationships
that could have appeared to influence the work reported in this paper.

\bibliographystyle{abbrv}
\bibliography{bib_beam_solid_preconditioner}

\begin{thebibliography}{10}

\bibitem{BaciURL}
{4C: A Comprehensive Multi-Physics Simulation Framework}.
\newblock https://www.4c-multiphysics.org.
\newblock Accessed: 2024-06-07.

\bibitem{TrilinosURL}
{The Trilinos Project}.
\newblock https://trilinos.github.io.
\newblock Accessed: 2024-06-07.

\bibitem{Agarwal2017a}
B.~D. Agarwal, L.~J. Broutman, and K.~Chandrashekhara.
\newblock {\em {Analysis and Performance of Fiber Composites}}.
\newblock Wiley, 2017.

\bibitem{Ao2022a}
J.~Ao, M.~Zhou, and B.~Zhang.
\newblock A dual mortar embedded mesh method for internal interface problems
  with strong discontinuities.
\newblock {\em International Journal for Numerical Methods in Engineering},
  123(22):5652--5681, 2022.

\bibitem{Baerland2019a}
T.~Baerland, M.~Kuchta, and K.-A. Mardal.
\newblock {Multigrid Methods for Discrete Fractional Sobolev Spaces}.
\newblock {\em SIAM Journal on Scientific Computing}, 41(2):A948--A972, 2019.

\bibitem{Bechet2009a}
E.~B\'echet, N.~Mo\"es, and B.~I. Wohlmuth.
\newblock A stable lagrange multiplier space for stiff interface conditions
  within the extended finite element method.
\newblock {\em International Journal for Numerical Methods in Engineering},
  78(8):931--954, 2009.

\bibitem{BergerVergiat2023a}
L.~Berger-Vergiat, C.~A. Glusa, G.~Harper, J.~J. Hu, M.~Mayr, A.~Prokopenko,
  C.~M. Siefert, R.~S. Tuminaro, and T.~A. Wiesner.
\newblock {MueLu User's Guide}.
\newblock Technical Report SAND2023-12265, Sandia National Laboratories,
  Albuquerque, NM (USA) 87185, 2023.

\bibitem{Broeker2002a}
O.~Br\"{o}ker and M.~J. Grote.
\newblock Sparse approximate inverse smoothers for geometric and algebraic
  multigrid.
\newblock {\em Applied Numerical Mathematics}, 41(1):61--80, 2002.

\bibitem{Budisa2024a}
A.~Budi\v{s}a, X.~Hu, M.~Kuchta, K.-A. Mardal, and L.~Zikatanov.
\newblock Algebraic multigrid methods for metric-perturbed coupled problems.
\newblock {\em SIAM Journal on Scientific Computing}, 46(3):A1461--A1486, 2024.

\bibitem{Cerroni2019a}
D.~Cerroni, F.~Laurino, and P.~Zunino.
\newblock {Mathematical analysis, finite element approximation and numerical
  solvers for the interaction of 3D reservoirs with 1D wells}.
\newblock {\em International Journal on Geomathematics}, 10:4, 2019.

\bibitem{Chacon2008a}
L.~Chac\'{o}n.
\newblock {An optimal, parallel, fully implicit Newton--Krylov solver for
  three-dimensional viscoresistive magnetohydrodynamics}.
\newblock {\em Physics of Plasmas}, 15:056103, 2008.

\bibitem{Chow2001a}
E.~Chow.
\newblock Parallel implementation and practical use of sparse approximate
  inverse preconditioners with a priori sparsity patterns.
\newblock {\em The International Journal of High Performance Computing
  Applications}, 15(1):56--74, 2001.

\bibitem{Chow2015a}
E.~Chow and A.~Patel.
\newblock {Fine-Grained Parallel Incomplete LU Factorization}.
\newblock {\em SIAM Journal on Scientific Computing}, 37(2):C169--C193, 2015.

\bibitem{Cyr2016a}
E.~C. Cyr, J.~N. Shadid, and R.~S. Tuminaro.
\newblock {Teko: A Block Preconditioning Capability with Concrete Example
  Applications in Navier--Stokes and MHD}.
\newblock {\em SIAM Journal on Scientific Computing}, 38(5):S307--S331, 2016.

\bibitem{Dimola2023a}
N.~Dimola, M.~Kuchta, K.-A. Mardal, and P.~Zunino.
\newblock {Robust Preconditioning of mixed-dimensional PDEs on 3d-1d domains
  coupled with Lagrange multipliers}.
\newblock In A.~Linninger, K.-A. Mardal, and P.~Zunino, editors, {\em
  Quantitative Approaches to Microcirculation: Mathematical Models,
  Computational Methods and Data Analysis}, pages 137--171. Springer Nature
  Switzerland, Cham, 2024.

\bibitem{Drucker1952a}
D.~C. Drucker and W.~Prager.
\newblock Soil mechanics and plastic analysis for limit design.
\newblock {\em Quarterly of Applied Mathematics}, 10(2):157--166, 1952.

\bibitem{Durville2007a}
D.~Durville.
\newblock Finite element simulation of textile materials at mesoscopic scale.
\newblock In {\em Finite element modelling of textiles ans textile composites},
  Saint Petersburg, 2007.

\bibitem{Elman2008a}
H.~C. Elman, V.~E. Howle, J.~N. Shadid, R.~Shuttleworth, and R.~S. Tuminaro.
\newblock {A taxonomy and comparison of parallel block multi-level
  preconditioners for the incompressible Navier--Stokes equations}.
\newblock {\em Journal of Computational Physics}, 227(3):1790--1808, 2008.

\bibitem{feingold1962a}
D.~Feingold and R.~Varga.
\newblock {Block diagonally dominant matrices and generalizations of the
  Gershgorin Theorem}.
\newblock {\em Pacific Journal of Mathematics}, 12(4):1241--1250, 1962.

\bibitem{Flemisch2007a}
B.~Flemisch and B.~I. Wohlmuth.
\newblock {Stable Lagrange multipliers for quadrilateral meshes of curved
  interfaces in 3D}.
\newblock {\em Computer Methods in Applied Mechanics and Engineering},
  196(8):1589--1602, 2007.

\bibitem{Grote1997a}
M.~J. Grote and T.~Huckle.
\newblock Parallel preconditioning with sparse approximate inverses.
\newblock {\em SIAM Journal on Scientific Computing}, 18(3):838--853, 1997.

\bibitem{Hagmeyer-TwoWay}
N.~Hagmeyer, M.~Mayr, and A.~Popp.
\newblock A fully coupled regularized mortar-type finite element approach for
  embedding one-dimensional fibers into three-dimensional fluid flow.
\newblock {\em International Journal for Numerical Methods in Engineering},
  published online ahead of print:e7435, 2024.

\bibitem{Hagmeyer2022a}
N.~Hagmeyer, M.~Mayr, I.~Steinbrecher, and A.~Popp.
\newblock {One-way coupled fluid-beam interaction: Capturing the effect of
  embedded slender bodies on global fluid flow and vice versa}.
\newblock {\em Advanced Modeling and Simulation in Engineering Sciences}, 9:9,
  2022.

\bibitem{Hautefeuille2012a}
M.~Hautefeuille, C.~Annavarapu, and J.~E. Dolbow.
\newblock {Robust imposition of Dirichlet boundary conditions on embedded
  surfaces}.
\newblock {\em International Journal for Numerical Methods in Engineering},
  90(1):40--64, 2012.

\bibitem{Heroux2005a}
M.~A. Heroux, R.~A. Bartlett, V.~E. Howle, R.~J. Hoekstra, J.~J. Hu, T.~G.
  Kolda, Lehoucq, K.~R. Long, R.~P. Pawlowski, E.~T. Phipps, A.~G. Salinger,
  H.~K. Thornquist, R.~S. Tuminaro, J.~M. Willenbring, A.~Williams, and K.~S.
  Stanley.
\newblock {An Overview of the Trilinos Project}.
\newblock {\em ACM Transactions on Mathematical Software}, 31(3):397--423,
  2005.

\bibitem{Jodlbauer2019a}
D.~Jodlbauer, U.~Langer, and T.~Wick.
\newblock Parallel block-preconditioned monolithic solvers for fluid-structure
  interaction problems.
\newblock {\em International Journal for Numerical Methods in Engineering},
  117(6):623--643, 2019.

\bibitem{Kakaletsis2023}
S.~Kakaletsis, E.~Lejeune, and M.~Rausch.
\newblock The mechanics of embedded fiber networks.
\newblock {\em Journal of the Mechanics and Physics of Solids}, 181:105456,
  2023.

\bibitem{Khristenko2021a}
U.~Khristenko, S.~Schu{\ss}, M.~Kr\"uger, F.~Schmidt, B.~Wohlmuth, and
  C.~Hesch.
\newblock Multidimensional coupling: A variationally consistent approach to
  fiber-reinforced materials.
\newblock {\em Computer Methods in Applied Mechanics and Engineering},
  382:113869, 2021.

\bibitem{Kuchta2019a}
M.~Kuchta, K.-A. Mardal, and M.~Mortensen.
\newblock {Preconditioning trace coupled 3d-1d systems using fractional
  Laplacian}.
\newblock {\em Numerical Methods for Partial Differential Equations},
  35(1):375--393, 2019.

\bibitem{Kuchta2016a}
M.~Kuchta, M.~Nordaas, J.~C.~G. Verschaeve, M.~Mortensen, and K.-A. Mardal.
\newblock {Preconditioners for Saddle Point Systems with Trace Constraints
  Coupling 2D and 1D Domains}.
\newblock {\em SIAM Journal on Scientific Computing}, 38(6), 2016.

\bibitem{Langer2016a}
U.~Langer and H.~Yang.
\newblock Robust and efficient monolithic fluid-structure-interaction solvers.
\newblock {\em International Journal for Numerical Methods in Engineering},
  108(4):303--325, 2016.

\bibitem{Le2017a}
B.~L\'e, G.~Legrain, and N.~Mo\"es.
\newblock Mixed dimensional modeling of reinforced structures.
\newblock {\em Finite Elements in Analysis and Design}, 128:1--18, 2017.

\bibitem{Lespagnol2023a}
F.~Lespagnol, C.~Grandmont, P.~Zunino, and M.~A. Fernandez.
\newblock A mixed-dimensional formulation for the simulation of slender
  structures immersed in an incompressible flow.
\newblock Technical Report 98/2023, MOX Politecnico di Milano, 2023.

\bibitem{Li2003a}
X.~S. Li and J.~W. Demmel.
\newblock {SuperLU\_DIST: A scalable distributed-memory sparse direct solver
  for unsymmetric linear systems}.
\newblock {\em {ACM} Transactions on Mathematical Software}, 29(2):110--140,
  2003.

\bibitem{Mardal2011a}
K.-A. Mardal and R.~Winther.
\newblock Preconditioning discretizations of systems of partial differential
  equations.
\newblock {\em Numerical Linear Algebra with Applications}, 18(1):1--40, 2011.

\bibitem{Mayr2022a}
M.~Mayr, L.~Berger-Vergiat, P.~Ohm, and R.~S. Tuminaro.
\newblock Non-invasive multigrid for semi-structured grids.
\newblock {\em SIAM Journal on Scientific Computing}, 44(4):A2734--A2764, 2022.

\bibitem{Mayr2015a}
M.~Mayr, T.~Kl\"oppel, W.~A. Wall, and M.~W. Gee.
\newblock {A Temporal Consistent Monolithic Approach to Fluid--Structure
  Interaction Enabling Single Field Predictors}.
\newblock {\em SIAM Journal on Scientific Computing}, 37(1):B30--B59, 2015.

\bibitem{Mayr2020a}
M.~Mayr, M.~Noll, and M.~W. Gee.
\newblock A hybrid interface preconditioner for monolithic fluid-structure
  interaction solvers.
\newblock {\em Advanced Modeling and Simulation in Engineering Sciences}, 7:15,
  2020.

\bibitem{Mayr2023a}
M.~Mayr and A.~Popp.
\newblock Scalable computational kernels for mortar finite element methods.
\newblock {\em Engineering with Computers}, 39:3691--3720, 2023.

\bibitem{Meier2015a}
C.~Meier, A.~Popp, and W.~A. Wall.
\newblock {A locking-free finite element formulation and reduced models for
  geometrically exact Kirchhoff rods}.
\newblock {\em Computer Methods in Applied Mechanics and Engineering},
  290:314--341, 2015.

\bibitem{Meier2019a}
C.~A. Meier, A.~Popp, and W.~A. Wall.
\newblock {Geometrically Exact Finite Element Formulations for Slender Beams:
  Kirchhoff--Love Theory Versus {\SimoReissner} Theory}.
\newblock {\em Archives of Computational Methods in Engineering},
  26(1):163--243, 2019.

\bibitem{Patankar1972a}
S.~Patankar and D.~Spalding.
\newblock A calculation procedure for heat, mass and momentum transfer in
  three-dimensional parabolic flows.
\newblock {\em International Journal of Heat and Mass Transfer},
  15(10):1787--1806, 1972.

\bibitem{Phillips2014a}
E.~G. Phillips, H.~C. Elman, E.~C. Cyr, J.~N. Shadid, and R.~P. Pawlowski.
\newblock {A Block Preconditioner for an Exact Penalty Formulation for
  Stationary MHD}.
\newblock {\em SIAM Journal on Scientific Computing}, 36(6):B930--B951, 2014.

\bibitem{Phillips2016a}
E.~G. Phillips, J.~N. Shadid, E.~C. Cyr, H.~C. Elman, and R.~P. Pawlowski.
\newblock {Block Preconditioners for Stable Mixed Nodal and Edge finite element
  Representations of Incompressible Resistive MHD}.
\newblock {\em SIAM Journal on Scientific Computing}, 38(6):B1009--B1031, 2016.

\bibitem{Popp2010a}
A.~Popp, M.~Gitterle, M.~W. Gee, and W.~A. Wall.
\newblock {A dual mortar approach for 3D finite deformation contact with
  consistent linearization}.
\newblock {\em International Journal for Numerical Methods in Engineering},
  83(11):1428--1465, 2010.

\bibitem{Puso2004c}
M.~A. Puso.
\newblock {A 3D mortar method for solid mechanics}.
\newblock {\em International Journal for Numerical Methods in Engineering},
  59(3):315--336, 2004.

\bibitem{Puso2004a}
M.~A. Puso and T.~A. Laursen.
\newblock A mortar segment-to-segment contact method for large deformation
  solid mechanics.
\newblock {\em Computer Methods in Applied Mechanics and Engineering},
  193(6--8):601--629, 2004.

\bibitem{Puso2004b}
M.~A. Puso and T.~A. Laursen.
\newblock A mortar segment-to-segment frictional contact method for large
  deformations.
\newblock {\em Computer Methods in Applied Mechanics and Engineering},
  193(45--47):4891--4913, 2004.

\bibitem{Raghavan2005a}
P.~Raghavan and S.~Ghosh.
\newblock A continuum damage mechanics model for unidirectional composites
  undergoing interfacial debonding.
\newblock {\em Mechanics of Materials}, 37(9):955--979, 2005.

\bibitem{Saad1994a}
Y.~Saad.
\newblock Ilut: A dual threshold incomplete lu factorization.
\newblock {\em Numerical Linear Algebra with Applications}, 1(4):387--402,
  1994.

\bibitem{Saad2003a}
Y.~Saad.
\newblock {\em {Iterative Methods for Sparse Linear Systems}}.
\newblock SIAM, Philadelphia, PA, USA, 2003.

\bibitem{Saad1986a}
Y.~Saad and M.~H. Schultz.
\newblock {GMRES: A Generalized Minimal Residual Algorithm for Solving
  Nonsymmetric Linear Systems}.
\newblock {\em SIAM Journal on Scientific and Statistical Computing},
  7(3):856--869, 1986.

\bibitem{Sedlacek2012a}
M.~Sedlacek.
\newblock {\em {Sparse Approximate Inverse for Preconditioning, Smoothing, and
  Regularization}}.
\newblock PhD thesis, Technische Universit\"{a}t M\"{u}nchen, 2012.

\bibitem{Steinbrecher2022b}
I.~Steinbrecher, N.~Hagmeyer, C.~Meier, and A.~Popp.
\newblock A consistent mixed-dimensional coupling approach for 1{D} cosserat
  beams and 2{D} solid surfaces.
\newblock preprint on arXiv: \url{https://arxiv.org/abs/2210.16010}, 2022.

\bibitem{Steinbrecher2020a}
I.~Steinbrecher, M.~Mayr, M.~J. Grill, J.~Kremheller, C.~Meier, and A.~Popp.
\newblock {A mortar-type finite element approach for embedding 1{D} beams into
  3{D} solid volumes}.
\newblock {\em Computational Mechanics}, 66(6):1377--1398, 2020.

\bibitem{MeshPyWebsite}
I.~Steinbrecher and A.~Popp.
\newblock {M}esh{P}y -- {A} general purpose {3D} beam finite element input
  generator.
\newblock \url{https://imcs-compsim.github.io/meshpy}, 2021.

\bibitem{Steinbrecher2021a}
I.~Steinbrecher, A.~Popp, and C.~Meier.
\newblock {Consistent coupling of positions and rotations for embedding 1{D}
  Cosserat beams into 3{D} solid volumes}.
\newblock {\em Computational Mechanics}, 69:701--732, 2022.

\bibitem{Steinbrecher2022a}
I.~S. Steinbrecher.
\newblock {\em Mixed-dimensional finite element formulations for beam-to-solid
  interaction}.
\newblock PhD thesis, Universit{\"a}t der Bundeswehr M{\"u}nchen, 2022.

\bibitem{Thomas2019a}
S.~J. Thomas, S.~Ananthan, S.~Yellapantula, J.~J. Hu, M.~Lawson, and M.~A.
  Sprague.
\newblock A comparison of classical and aggregation-based algebraic multigrid
  preconditioners for high-fidelity simulation of wind turbine incompressible
  flows.
\newblock {\em SIAM Journal on Scientific Computing}, 41(5):S196--S219, 2019.

\bibitem{VanDoormaal1984a}
J.~P. Van~Doormaal and G.~D. Raithby.
\newblock {Enhancements of the SIMPLE Method for Predicting Incompressible
  Fluid Flows}.
\newblock {\em Numerical Heat Transfer}, 7(2):147--163, 1984.

\bibitem{Vanek1996a}
P.~Van\v{e}k, J.~Mandel, and M.~Brezina.
\newblock {Algebraic Multigrid By Smoothed Aggregation For Second And Fourth
  Order Elliptic Problems}.
\newblock {\em Computing}, 56:179--196, 1996.

\bibitem{Wiesner2021a}
T.~A. Wiesner, M.~Mayr, A.~Popp, M.~W. Gee, and W.~A. Wall.
\newblock Algebraic multigrid methods for saddle point systems arising from
  mortar contact formulations.
\newblock {\em International Journal for Numerical Methods in Engineering},
  122(15):3749--3779, 2021.

\end{thebibliography}

\end{document}